\documentclass[aps,prd,superscriptaddress,eqsecnum,nofootinbib,preprint,tightenlines,floatfix,titlepage]{revtex4-2}

\usepackage{bm}
\usepackage{graphicx}
\usepackage{rotating}
\usepackage{float}
\usepackage[pdftex]{xcolor}
\usepackage{amsmath, amssymb}
\usepackage{array}
\usepackage{rotating}
\usepackage{mathtools}
\usepackage{multirow}
\usepackage{changepage}
\usepackage[normalem]{ulem}
\usepackage{verbatim}
\usepackage{physics}
\usepackage{tabularx}
\usepackage{wasysym}

\usepackage[sort&compress]{natbib}

\usepackage[hyperfootnotes=false]{hyperref}
\hypersetup{
    colorlinks=true,     
    linkcolor=blue,      
    citecolor=blue,      
    filecolor=blue,      
    urlcolor=blue,        
    linktoc=page
}

\newcolumntype{C}{>{\centering\arraybackslash}X}
\newcolumntype{L}{>{\raggedright\arraybackslash}X}
\newcolumntype{R}{>{\raggedleft\arraybackslash}X}

\usepackage[capitalise]{cleveref}

\crefname{section}{Sec.}{Secs.}
\Crefname{section}{Section}{Sections}
\crefrangeformat{equation}{Eqs.~(#3#1#4)--(#5#2#6)}
\Crefrangeformat{equation}{Equations~(#3#1#4)--(#5#2#6)}
\crefrangeformat{figure}{Figs.~#3#1#4--#5#2#6}
\Crefrangeformat{figure}{Figures~#3#1#4--#5#2#6}

\newcommand{\panel}[1]{\textit{{#1}}}

\newcommand{\bi}{\begin{itemize}}
\newcommand{\ei}{\end{itemize}}

\newcommand{\ben}{\begin{enumerate}}
\newcommand{\een}{\end{enumerate}} 

\newcommand{\be}{\begin{equation}}
\newcommand{\ee}{\end{equation}}

\newcommand{\bea}{\begin{eqnarray}}
\newcommand{\eea}{\end{eqnarray}}

\newcommand{\MSbar}{\ensuremath{\overline{\rm MS}}}

\newcommand{\chpt}{$\chi$PT}

\newcommand{\gmTwo}{\ensuremath{(g-2)_\mu}}

\newcommand{\cov}{\operatorname{Cov}}

\newcommand{\e}{\ensuremath{\text{e}}}
\newcommand{\iunit}{\ensuremath{\text{i}}}
\renewcommand{\case}[2]{\ensuremath{{\textstyle\frac{#1}{#2}}}}
\newcommand{\half}{\ensuremath{{\case{1}{2}}}}
\newcommand{\sixth}{\ensuremath{{\case{1}{6}}}}
\newcommand{\third}{\ensuremath{{\case{1}{3}}}}
\newcommand{\Sh}{\ensuremath{\widetilde{\mathcal{S}\kern-0.15em\mathit{h}}}}
\newcommand{\Ch}{\ensuremath{\widetilde{\mathcal{C}\kern-0.15em\mathit{h}}}}

\newcommand{\amu}{a_\mu}
\newcommand{\amuHVP}{a_\mu^{\mathrm{HVP,LO}}}
\newcommand{\amuW}{a^{\mathrm W}_{\mu}}
\newcommand{\amuSD}{a^{\mathrm{SD}}_{\mu}}
\newcommand{\amuLD}{a^{\mathrm{LD}}_{\mu}}
\newcommand{\amuWTwo}{a^{\mathrm{W2}}_{\mu}}

\newcommand{\nconf}{N_{\text{conf}}}
\newcommand{\nloose}{N_{\text{loose}}}
\newcommand{\nsrc}{N_{\text{src}}}
\newcommand{\neig}{N_{\text{eig}}}

\newcommand{\amuL}{a^{ll}_{\mu}(\mathrm{conn.})}
\newcommand{\amuS}{a^{ss}_{\mu}(\mathrm{conn.})}
\newcommand{\amuC}{a^{cc}_{\mu}(\mathrm{conn.})}
\newcommand{\amuB}{a^{bb}_{\mu}(\mathrm{conn.})}
\newcommand{\amuD}{a_{\mu}(\mathrm{disc.})}
\newcommand{\amuSIB}{\Delta a_{\mu}^{ud}(\mathrm{SIB})}
\newcommand{\amuSIBC}{\Delta a_{\mu}^{ud}(\mathrm{SIB,conn.})}
\newcommand{\amuSIBD}{\Delta a_{\mu}^{ud}(\mathrm{SIB,disc.})}
\newcommand{\amuQED}{\Delta a_{\mu}(\mathrm{QED})}

\newcommand{\amuLW}{a^{ll,\,{\mathrm W}}_{\mu}(\mathrm{conn.})}
\newcommand{\amuSW}{a^{ss,\,{\mathrm W}}_{\mu}(\mathrm{conn.})}
\newcommand{\amuCW}{a^{cc,\,{\mathrm W}}_{\mu}(\mathrm{conn.})}
\newcommand{\amuBW}{a^{bb,\,{\mathrm W}}_{\mu}(\mathrm{conn.})}
\newcommand{\amuDW}{a^{{\mathrm W}}_{\mu}(\mathrm{disc.})}
\newcommand{\amuSIBW}{\Delta a_{\mu}^{ud,\,{\mathrm W}}(\mathrm{SIB})}
\newcommand{\amuSIBCW}{\Delta a_{\mu}^{ud, \textrm{W}}(\mathrm{SIB,conn.})}
\newcommand{\amuSIBDW}{\Delta a_{\mu}^{ud, \textrm{W}}(\mathrm{SIB,disc.})}
\newcommand{\amuQEDW}{\Delta a_{\mu}^{{\mathrm W}}(\mathrm{QED})}

\newcommand{\amuLSD}{a^{ll,\,{\mathrm {SD}}}_{\mu}(\mathrm{conn.})}
\newcommand{\amuSSD}{a^{ss,\,{\mathrm {SD}}}_{\mu}(\mathrm{conn.})}
\newcommand{\amuCSD}{a^{cc,\,{\mathrm {SD}}}_{\mu}(\mathrm{conn.})}
\newcommand{\amuBSD}{a^{bb,\,{\mathrm {SD}}}_{\mu}(\mathrm{conn.})}
\newcommand{\amuDSD}{a^{{\mathrm {SD}}}_{\mu}(\mathrm{disc.})}
\newcommand{\amuSIBSD}{\Delta a_{\mu}^{ud,\,{\mathrm{SD}}}(\mathrm{SIB})}
\newcommand{\amuSIBCSD}{\Delta a_{\mu}^{ud, \textrm{SD}}(\mathrm{SIB,conn.})}

\newcommand{\amuQEDSD}{\Delta a_{\mu}^{{\mathrm {SD}}}(\mathrm{QED})}
\newcommand{\amuQEDLSD}{\Delta a_{\mu}^{ll,\,{\mathrm {SD}}}(\mathrm{QED,conn.})}
\newcommand{\amuQEDSSD}{\Delta a_{\mu}^{ss,\,{\mathrm {SD}}}(\mathrm{QED,conn.})}

\newcommand{\amuLSDRes}{a^{ll,\,{\mathrm{SD}}}_{\mu}(\mathrm{conn.}) = 48.139(11)(91)[92] \times 10^{-10}}
\newcommand{\amuLSDResABS}{a^{ll,\,{\mathrm{SD}}}_{\mu}(\mathrm{conn.}) = 48.139(11)_{\mathrm{stat}}(91)_{\mathrm{syst}}[92]_{\mathrm{total}} \times 10^{-10}}

\newcommand{\amuSSDRes}{a^{ss,\,{\mathrm{SD}}}_{\mu}(\mathrm{conn.}) = 9.111(3)(16)[17] \times 10^{-10}}
\newcommand{\amuCSDRes}{a^{cc,\,{\mathrm{SD}}}_{\mu}(\mathrm{conn.}) = 11.46(0)(17)[17] \times 10^{-10}}
\newcommand{\amuDSDRes}{a^{{\mathrm{SD}}}_{\mu}(\mathrm{disc.}) = -0.0019(4)(26)[26] \times 10^{-10}} 
\newcommand{\amuSIBCSDRes}{\Delta a^{ud,{\mathrm{SD}}}_{\mu}(\mathrm{SIB,conn.}) = -0.0049(26)(24)[35] \times 10^{-10}}
\newcommand{\amuSIBDSDRes}{\Delta a^{ud,{\mathrm{SD}}}_{\mu}(\mathrm{SIB,disc.}) = 0.015(1)(12)[12]\times 10^{-10}}
\newcommand{\amuQEDSDRes}{\amuQEDSD = 0.028(28) \times 10^{-10}}

\newcommand{\amuSDResABS}{a^{{\mathrm{SD}}}_{\mu} = 69.05(1)_{\mathrm{stat}}(21)_{\mathrm{syst}}[21]_{\mathrm{total}} \times 10^{-10}}

\newcommand{\amuLWRes}{a^{ll,\,{\mathrm{W}}}_{\mu}(\mathrm{conn.}) = 206.90(14)(61)[63] \times 10^{-10}}
\newcommand{\amuLWResABS}{a^{ll,\,{\mathrm{W}}}_{\mu}(\mathrm{conn.}) =  206.90(14)_{\mathrm{stat}}(61)_{\mathrm{syst}}[63]_{\mathrm{total}} \times 10^{-10}}

\newcommand{\amuSWRes}{a^{ss,\,{\mathrm{W}}}_{\mu}(\mathrm{conn.}) = 27.20(1)(13)[13] \times 10^{-10}}
\newcommand{\amuCWRes}{a^{cc,\,{\mathrm{W}}}_{\mu}(\mathrm{conn.}) = 2.624(0)(87)[87] \times 10^{-10}}
\newcommand{\amuDWRes}{a^{{\mathrm{W}}}_{\mu}(\mathrm{disc.}) = -0.85(6)(19)[20] \times 10^{-10}} 
\newcommand{\amuSIBCWRes}{\Delta a^{ud,{\mathrm{W}}}_{\mu}(\mathrm{SIB,conn.}) = 0.73(7)(11)[13] \times 10^{-10}}
\newcommand{\amuSIBDWRes}{\Delta a^{ud,{\mathrm{W}}}_{\mu}(\mathrm{SIB,disc.}) = -0.175(15)(13)[20] \times 10^{-10}}
\newcommand{\amuQEDWRes}{\amuQEDW = 0.0(2) \times 10^{-10}}

\newcommand{\amuWResABS}{a^{{\mathrm{W}}}_{\mu} = 236.45(17)_{\mathrm{stat}}(83)_{\mathrm{syst}}[85]_{\mathrm{total}} \times 10^{-10}} 

\newcommand{\gmtwo}{$g-2$} 

\newcommand{\pr}{\operatorname{pr}}
\newcommand{\chidof}{\chi^2/\text{d.o.f.}}

\newcommand{\coloaf}{Department of Physics, University of Colorado, Boulder, Colorado 80309, USA}
\newcommand{\fnalaf}{Theory Division, Fermi National Accelerator Laboratory, Batavia, Illinois, 60510, USA}
\newcommand{\iuaf}{Department of Physics, Indiana University, Bloomington, Indiana 47405, USA}

\newcommand{\msuaf}{Department of Computational Mathematics, Science and Engineering, and Department of Physics and Astronomy, Michigan State University, East Lansing, Michigan 48824, USA}

\newcommand{\ugraf}{CAFPE and Departamento de Física Teórica y del Cosmos, Universidad de Granada, \\E-18071 Granada, Spain}
\newcommand{\uiucaf}{Department of Physics, University of Illinois Urbana-Champaign, Urbana, IL 61801, USA}
\newcommand{\icasuuiaf}{Illinois Center for Advanced Studies of the Universe, University of Illinois \\Urbana-Champaign, Urbana, IL 61801, USA}
\newcommand{\unizar}{Departmento de Física Teórica, Universidad de Zaragoza, 50009 Zaragoza, Spain}
\newcommand{\utahaf}{Department of Physics and Astronomy, University of Utah, Salt Lake City, UT 84112, USA}
\newcommand{\glasaf}{SUPA, School of Physics and Astronomy, University of Glasgow, Glasgow, G12 8QQ, \\United Kingdom}
\newcommand{\cornaf}{Laboratory for Elementary-Particle Physics, Cornell University, Ithaca, NY 14853, USA}
\newcommand{\plyaf}{Centre for Mathematical Sciences, University of Plymouth, Plymouth PL4 8AA, \\United Kingdom}
\newcommand{\syracuseaf}{Department of Physics, Syracuse University, NY 13244, USA}
\newcommand{\csuaf}{Department of Physics, Colorado State University, Fort Collins, CO 80523, USA}
\newcommand{\capa}{Center for Astroparticles and High Energy Physics (CAPA), Calle Pedro Cerbuna 12, 50009 Zaragoza, Spain}

\begin{document}
\count\footins = 1000 

\preprint{FERMILAB-PUB-24-0835-T}

\title{Hadronic vacuum polarization for the muon \texorpdfstring{\boldmath\gmtwo}{g-2} from lattice QCD: Complete short and intermediate windows}

\author{Alexei~Bazavov}\affiliation{\msuaf}
\author{David~A.~Clarke}\email{clarke.davida@gmail.com}\affiliation{\utahaf}
\author{Christine~T.~H.~Davies}\affiliation{\glasaf}
\author{Carleton~DeTar}\affiliation{\utahaf}
\author{Aida~X.~El-Khadra}\affiliation{\uiucaf}\affiliation{\icasuuiaf}
\author{Elvira~G\'amiz}\affiliation{\ugraf}
\author{Steven~Gottlieb}\affiliation{\iuaf}
\author{Anthony~V.~Grebe}\affiliation{\fnalaf}
\author{Leon~Hostetler}\affiliation{\iuaf}
\author{William~I.~Jay}\affiliation{\csuaf}
\author{Hwancheol~Jeong}\affiliation{\iuaf}
\author{Andreas~S.~Kronfeld}\affiliation{\fnalaf}
\author{Shaun~Lahert}\email{shaun.lahert@gmail.com}\affiliation{\utahaf}
\author{Jack~Laiho}\affiliation{\syracuseaf}
\author{G.~Peter~Lepage}\affiliation{\cornaf}
\author{Michael~Lynch}\email{ml11@illinois.edu}\affiliation{\uiucaf}\affiliation{\icasuuiaf}
\author{Andrew~T.~Lytle}\affiliation{\uiucaf}\affiliation{\icasuuiaf}
\author{Craig~McNeile}\affiliation{\plyaf}
\author{Ethan~T.~Neil}\affiliation{\coloaf}
\author{Curtis~T.~Peterson}\affiliation{\msuaf}
\author{James~N.~Simone}\affiliation{\fnalaf}
\author{Jacob~W.~Sitison}\email{jacob.sitison@colorado.edu}\affiliation{\coloaf}
\author{Ruth~S.~\surname{Van~de~Water}}\affiliation{\fnalaf}
\author{Alejandro~Vaquero}\affiliation{\utahaf}\affiliation{\unizar}\affiliation{\capa}

\collaboration{Fermilab Lattice, HPQCD, and MILC Collaborations}
\noaffiliation

\date{\today}

\begin{abstract}
We present complete results for the hadronic vacuum polarization (HVP) contribution to the muon anomalous magnetic moment $a_\mu$ in the short- and intermediate-distance window regions, which account for roughly 10\% and 35\% of the total HVP contribution to $a_\mu$, respectively. In particular, we perform lattice-QCD calculations for the isospin-symmetric connected and disconnected contributions, as well as corrections due to strong-isospin breaking. For the short-distance window observables, we investigate the so-called log-enhancement effects as well as the significant oscillations associated with staggered quarks in this region. For the dominant, isospin-symmetric light-quark connected contribution, we obtain $\amuLSDResABS$ and $\amuLWResABS$. We use Bayesian model averaging to fully estimate the covariance matrix between the individual contributions. Our determinations of the complete window contributions are $\amuSDResABS$ and $\amuWResABS$. This work is part of our ongoing effort to compute all contributions to HVP with an overall uncertainty at the few permille level.
\end{abstract}

\maketitle

\raggedbottom
\allowdisplaybreaks

\newpage

\tableofcontents

\newpage

\section{Introduction}\label{sec:intro}

The anomalous magnetic moment of the muon, $\amu \equiv\gmTwo /2$, provides one of the most precise tests of the Standard Model (SM) of particle physics.  It has been measured to exquisite precision in a series of experiments, most recently at Fermilab by the Muon $g-2$ Collaboration \cite{Muong-2:2021ojo,Muong-2:2023cdq}, with the latest results reported at a precision of 200 parts per billion (ppb).  Final results from the Fermilab experiment incorporating their complete data set are anticipated to improve the current precision by close to another factor of two~\cite{Muong-2:2023cdq}.  In addition, the Muon g-2/EDM experiment at J-PARC \cite{Abe:2019thb,E34webpage} will further improve our experimental knowledge of this quantity. 

Experimental measurements of $\amu$ have attracted particular interest because of their long-standing tension with Standard Model (SM) expectations, providing an enduring hint for the possibility that this quantity is influenced by new, as-yet unknown particles. The uncertainty on the SM prediction~\cite{Aoyama:2020ynm,Colangelo:2022jxc} is dominated by contributions from Quantum Chromodynamics (QCD), namely hadronic vacuum polarization (HVP) and hadronic light-by-light scattering (HLbL).  In this paper, we focus on HVP, which is the larger contribution to $\amu$ and the largest source of error in its SM prediction. There are two independent approaches, a data-driven one and lattice QCD. The former uses a dispersive framework together with experimental inputs, primarily taken from measurements of $e^+e^-$ annihilation cross sections into hadronic channels. This approach was the basis of the SM prediction of Ref.~\cite{Aoyama:2020ynm}. However, a recent measurement by the CMD-3 collaboration \cite{CMD-3:2023alj} of the low-energy $e^+e^- \to \pi^+\pi^-$ cross section disagrees with all previous experimental results for this important channel, which accounts for roughly $75\%$ of the leading-order (LO) HVP, $\amuHVP$ in the data-driven approach.  

The second approach, lattice QCD, offers a determination of $\amuHVP$ based entirely on SM theory.\footnote{To set the quark masses and strong coupling, $N_f+1$ experimental inputs are needed, taken here to be precisely measured hadron masses \cite{FlavourLatticeAveragingGroupFLAG:2024oxs}.}
While published lattice-QCD results for $\amuHVP$ were not precise enough at the time to be used in the SM prediction of Ref.~\cite{Aoyama:2020ynm}, the situation has changed significantly in recent years, thanks to dedicated efforts by several lattice collaborations \cite{Borsanyi:2020mff,Aubin:2022hgm,Alexandrou:2022amy,Ce:2022kxy,FermilabLatticeHPQCD:2023jof,Blum:2023qou,Boccaletti:2024guq,RBC:2024fic,Spiegel:2024dec,Djukanovic:2024cmq}. Hence, we expect that lattice-QCD results will play an important role in obtaining a new SM prediction of $\amu$ and in the subsequent interpretation of the experimental measurements.  Especially in light of the puzzle created in the data-driven approach by the new CMD-3 $e^+e^- \to \pi^+\pi^-$ cross section measurement, continued efforts to further improve lattice-QCD calculations to match the expected precision of the $g-2$ experiments are clearly important.

In lattice QCD, the HVP can be computed from Euclidean time integrals of correlation functions \cite{Bernecker:2011gh}, where it is advantageous to study the different regions of Euclidean time separately to refine control over systematic effects. The most common division is into three time complementary regions, known as ``windows'' \cite{RBC:2018dos}: the ``short-distance'' contribution $\amuSD$, the ``intermediate-distance'' contribution $\amuW$, and the ``long-distance'' contribution $\amuLD$. By construction, the sum of these three terms gives the total HVP contribution to $\amu$.  
 
The smallest component, $\amuSD$, at roughly 10\% of the total, is the most sensitive to lattice discretization effects. The largest component, $\amuLD$, roughly 55\% of the total, is the most challenging, as it is sensitive to the finite lattice volume effects and suffers from large statistical errors due to well-known signal-to-noise issues in the underlying correlation functions at large Euclidean times. The intermediate window observable, $\amuW$, is relatively insensitive to these two extremes and hence can be obtained reasonably straightforwardly. Both $\amuSD$ and $\amuW$ can be obtained with good statistical precision. The evaluation of windowed HVP observables also provides more fine-grained comparisons with data-driven results, which will be crucial if current tensions between data-driven evaluations and lattice calculations persist after the disagreements between experimental cross-section measurements are understood or resolved. In this paper, we focus on the short- and intermediate-distance window contributions, deferring a discussion of our calculation of $\amuLD$ to a companion paper. 

Besides the division into windows with different characteristic energy scales, the HVP (either the total or within a given window) can also be divided into contributions from individual quark flavors for quark-line connected and disconnected diagrams, as well as corrections due to isospin symmetry breaking (including strong-isospin breaking effects due to $m_u \neq m_d$ as well as QED).
We have previously presented results for the intermediate window HVP~\cite{FermilabLatticeHPQCD:2023jof}, in good agreement with other, existing lattice-QCD calculations \cite{Borsanyi:2020mff,Aubin:2022hgm,Ce:2022kxy,Alexandrou:2022amy,Wang:2022lkq,Blum:2023qou}. However, this previous work considered only $\amuLW$ (the ``light-quark connected'' $\amuW$), omitting contributions from other quark flavors, as well as those from disconnected diagrams and isospin-breaking corrections. In this work, we present results from our lattice-QCD calculations of these additional contributions, as well as a refined calculation of $\amuLW$. We note, however, that a discussion of our direct lattice-QCD calculation of the QED corrections is deferred to a separate paper. Instead, in this paper, we use a combination of perturbative QCD (pQCD), phenomenological estimates and comparisons with previous direct lattice-QCD calculations of QED corrections to estimate their contributions to the  short- and intermediate-distance window observables. 

Our results for these individual contributions and corrections provide a more fine-grained view into the HVP at short and intermediate distances, enabling comparisons with the corresponding results from other lattice groups as available.  We also obtain complete physical results for the total short- and intermediate-distance HVP. 

This paper is organized as follows. First, \cref{subsec:window} provides analytic expressions for $\amuSD$ and $\amuW$ in terms of the Euclidean-time vector-current correlation function, as well as the definitions of the individual contributions to these observables. Next, \cref{subsec:inputs} defines the isospin-symmetric QCD limit employed in the majority of these calculations. The inputs for QCD with strong-isospin breaking are also described. The ensembles employed in this work and the correlation function datasets generated from them are described in \cref{subsec:lattice}. \Cref{sec:analysis} presents a detailed description of the general analysis procedure employed in this work: our blinding strategy (\cref{sec:blind}), the systematic corrections applied to the lattice observables (\cref{sec:latticeCorrections}), our approach to taking the continuum limit (\cref{sec:contextrap}), and the subsequent estimation of uncertainties from these systematics using Bayesian Model averaging (BMA) (\cref{sec:BMA}). In \cref{sec:windowobs}, we present our analyses for the individual window observables; this section is broken up into subsections describing the light-connected (\cref{subsec:light_analysis}), strange- and charm-connected (\cref{subsec:sc_analysis}), disconnected (\cref{subsec:disc_analysis}) and the strong-isospin-breaking (\cref{subsec:sib_analysis}) contributions. Perturbative QCD is used to estimate the short-distance window observables, which we compare with our lattice determinations, in \cref{subsec:pQCD_analysis}, while \cref{sec:results} presents our final results for the short-distance (\cref{subsec:sd_results}) and intermediate-distance window observables (\cref{subsec:w_results}). \Cref{sec:conclusion} concludes with a summary of our results, including comparisons with previous lattice-QCD calculations and data-driven evaluations, and an outlook in light of our ongoing efforts to obtain the total HVP at few-permille-level precision. \Cref{sec:oscEffects,sec:logEffects} examine discretization effects important in short-distance window observables, namely the log-enhanced discretization terms that arise in the correlation functions at small Euclidean times and the oscillatory contributions to correlation functions constructed with staggered quarks. Finally, our procedure for obtaining a complete correlation matrix in the BMA framework is given in \cref{sec:bma_cov}.

\section{Background and Simulation Details}\label{sec:background}

\subsection{Definitions of window observables}
\label{subsec:window}

The leading-order hadronic-vacuum-polarization contribution to the muon anomalous magnetic moment is obtained via
\begin{equation}
    \amuHVP = 4\alpha^{2} \int_{0}^{\infty} \dd{t} C(t)  \tilde{K}(t),
    \label{eq:amuTint}
\end{equation}
where $\alpha$ is the fine-structure constant, the kernel $\tilde{K}$ stems from QED, and $C(t)$ is the Euclidean-time two-point correlation function of the electromagnetic current, 
\begin{align}
    C(t) &= \frac{1}{3} \sum_{k=1}^3 \sum_{\bm{x}}\left\langle J^{k}(\bm{x}, t) J^{k}(0)\right\rangle ,
    \label{eq:corrFunc2pt} \\
    J^{\mu}(x)&= \sum_{f} q_{f} \bar{\psi}_{f}(x) \gamma^{\mu} \psi_{f}(x). \label{eq:vecCurrent}
\end{align} 
The current $J^{\mu}(x)$ is summed over all quark flavors $f\in\{u,d,s,c,b,t\}$ of electric charge $q_f$.
In lattice QCD, the calculation of $C(t)$ is straightforward, although controlling all uncertainties at the needed precision level is challenging.
The QED kernel is given by
\begin{equation}
    \tilde{K}(t) = 2 \int_{0}^{\infty} \frac{\dd Q}{Q}\, K_{E}(Q^{2})
        \left[Q^{2} t^{2}-4 \sin ^{2}\left(\frac{Q t}{2}\right)\right],
    \label{eq:Ktilde}
\end{equation}
where the energy-momentum kernel \cite{Blum:2002ii}, $K_E(Q^2)$, depends on the muon mass:
\begin{align}
    K_E\left(Q^2\right) = \frac{m_\mu^2Q^2Z^3 (1-Q^2Z)}{1+m_\mu^2Q^2Z^2},  \quad\quad
    Z = \frac{(Q^4 + 4m_\mu^2 Q^2)^{1/2}-Q^2}{2m_\mu^2Q^2}.
    \label{eq:KE}
\end{align}
Higher-order HVP can be obtained from \cref{eq:amuTint} by substituting a different kernel~\cite{Kurz:2014wya,Balzani:2021del,Balzani:2024gmu,Nesterenko:2021byp,Chakraborty:2018iyb}.

Lattice-QCD calculations of $\amuHVP$ separately compute the contributions from each quark flavor and from connected and disconnected Wick contractions. Additionally, these calculations are performed in the isospin-symmetric limit in pure QCD (see \cref{subsec:inputs}) with the effects of strong-isospin breaking and quantum electrodynamics (QED) added as corrections to obtain the complete $a_\mu^{\mathrm{HVP}, \mathrm{LO}}$. The complete result is the sum of the following individual contributions,
\begin{align}
    a_\mu^{\mathrm{HVP}, \mathrm{LO}} &= \amuL + \amuS+ \amuC + \amuB + \ldots \nonumber \\
    & \hspace{1em} + \amuD + \amuSIB +\amuQED. \label{eq:amuBreakdown}
\end{align}
The contributions listed on the first line are collectively called the connected contributions and correspond to connected Wick contractions of \cref{eq:corrFunc2pt} for each individual quark flavor of \cref{eq:vecCurrent}.\footnote{The light-quark charge factor is given by $q_l^2=q_u^2 + q_d^2 = 5/9$.} The first contribution on the second line is from the disconnected Wick contractions of \cref{eq:corrFunc2pt}, which contains quark-flavor mixing. In this work, we calculate only the light- and strange-disconnected contribution. The final two terms are the strong-isospin breaking (SIB) and QED corrections defined by
\begin{align}
    \Delta a_\mu^{u d}(\mathrm{SIB})&= a_\mu^{u d}-a_\mu^{l l} \label{eq:sib-def},  \\
    \Delta a_\mu^{f}(\mathrm{QED}) &= a_\mu^{ff}( \mathrm{QCD+QED}) - a_\mu^{ff}. \label{eq:amuQED}
\end{align}
For the correction due to strong-isospin breaking, $a_\mu^{u d}$ and $a_\mu^{l l}$ contain both the connected and disconnected contractions, which we compute separately (see \cref{subsec:lattice,subsec:sib_analysis}). 

By limiting the Euclidean-time region over which $C(t)$ is integrated, one obtains the so-called one- and two-sided window observables~\cite{RBC:2018dos}:
\begin{subequations} \label{eq:amuTintWin}
\begin{align}
    a_\mu^{\mathrm{win}_1(t_0,      \Delta)} &= 4 \alpha^{2} \int_{0}^{\infty} \dd{t} C(t)\tilde{K}(t) \mathcal{W}_1\left(t, t_0,      \Delta\right), \label{eq:amuTintWin1}\\
    a_\mu^{\mathrm{win}_2(t_0, t_1, \Delta)} &= 4 \alpha^{2} \int_{0}^{\infty} \dd{t} C(t)\tilde{K}(t) \mathcal{W}_2\left(t, t_0, t_1, \Delta\right), \label{eq:amuTintWin2}
\end{align}
\end{subequations}
where the window functions are\footnote{In earlier work~\cite{FermilabLatticeHPQCD:2023jof}, we used a slightly different definition of the window functions.  To the accuracy given there, none of the reported digits would change.  Here, we use the standard definitions, namely Eq.~(3.26) of Ref.~\cite{Aoyama:2020ynm}.}
\begin{subequations}  \label{eq:windofunc}
\begin{align}
    \mathcal{W}_1(t, t_0,      \Delta) &= \frac{1}{2}\left[1 - \tanh \left(\frac{t-t_0}{\Delta}\right)\right] . \label{eq:windofunc1} \\
    \mathcal{W}_2(t, t_0, t_1, \Delta) &= \frac{1}{2}\left[\tanh \left(\frac{t-t_0}{\Delta}\right)-\tanh \left(\frac{t-t_1}{\Delta}\right)\right] . \label{eq:windofunc2}
\end{align}
\end{subequations}
The parameters $t_0$ and $t_1$ of the $\mathcal{W}_i$ control the location of the window's boundaries, while $\Delta$ controls the sharpness of its edges. The window functions satisfy $\mathcal{W}_1(t,t_0,\Delta)+\mathcal{W}_2(t,t_0,t_1,\Delta)=\mathcal{W}_1(t,t_1,\Delta)$, $\mathcal{W}_2(t,t_0,t_1,\Delta)+\mathcal{W}_2(t,t_1,t_2,\Delta)=\mathcal{W}_2(t,t_0,t_2,\Delta)$.

In this work, we consider two such windows, the short-distance window SD (defined with $t_0 = 0.4$~fm, $\Delta = 0.15$~fm), yielding
\begin{align}
    \amuSD &\equiv a_\mu^{\mathrm{win}_1(\textrm{0.4, 0.15})} \label{eqn:SDDef} , 
\end{align} 
and the intermediate-distance window W (defined with $t_0 = 0.4$~fm, $t_1 = 1.0$~fm, $\Delta = 0.15$~fm), yielding
\begin{align}
    \amuW &\equiv a_\mu^{\mathrm{win}_2(\textrm{0.4, 1.0, 0.15})} \label{eqn:WDef} ,
\end{align}
with the parameters in~fm. 
\begin{figure}
\centering
\includegraphics[scale=0.73]{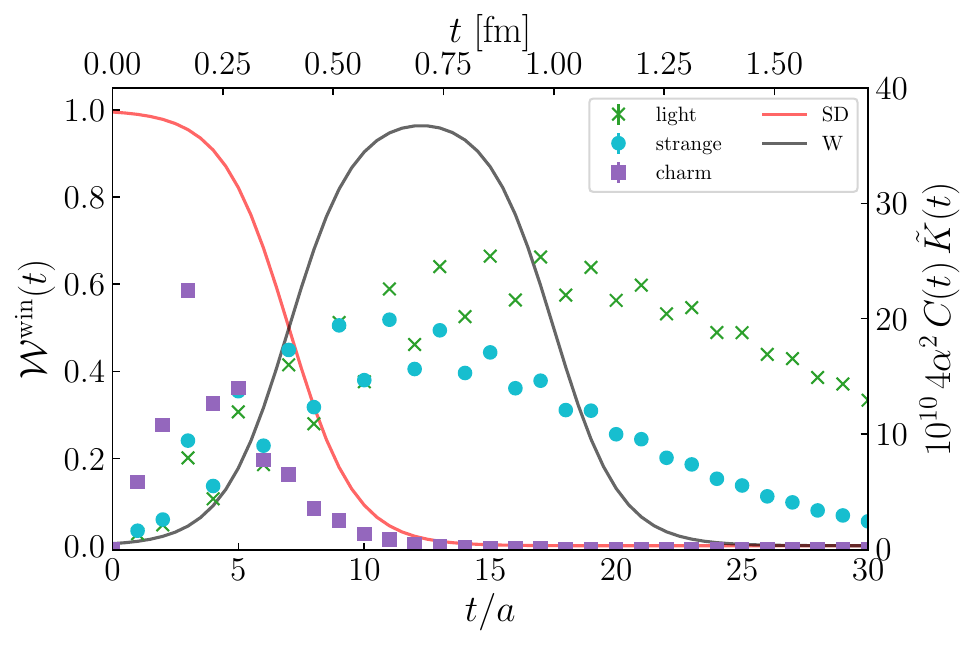}
\vspace{-5mm}
\caption{The SD (red) and W (black) window functions corresponding to the parameters in \cref{eqn:SDDef,eqn:WDef}, overlaid with raw lattice data for the integrand of \cref{eq:amuTint} for light- (green crosses), strange- (blue circles), and charm-connected (purple squares) contributions obtained on our ensemble with $a\approx 0.06 $~fm (see \cref{table:ensParams}).}
\label{fig:window06Data}
\end{figure}
We plot the $\mathcal{W}_i$ in \cref{eq:windofunc} for these window regions in \cref{fig:window06Data}. Also shown are the light-quark,  strange-quark, and charm-quark connected contributions to the integrand, ${\tilde{K}}(t)C(t)$ in \cref{eq:amuTint}, using the lattice correlation functions obtained on our $0.06$~fm ensemble (see \cref{subsec:lattice}). The light-quark contribution dominates both windows, as can be seen by the scaling factor of six applied to the strange and charm integrands in the plot. After the light, the charm is the leading contribution in the short-distance window, while the strange is the leading contribution in~W. The remaining contributions in \cref{eq:amuBreakdown} are sub-leading for these two windows and are at the sub-percent level on the complete results.

\subsection{Physical inputs}
\label{subsec:inputs}

Our lattice calculations make use of several ensembles of four-flavor gauge-field configurations generated by the MILC collaboration (details are given below in \cref{subsec:lattice}).  With one exception, these ensembles are generated in the limit of isospin symmetry (that is, with the up- and down-quark masses set to be equal) and without dynamical QED. Converting our numerical results into physical units requires the selection of a set of physical inputs which determine the quark masses and the overall energy scale.

Our choices for the inputs for the quark masses follow the prescription introduced in Refs.~\cite{HPQCD:2004hdp,MILC:2004qnl} and extended to charm in \cite{FermilabLattice:2014tsy}.  The masses of the $u$, $d$, and $s$ quarks are determined by the masses of the $\pi^+, K^0$, and $K^+$ mesons, and the $c$ quark mass is fixed by the mass of the $D_s^+$ meson. While in Refs.~\cite{HPQCD:2004hdp,MILC:2004qnl,FermilabLattice:2014tsy} the energy scale is fixed by the pion decay constant $f_{\pi^+}$, here we employ the $\Omega^-$ baryon mass.

To match the ``pure-QCD world'' of our simulations, QED effects are removed from the values of these masses used for scale setting \cite{Grebe:2025}.  For all meson masses used to determine quark masses, we adopt the values given in the Flavor Lattice Averaging Group (FLAG) 2024 review \cite{FlavourLatticeAveragingGroupFLAG:2024oxs}. These choices define the following separation scheme:

\begin{equation}
    (\textrm{pure-QCD world}) \Rightarrow \begin{cases}
        M_\Omega &\equiv 1671.26\ {\rm MeV}, \\
        M_{\pi^+} &\equiv 135.0\ {\rm MeV}, \\
        M_{K^+} &\equiv 491.6\ {\rm MeV}, \\
        M_{K^0} &\equiv 497.6\ {\rm MeV}, \\
        M_{D_s^+} &\equiv 1967\ {\rm MeV}.
\end{cases}
\label{eq:iso-broken-pure-QCD}
\end{equation}
Setting the three meson masses $\{M_{\pi^+}, M_{K^+}, M_{K^0}\}$ to their physical values fixes the three light quark masses $\{m_u, m_d, m_s\}$. In practice, we implement this scheme with the equivalent set 
$\{M_{\pi^+}, M_{K}, \Delta M_K^2\}$, with
\begin{align}
\label{eq:MK}
M_K^2 &\equiv \frac{1}{2} (M_{K^0}^2 + M_{K^+}^2) = (494.6\ {\rm MeV})^2, \\
\label{eq:DeltaK}
    \Delta M_K^2 &\equiv M_{K^0}^2 - M_{K^+}^2 = {5935}\ {\rm MeV}^2.
\end{align}
In this approach, adjusting $M_{\pi^+}$ and the root-mean-squared (RMS) average mass $M_{K}$ to the physical point given in \cref{eq:iso-broken-pure-QCD} fixes $m_l\equiv (m_u + m_d) / 2$ and $m_s$, and adjusting $\Delta M_K^2$ determines the up-down mass splitting $m_d - m_u$. This implementation is convenient given the data available in the analyses below.

To define the separation scheme for the isospin-symmetric limit, where the up and down masses are both set to the average value $m_l = (m_u + m_d)/2$, we replace $M_{K^+}$ and $M_{K^0}$ with the RMS average $M_K$ given above:
\begin{equation}
    (\textrm{isospin-symmetric pure-QCD world}) \Rightarrow \begin{cases}
M_\Omega &\equiv 1671.26\ {\rm MeV} , \\
M_{\pi^+} &\equiv 135.0\ {\rm MeV}, \\
M_{K} &\equiv 494.6\ {\rm MeV}, \\
M_{D_s^+} &\equiv 1967 \ {\rm MeV}.
\end{cases}
\label{eq:iso-QCD}
\end{equation}

For the analysis of strong-isospin breaking, in particular, it is helpful to define fictitious physical pion masses with both constituent quark masses set to $m_u$, $m_d$, or $m_l$.  In principle, this could be done from the pion and kaon mass inputs above by using the leading-order Gell-Mann--Oakes--Renner \cite{Gell-Mann:1968hlm} (GMOR) relations, {\it i.e.},~leading-order chiral perturbation theory, which states that the squared pion and kaon masses scale linearly with quark mass.  Although this statement is locally true to good approximation, inclusion of both pions and kaons simultaneously in leading-order chiral perturbation theory relations can lead to deviations due to the presence of significant sub-leading order corrections.

Instead of assuming purely GMOR scaling, we adapt the results from the MILC Collaboration~\cite{MILC:2018ddw}, which computes the up-down quark mass ratio to be $m_u / m_d = 0.4529(48)_{\rm stat}\left(\begin{smallmatrix}+150\\-67\end{smallmatrix}\right)_{\rm syst}$.  Following the procedures of FLAG \cite{FlavourLatticeAveragingGroupFLAG:2021npn} for symmetrizing two-sided errors and combining statistical and systematic errors in quadrature, this becomes $m_u / m_d = 0.4550(138)$.  To account for the slight difference between \cref{eq:iso-broken-pure-QCD} and the scheme used in Refs.~\cite{MILC:2018ddw,Bazavov:2017lyh}---namely $M_{K^+}=491.6$ MeV versus $M_{K^+}=491.405$ MeV, respectively---we include an additional systematic uncertainty that conservatively accounts for the implied difference in $\epsilon$ (the leading correction to Dashen's theorem defined in Ref.~\cite{MILC:2018ddw}) between the two schemes, which gives 
\begin{equation}
m_u/m_d = 0.455(18).
\end{equation}
Adopting this as our main result and using GMOR scaling only in the vicinity of the pion mass, we have the results
\begin{align}
    M_{\pi}^{uu} &= M_{\pi^+} \sqrt{\frac{2}{1+m_d/m_u}} = 106.7(1.5)\ {\rm MeV}, \\
    M_{\pi}^{dd} &= M_{\pi^+} \sqrt{\frac{2}{1+m_u/m_d}} = 158.25(98)\ {\rm MeV}.
\end{align}

\subsection{Lattice ensembles and correlation functions}\label{subsec:lattice}

\begin{table*}
\centering
\caption{Ensemble parameters used in this work. The first column lists the approximate lattice spacings in~fm. (In previous work, the ensemble labeled ``0.04" was labeled ``0.042.") The second column gives the spatial length $L$ of the lattices in~fm. The third column lists the volumes of the lattices in number of space-time points. The fourth to seventh columns gives the sea-quark masses in lattice-spacing units. We note here that the only difference between the $0.15^\prime$ and $0.15$ ensembles is that sea-quark masses of $0.15^\prime$ include strong-isospin breaking $m_u \neq m_d$. Additionally, the only difference between the $0.09$ and $0.09^\star$ ensembles is that the sea-quark masses of $0.09^\star$ are better tuned to the physical point. 
The eighth column lists the $aM_\Omega$ values on each ensemble, these were first given in Ref.~\cite{Bazavov:2024eou}, except for the value on the 0.09 ensemble. To convert simulation results to physical units, we take $M_\Omega= 1.67126(32)$~GeV, also first given in Ref.~\cite{Bazavov:2024eou}.}
\label{table:ensParams}
\begin{tabularx}{\linewidth}{lcCLLLLl}
\hline\hline
$\approx a/\mathrm{fm}$ & $L/\mathrm{fm}$ & $N_s^3 \times N_t$ & $a m_{u}^\text{sea}$ &  $a m_{d}^\text{sea}$  &  $a m_{s}^\text{sea}$  &  $a m_{c}^\text{sea}$ & $aM_\Omega$ \\ 
\hline
$0.15^\prime$ & $5.00$ & $ 32^3 \times 48$ & 0.001524 & 0.003328 & 0.0673 & 0.8447 & 1.3246(26) \\ 
$0.15$ & $5.00$ & $ 32^3 \times 48$ & 0.002426 & 0.002426 & 0.0673 & 0.8447 & 1.3246(26)\\  
$0.12$ & $5.95$ & $48^3 \times 64$ & 0.001907 & 0.001907 & 0.05252 & 0.6382 & 1.0494(17)\\ 
$0.09$ & $5.67$ & $64^3 \times 96$ &  0.00120 & 0.00120 & 0.0363 & 0.432 & 0.7506(14)\\ 
$0.09^\star$ & $5.70$ & $64^3 \times 96$ & 0.001326 & 0.001326 & 0.03636 & 0.4313 & 0.75372(97)\\ 
$0.06$ & $5.48$ & $96^3\times128$ & 0.0008 & 0.0008 & 0.022 & 0.260 & 0.4834(11)\\ 
$0.04$ & $6.13$ & $144^3\times288$ &   0.000569 & 0.000569 & 0.01555 & 0.1827  & 0.3608(25)\\ 
\hline \hline
\end{tabularx}
\end{table*}

\begin{table*}
\centering
\caption{Pseudoscalar meson masses calculated on each of the gauge ensembles given in \cref{table:ensParams} for various combinations of valence quark masses.  Results are shown in lattice units (column 5).  The label $l$ denotes the symmetrized light-quark mass $m_l = (m_u + m_d) / 2$.  On the 0.09~fm ensemble, the valence $m_l$ is mistuned to be somewhat lighter than physical; the retuned valence mass is labeled $r$.  The 0.09$^\star$~fm ensemble includes a lighter-than-physical up-quark mass, denoted $h$, as described in the text.  Meson masses from previous publications are marked with $\S$ (\cite{Davies:2019efs}), $\dagger$ (\cite{FermilabLattice:2022gku}), and $\ddagger$ (\cite{FermilabLatticeHPQCD:2023jof}). Column 2 lists the valence quark masses $am_q$ only if they differ from the sea-quark masses listed in \cref{table:ensParams}.}
\label{table:mesonMasses}
\begin{tabularx}{\linewidth}{LCCR}
\hline \hline
$\approx a/\mathrm{fm}$ & $am_q$ & $qq^\prime$  & $aM_{\gamma_5 \otimes \gamma_5}$ \\ 
\hline
\multirow{6}{*}{$0.15^\prime$} 
 & & $uu$  & 0.082368(18) \\
 & & $ll$ & 0.103420(18) \\
 & & $dd$ & 0.120665(17) \\
 & & $us$ & 0.376023(22) \\
 & & $ls$ & 0.37847(20) \\
 & & $ds$ & 0.380749(19) \\
 \hline
\multirow{5}{*}{$0.15$} 
 & 0.001524 & $uu^\ddagger$ & 0.082362(11) \\
 & & $ll^\S$ & 0.103414(11) \\
 & 0.003328 & $dd^\ddagger$ & 0.120652(11) \\
 & & $ls^\dagger$ & 0.37847(11) \\
 & & $sc^\dagger$ & 1.50544(19)  \\
 \hline
\multirow{5}{*}{$0.12$} 
 & 0.001190 & $uu^\ddagger$ & 0.0659338(57) \\
 & & $ll^\dagger$ & 0.0830651(63) \\
 & 0.002625 & $dd^\ddagger$ & 0.0970979(67) \\
 & & $ls^\dagger$ & 0.303949(77) \\
 & & $sc^\dagger$ & 1.209384(78)  \\
 \hline
\multirow{5}{*}{$0.09$} 
 & & $ll^\dagger$ & 0.057193(16) \\
 & 0.00133 & $rr$ & 0.060136(16) \\
 & 0.00183 & $dd$ & 0.070267(19) \\
 & & $ls^\dagger$ & 0.219482(70) \\
 \hline
\multirow{5}{*}{$0.09^\star$} 
 &  0.000514 & $hh$ & 0.037831(13) \\
 & & $ll$ & 0.060124(16) \\
 & 0.001802 & $dd$ & 0.069830(17) \\
 & & $ls$ & 0.220231(40) \\
 & & $sc$ & 0.875909(63) \\
 \hline
\multirow{3}{*}{$0.06$} 
& & $ll^\dagger$ & 0.038842(29) \\
& & $ls^\dagger$ & 0.142607(51) \\
& & $sc^\dagger$ &  0.565279(49)\\
 \hline
\multirow{3}{*}{$0.04$} 
& & $ll^\dagger$ & 0.028981(18) \\
& & $ls^\dagger$ & 0.106297(39) \\
& & $sc^\dagger$ & 0.42352(13)\\
\hline \hline
\end{tabularx}
\end{table*}

We employ seven of the MILC Collaboration's four-flavor lattice-QCD configurations  with dynamical up, down, strange, and charm quarks in this work \cite{MILCConfigsGitHub} (see \cref{table:ensParams}). The ensembles use the highly improved staggered quark (HISQ) action~\cite{Follana:2006rc} for the sea quarks, a Symanzik-improved gauge action~\cite{Symanzik:1983gh,Luscher:1984xn,Luscher:1985zq,Alford:1995hw,Hart:2008sq} that includes the plaquette, the $1\times2$ rectangle, and the so-called bent-chair six-link term for the gluon fields as well as tadpole improvement~\cite{Lepage:1992xa} based on the plaquette. Details of the configuration generation can be found in Refs.~\cite{MILC:2010pul,MILC:2012znn,Bazavov:2017lyh}.\footnote{The $0.09^\star$~fm ensemble was, in-part, generated by the CalLat collaboration \cite{Miller:2020evg} using retuned values of the quark masses determined by MILC \cite{FermilabLattice:2014tsy}. Of the 1000 configurations used in this work, half were generated by CalLat, and half were generated by the MILC Collaboration.} Our physical mass ensemble set includes five lattice spacings spanning the range $a\approx 0.15$--0.04~fm.
The sea-quark masses listed in \cref{table:ensParams} are tuned to the isospin-symmetric physical point, defined in \cref{eq:iso-QCD}, except for the $0.15^\prime$~fm ensemble, in which the sea-quark masses are tuned to the isospin-broken physical point of \cref{eq:iso-broken-pure-QCD}.\footnote{The 0.15$^\prime$~fm ensemble, which 
has light sea quarks with $m_u \neq m_d$, was used previously~\cite{FermilabLattice:2017wgj} to test for sea isospin-breaking effects, and we use it here for the same purpose and with improved statistics.} Refined values of the tuned sea-quark masses were obtained from an analysis in Ref.~\cite{Bazavov:2017lyh}, in which pseudoscalar-meson masses and decay constants were computed using 24 gauge ensembles with six lattice spacings ranging from $a\approx 0.15$--0.03~fm.

\Cref{table:mesonMasses} lists the pseudoscalar meson masses, labeled by their valence-quark content $qq^\prime$, obtained on the seven ensembles. On all ensembles (except $0.15^\prime$~fm), the valence quarks denoted $l,s,c$ are generated with the same masses as the sea quarks listed in \cref{table:ensParams}. On the ensembles with $a\gtrapprox 0.09$~fm we include correlation functions generated at additional light valence-quark masses. The valence quarks denoted by $u,d$ are tuned according to \cref{eq:iso-broken-pure-QCD}.\footnote{On the $0.15^\prime$~fm ensemble, the $u,d,s,c$ valence quark masses match the sea quarks, while the valence quark denoted  $l$ is adjusted so that the $ll$ pion mass takes on the physical value listed in \cref{eq:iso-QCD,eq:iso-broken-pure-QCD}.} 
To compensate for the mistuning of the light sea quarks in the 0.09~fm ensemble, a retuned light-valence quark mass (see the $rr$ entry in \cref{table:mesonMasses}) is included in the analysis. Finally, the 0.09$^{\star}$~fm data set includes $hh$ correlators generated at a valence-quark mass where the ``harmonic mean square'' of the full set of taste-pion masses, defined in Ref.~\cite{Borsanyi:2020mff}, equals the physical pion mass of \cref{eq:iso-QCD}. The resulting $hh$ Goldstone pion mass is listed in \cref{table:mesonMasses}.

\begin{figure}
\centering
\includegraphics[scale=0.79]{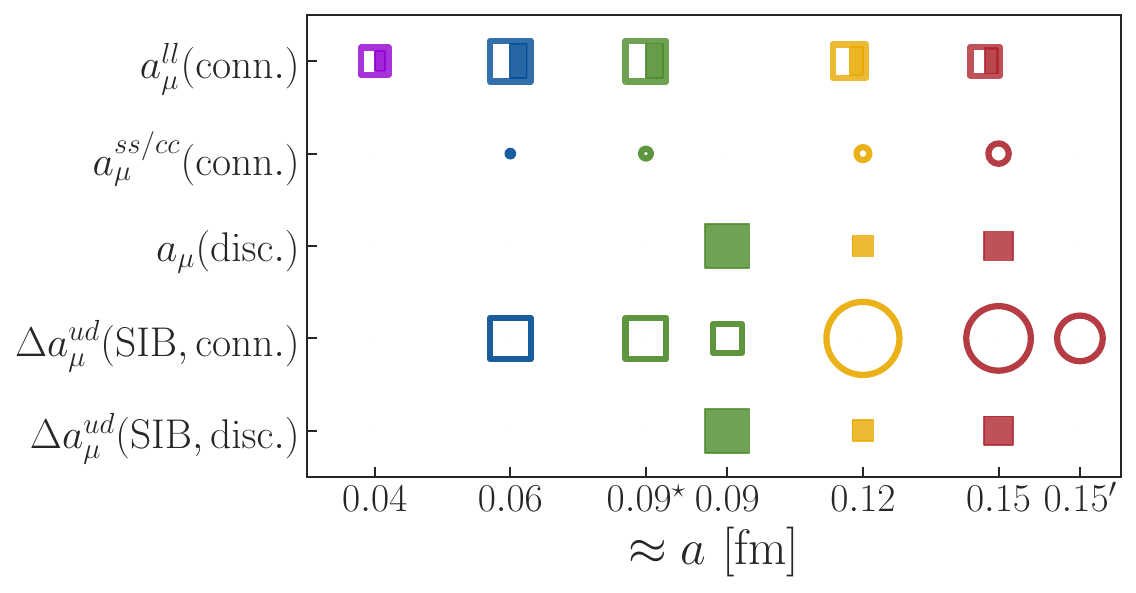}
\vspace{-5mm}
\caption{Visual representation of the correlation-function datasets described in \cref{table:lightData,table:strangeCharmData,table:discData,table:sibData} on the ensembles listed in \cref{table:ensParams} used in this work. Circles correspond to datasets computed with the truncated solver method; squares are datasets in which the low-modes of the Dirac operator are treated exactly with eigenvector methods.
Filled colors are one-link vector-current datasets; unfilled colors are local-operator datasets. The area of each symbol corresponds to the statistics of each ensemble. We stress that due to the different ways these datasets are generated, only statistical comparisons within specific observables are sensible.}
\label{fig:ensBreakdown}
\end{figure}

\begin{table*}
\centering
\caption{Renormalization factors for the local and one-link vector current used in this work. 
Results for the local current from the form-factor method are given in the first column~\cite{Chakraborty:2017hry}.
RI-SMOM results for the local and one-link currents are taken from \cite{Hatton:2019gha} and are given in columns three and four.
The 0.04~fm one-link result is obtained from the fit described in that work.}
\label{table:zvfactors}
\begin{tabularx}{\linewidth}{LLLR}
\hline \hline
$\approx a/\mathrm{fm}$ & $Z^{\textrm{FF}}_{V, \text{ local}}$ &  $Z^{\textrm{RI-SMOM}}_{V, \text{ local}}$ &  $Z^{\textrm{RI-SMOM}}_{V, \text{ one-link}}$  \\ 
\hline
$0.15$ &  0.9881(10) &  0.95932(18) &  1.14017(20)  \\ 
$0.12$ &  0.9922(4) &  0.97255(22) &  1.13784(24) \\ 
$0.09$ &  0.9940(5) &  0.98445(11) &  1.13428(13)\\ 
$0.06$ &  0.9950(6) &  0.99090(36) &  1.12518(39) \\ 
$0.04$ &  0.9949(7) &  0.99203(108)  &  1.11811(68)\\ 
\hline \hline
\end{tabularx}
\end{table*}

Two different discretizations of the vector current are employed, local and one-link, 
\begin{subequations} \label{eq:currents}
\begin{align}
    J_k^{\rm local}(x) &= \sum_f q_f \bar{\chi}_f(x) (-)^{x_k} \chi_f(x) ,
    \label{eq:vi:local} \\
    J_k^{\rm 1-link}(x) &= \sum_f q_f \half\bar{\chi}_f(x) \eta_k(x) \left( U_k(x) \chi_f(x+\hat{\bm{k}}a) + U_k^\dagger(x-\hat{\bm{k}}a) \chi_f(x-\hat{\bm{k}}a) \right),
    \label{eq:vi:1link}
\end{align}
\end{subequations}
where $\chi_f$ is the field of flavor~$f$ in the HISQ action and $\eta_k$ is the staggered sign factor that appears in place of Dirac matrices~\cite{Follana:2006rc}.
Local and one-link bilinears correspond to the taste vector and taste singlet, respectively. The resulting correlators and window observables have different discretization effects (see \cref{sec:logEffects} for a discussion of the discretization effects that arise at short distances). This is useful for constraining continuum extrapolations, as discussed in \cref{sec:windowobs}.

For each of the contributions in \cref{eq:amuBreakdown}, the vector-current correlation-function datasets that are generated on these ensembles are shown in \cref{fig:ensBreakdown}. In the figure, circles correspond to correlation function data constructed from propagators computed using the truncated solver method (TSM) \cite{Collins:2007mh,Bali:2009hu}. Specifically, using random-wall sources, we compute one fine-residual conjugate gradient solve and a number of loose-residual solves ($\nloose$). The size of the circle corresponds to the total number of statistics $\nconf\times\nloose$. The exact numbers for these, for each contribution, are detailed in the respective sections in \cref{sec:analysis}. The squares correspond to correlation function data generated using low-mode averaging (LMA)~\cite{DeGrand:2004qw,Giusti:2004yp,Blum:2012uh}, as explained below. In order to minimize autocorrelations in our LMA calculation, we select an evenly-spaced subset of the stored configurations in each ensemble that includes, at most, every other configuration.
Again, the size of the square corresponds to the total statistics, $\nconf\times\nsrc$, with specific numbers given in relevant subsections of \cref{sec:analysis}. The unfilled symbols are datasets generated using the local current, while filled symbols indicate data generated using the one-link current. In the case of the $\amuL$, both currents are employed. For all our datasets, we perform extensive tests for autocorrelations, specifically both measuring the integrated autocorrelation time and also examining how the statistical errors change with binning. For the strange and charm datasets, we account for the significant autocorrelations identified in this procedure by binning with bin sizes up to ten, leading to the small numbers of independent configurations seen in \cref{fig:ensBreakdown}. We see no indications of autocorrelations in any of the other data sets.

For the LMA procedure, we use color-diagonal random wall (RW) sources to approximate the high-mode remainder. Specifically, 2,000 exact eigenvector low-mode pairs enter the correlation-function generation in three important ways. First, they are used to calculate the low-mode all-to-all (A2A,LL) contribution to the correlation function, which gives the exact low-mode result for a given lattice configuration. Second, low modes are used to deflate the conjugate gradient solver for each RW source, significantly reducing the iterations required to reach the target residual of the solver. Finally, the low-mode contribution to the RW correlation function (RW,LL) is obtained as a byproduct of the deflation process; see \cref{eq:rwll}.

For the connected contribution we use
\begin{align} \label{eq:lma}
    C^{ff}_{\textrm{LMA, conn.}}(t) &= \frac{1}{3}\sum_{k=1}^{3}\left\langle C^{ff,k}_{\textrm{RW}}(U,t_0,t_1)-C^{ff,k}_{\textrm{RW,LL}}(U,t_0,t_1)+C^{ff,k}_{\textrm{A2A,LL}}(U,t_0,t_1) \right\rangle_U \,,
\end{align}
where $t = t_1 - t_0$,  $\langle\dots\rangle_U$ denotes averaging over the gauge-field ensemble, $f$ labels the quark flavor, and
\begin{subequations}     
\begin{align}
    C^{ff,k}_{\textrm{RW}}(U, t_0,t_1) &= \frac{1}{4N_S^3}\; q_f^2\Tr{M_f^{-1}(U)\Gamma^k_{t_0}\xi_{t_0}\xi_{t_0}^\dagger M_f^{-1}(U)\Gamma^k_{t_1} } , \\
    C^{ff,k}_{\textrm{RW,LL}}(U, t_0,t_1) &=\frac{1}{4N_S^3}\; q_f^2\Tr{M_{L,f}^{-1}(U)\Gamma^k_{t_0}\xi_{t_0}\xi_{t_0}^{\dagger} M_{L,f}^{-1}(U)\Gamma^k_{t_1}} \, , \label{eq:rwll}\\
    C^{ff,k}_{\textrm{A2A,LL}}(U, t_0,t_1) &=\frac{1}{4N_S^3}\; q_f^2\Tr{M_{L,f}^{-1}(U)\Gamma^k_{t_0} M_{L,f}^{-1}(U)\Gamma^k_{t_1}}\, . \label{eq:LMA:A2A}
\end{align}
\end{subequations}
Here, the trace is over color and space-time components; $N_S^3$ is the spatial volume; the $1/4$ is a rooting factor; $\xi_t$ denotes a random wall (RW) source at time $t$, the average over which is implicit; $M_f(U)$ is the Dirac matrix for the configuration for flavor $f$; $M^{-1}_{L,f}(U)=\sum_{i}\frac{1}{\lambda_{i,f}(U)}v_{i}(U)v^{\dagger}_{i}(U)$, where $v_{i}(U)$ and $\lambda_{i,f}(U)$ are the eigenvectors and eigenvalues of the Dirac matrix satisfying $M_f(U)v_i(U)=\lambda_{i,f}(U)v_{i}(U)$; and the $\Gamma_t$ matrix incorporates the staggered phases and shifts in the definition of the currents given in \cref{eq:vi:local,eq:vi:1link} on a selected time slice $t$, as follows for $x = (\mathbf{x},t)$ and $x^\prime = (\mathbf{x^\prime},t^\prime)$:

\begin{subequations}     
\begin{align}
    \Gamma^k_{t_0,\mathrm{local}}(x,x^\prime) &= (-)^{x_k}\delta_{\mathbf x, x^\prime} \delta_{t,t_0}\delta_{t_0,t^\prime}\label{eq:gammaLocal} \, ,\\
    \Gamma^k_{t_0,\rm{1-link}}(x,x^\prime) &= \eta_k(x) U_k(x)\delta_{{\mathbf x} + \mathbf{i}a, \mathbf{x^\prime}} \delta_{t,t_0}\delta_{t_0,t^\prime} + \mathrm{h.c} \, .\label{eq:gammaOneLink}
\end{align}
\end{subequations}

For the disconnected contribution, we use
\begin{align} \label{eq:lmadisc}
    C_{\textrm{LMA, disc.}}(t) &= \frac{1}{3}\sum_{k=1}^{3}\left\langle C^{k}_{\textrm{RV}}(U,t_0,t_1)-C^{k}_{\textrm{RV,LL}}(U,t_0,t_1)+C^{k}_{\textrm{A2A,LL}}(U,t_0,t_1) \right\rangle_U \,,
\end{align}
and
\begin{subequations}     
\begin{align}
    C^k_{\textrm{RV}}(U,t_0,t_1) &=-\frac{1}{16N_S^3}\sum_{f,f^\prime}q_fq_{f^\prime}\Tr{M_f^{-1}(U)\Gamma^k_{t_0}\xi\xi^{\dagger}}\Tr{\xi^\prime\xi^{\prime\dagger} M_{f^\prime}^{-1}(U)\Gamma^k_{t_1}} \, , \\
    C^k_{\textrm{RV,LL}}(U,t_0,t_1)&=-\frac{1}{16N_S^3}\sum_{f,f^\prime}q_fq_{f^\prime}\Tr{M_{L,f}^{-1}(U)\Gamma^k_{t_0}\xi\xi^{\dagger}}\Tr{\xi^\prime\xi^{\prime\dagger} M_{L,f^\prime}^{-1}(U)\Gamma^k_{t_1}}  \, ,\\
    C^k_{\textrm{A2A,LL}}(U,t_0,t_1)&=-\frac{1}{16N_S^3}\sum_{f,f^\prime}q_fq_{f^\prime}\Tr{M_{L,f}^{-1}(U)\Gamma^k_{t_0}}\Tr{ M_{L,f^\prime}^{-1}(U)\Gamma^k_{t_1}}\,, \label{eq:LMA:A2Adisc}
\end{align}
\end{subequations}
where taste-symmetry requires the one-link current. Here, the random sources $\xi,\xi^\prime$ have support over the entire lattice volume (RV). The flavor sum is over $f,f^\prime = u,d,s$, where in the isospin symmetric case $M_{u}(U) = M_{d}(U) = M_{l}(U)$. On some of the ensembles, we use the TSM method to reduce the cost of the inversions. Further details, including numbers of random sources, are discussed below and in Refs.~\cite{Yamamoto:2018cqm, YamamotoPhD}.

Two different schemes are employed to renormalize the currents. The first set of renormalization factors, $Z^{\textrm{FF}}_V$, are obtained from a form factor approach~\cite{Chakraborty:2017hry}. They were used in previous Fermilab, HPQCD, and MILC HVP calculations~\cite{Davies:2019efs,FermilabLatticeHPQCD:2023jof} in which the local current was employed. The second approach employs the RI-SMOM scheme, and the factors, $Z^{\textrm{RI-SMOM}}_V$, are given in Refs.~\cite{Hatton:2019gha,Hatton:2020qhk}. The RI-SMOM scheme has a residual dependence on the matching scale $\mu$, even for $Z_V$, and we take the $Z^{\textrm{RI-SMOM}}_V$ values at $\mu=2$~GeV from those works. The value of $Z^{\textrm{RI-SMOM}}_V$ for the one-link current at 0.04~fm was not calculated in those works; hence, we obtain it from the fit procedure described in Ref.~\cite{Hatton:2019gha}.
\Cref{table:zvfactors} tabulates the values of all vector-current renormalization factors $Z_V$ needed in the analysis explained in the subsequent sections.

\section{Analysis Procedure}\label{sec:analysis}

Our overarching analysis approach follows the strategy described in Ref.~\cite{FermilabLatticeHPQCD:2023jof}. Here, we recap the main steps of our analysis strategy, with a focus on the procedures specific to this work. The window observables are obtained according to \cref{eqn:SDDef,eqn:WDef} from weighted Euclidean-time integrals of the correlation functions $C(t)$. 
In Ref.~\cite{FermilabLatticeHPQCD:2023jof}, we examined the effects of the oscillating contributions, which arise in our staggered correlation functions, on intermediate-distance $\amu$ observables in the continuum limit.  Comparing results obtained from fit reconstructions that remove the oscillating terms versus direct integration of the correlation functions, we found that the oscillating contributions have no effect on the observables after continuum limit extrapolation. Extending this study to the short-distance window observables in \cref{sec:oscEffects}, we find the same result. Hence, in this work, we use only direct integration to obtain the short- and intermediate-distance window observables. 

The following subsections describe our strategies for each analysis step. First, \cref{sec:blind} provides a description of the blinding applied to each individual window observable. The lattice corrections for finite-volume effects, pion-mass mistunings and in some cases taste-breaking effects are detailed in \cref{sec:latticeCorrections}. These corrections are applied only to observables containing light-quark contributions ($\amuLSD$, $\amuLW$, $\amuDSD$, $\amuDW$, $\amuSIBSD$, $\amuSIBW$).
Subsequently, all observables are extrapolated to the continuum, which is described in \cref{sec:contextrap}.
In the final step, the analysis variations for the corrections and continuum extrapolations are input into a BMA, as discussed in \cref{sec:BMA}. 
We consider a broad range of analysis variations, to encompass \emph{all} reasonable choices, and obtain the corresponding systematic uncertainty using BMA; in addition, detailed cross-checks of analysis methods and code are provided by at least two different people.

\subsection{Blinding} \label{sec:blind}

As an overview, first all window observables are blinded including a multiplicative blinding factor, with some including an additional additive blind:
\begin{equation}
    \amu^{(X),{\rm blind}} = \alpha^{(X)} \amu^{(X)} + \beta^{(X)}.
\end{equation}
The additive blind is present for the local current light-quark connected dataset on the 0.15~fm ensemble (see \cref{table:ensParams}), where previously used data are used in this analysis; all the other correlation function data used in the light-quark connected analysis are new.  In addition, we use an 
additive blind for the connected $\amuSIB$ analysis to avoid cancellation of blinding factors when constructing the ratio to $\amuL$.  

For the light-quark connected analysis (\cref{subsec:light_analysis}), the blinding is applied to the renormalization factor $Z_V$, which is then squared in the observable $\amu$.  The factors multiplying $Z_V$ are drawn uniformly over the range $[0.8, 1.2]$, resulting in a multiplicative blind $\alpha$ in the range $[0.64, 1.44]$.  The additive blind $\beta$ for the 0.15~fm ensemble is set equal to the standard deviation of that $\amu$ result times a random number drawn uniformly from the range $[-1,1]$.  The strange and charm analyses (\cref{subsec:sc_analysis}) are blinded in exactly the same way.  For the disconnected analysis (\cref{subsec:disc_analysis}), $\alpha$ is drawn from the range $[0.8, 1.2]$, and no additive blind is used.  Finally, for the SIB analysis (\cref{subsec:sib_analysis}), $\alpha$ is drawn from the range $[0.95, 1.05]$ (connected) and $[0.6, 1.4]$ (disconnected), and the connected observables include an additional additive blind $\beta$ drawn uniformly from $[-2, 2] \times 10^{-10}$.  

The blinds are held in place until all individual analyses are complete. The analyses are then frozen, the blinds are removed, and the analysis scripts are rerun. All blinds are implemented at the software level using a blinding flag rather than modifying any stored data, so that there is no possibility of incomplete removal of blinds when obtaining our final analysis results.

\subsection{Lattice corrections}\label{sec:latticeCorrections}

We perform our corrections for finite volume (FV), pion-mass mistuning, and taste breaking (TB) by expressing the physical point as
\begin{equation}
    a_{\mu}\left(L_{\infty}, M_{\pi_{\text {phys. }}}\right)=a_{\mu}\left(L_{\text {latt.}}, M_{\pi_{\text {latt.},\, \xi_{1}}}, \ldots, M_{\pi_{\text {latt.},\,  \xi_{16}}}\right)+\Delta_{\mathrm{FV}}+\Delta_{M_{\pi}}+\Delta_{\mathrm{TB}},  \label{eq:corrScheme}
\end{equation}
where the first term on the RHS corresponds to lattice data and
\begin{flalign}
\Delta_{\mathrm{FV}}&=a_{\mu}\left(L_{\infty}, M_{\pi_{\mathrm{latt.},\, \xi_{1}}}, \ldots, M_{\pi_{\mathrm{latt.},\, \xi_{16}}}\right)-a_{\mu}\left(L_{\text {latt.}}, M_{\pi_{\mathrm{latt.},\, \xi_{1}}}, \ldots, M_{\pi_{\mathrm{latt.},\, \xi_{16}}}\right), \label{eq:corrFV} \\
\Delta_{M_{\pi}}&=a_{\mu}\left(L_{\infty}, M_{\pi_{\mathrm{phys.},\, \xi_{1}}}, \ldots, M_{\pi_{\mathrm{phys.},\, \xi_{16}}}\right)-a_{\mu}\left(L_{\infty}, M_{\pi_{\mathrm{latt.},\, \xi_{1}}}, \ldots, M_{\pi_{\mathrm{latt.},\, \xi_{16}}}\right), \label{eq:corrMpi} \\
\Delta_{\mathrm{TB}}&=a_{\mu}\left(L_{\infty}, M_{\pi_{\mathrm{phys.}}}\right)-a_{\mu}\left(L_{\infty}, M_{\pi_{\mathrm{phys.},\, \xi_{1}}}, \ldots, M_{\pi_{\mathrm{phys.},\, \xi_{16}}}\right).  \label{eq:corrTB}
\end{flalign}
In particular, we correct the light-quark containing observables, where our estimates of the corrections are based on the use of effective field theory (EFT) and EFT-inspired correction schemes that capture the dominant low-energy, two-pion physics contribution to $\amuHVP$. We consider variations obtained from five different approaches: next-to-leading-order (NLO) and next-to-next-to-leading order (NNLO) chiral perturbation theory (\chpt)  \cite{Aubin:2015rzx,Bijnens:2017esv,Aubin:2020scy,Aubin:2019usy,Borsanyi:2020mff,Aubin:2022hgm}, the chiral model (CM) \cite{Chakraborty:2016mwy} employed in Ref.~\cite{Davies:2019efs}, the Meyer-Lellouch-L\"uscher-Gounaris-Sakurai (MLLGS) approach \cite{Gounaris:1968mw, Luscher:1991cf, Luscher:1990ux,Lellouch:2000pv, Lin:2001ek, Meyer:2011um,Francis:2013fzp,DellaMorte:2017dyu}, and the Hansen-Patella (HP) scheme~\cite{Hansen:2020whp}. These low-energy effects are heavily suppressed in the short-distance window region, which is also outside the range of validity of most of these EFT-based schemes. The HP scheme, however, is a complete resummation of the perturbative theory of relativistic pions, hence is convergent at any energy scale.  Therefore, we employ only the HP scheme for short-distance window observables.  In order to simplify our analysis, we have chosen throughout this work not to mix different EFT schemes for FV, $M_\pi$, and TB corrections.  Because the HP scheme yields an explicit finite-volume difference, it is not straightforward to obtain infinite-volume pion-mass corrections from it. Hence we omit the HP scheme from analyses where this mass correction is required, namely $\amuLW$ (\cref{subsec:light_analysis}) and $\amuDW$ (\cref{subsec:disc_analysis}).  For these quantities, the predictions of HP for FV corrections are closely consistent with \chpt\ as expected (see the end of Sec. 2.1 in Ref.~\cite{Hansen:2020whp} for a discussion), so including the HP scheme would not significantly alter our estimates of finite-volume effects.

\begin{figure}[t]
\centering
\includegraphics[width=0.7\textwidth]{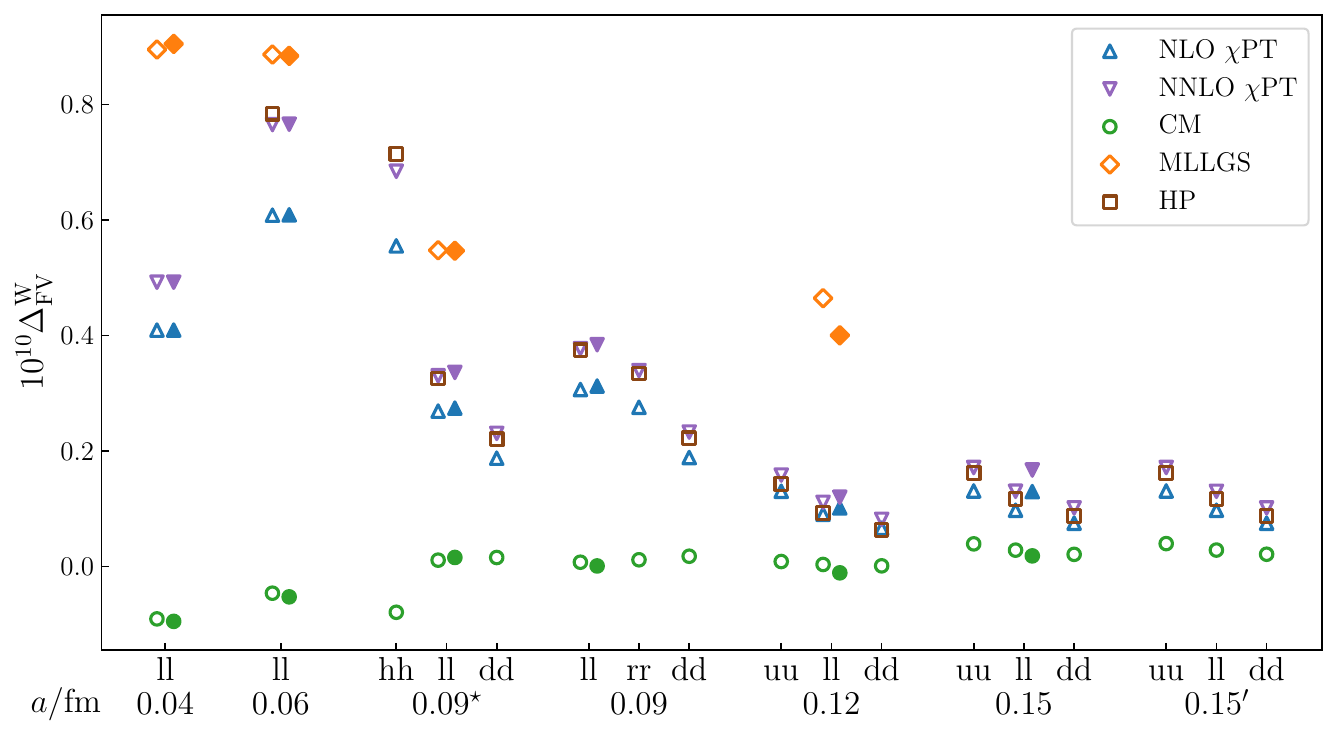}
\vspace{-5mm}
\caption{Finite volume corrections for each ensemble and valence-quark content for light-quark contributions in the intermediate window region. The unfilled symbols correspond to the local current and the filled to the one-link.}
\label{fig:FVCorr}
\end{figure}

\begin{figure}[t]
\centering
\includegraphics[width=0.5\textwidth]{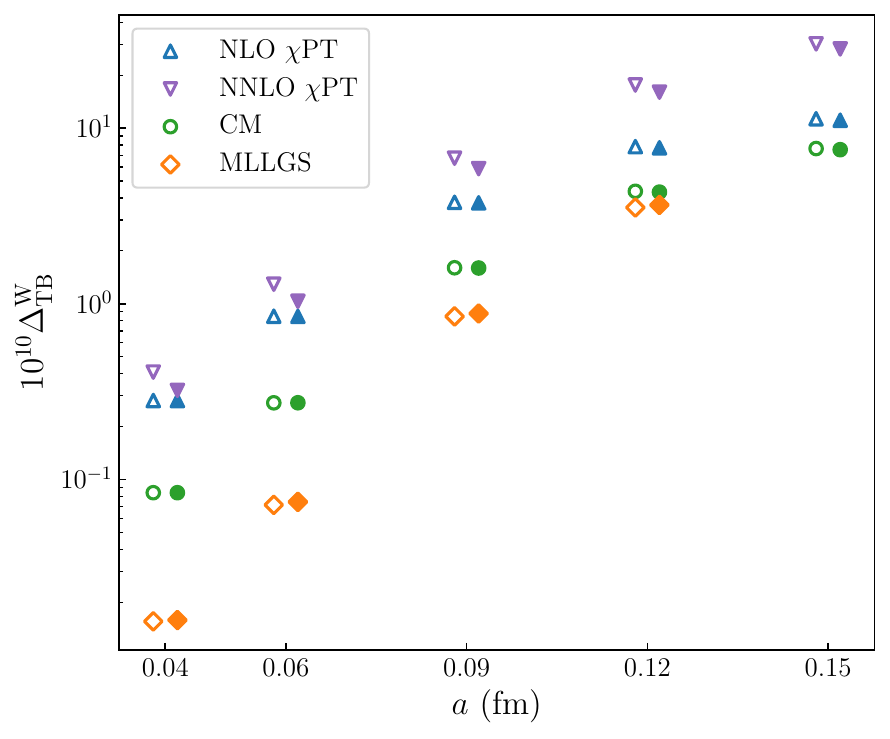}
\vspace{-5mm}
\caption{Taste breaking corrections for light-quark contributions in the intermediate window region for each lattice spacing at the light valence-quark value.  The unfilled symbols correspond to the local current and the filled to the one-link.}
\label{fig:TBCorr}
\end{figure}

\begin{figure}[t]
\centering
\includegraphics[width=0.7\textwidth]{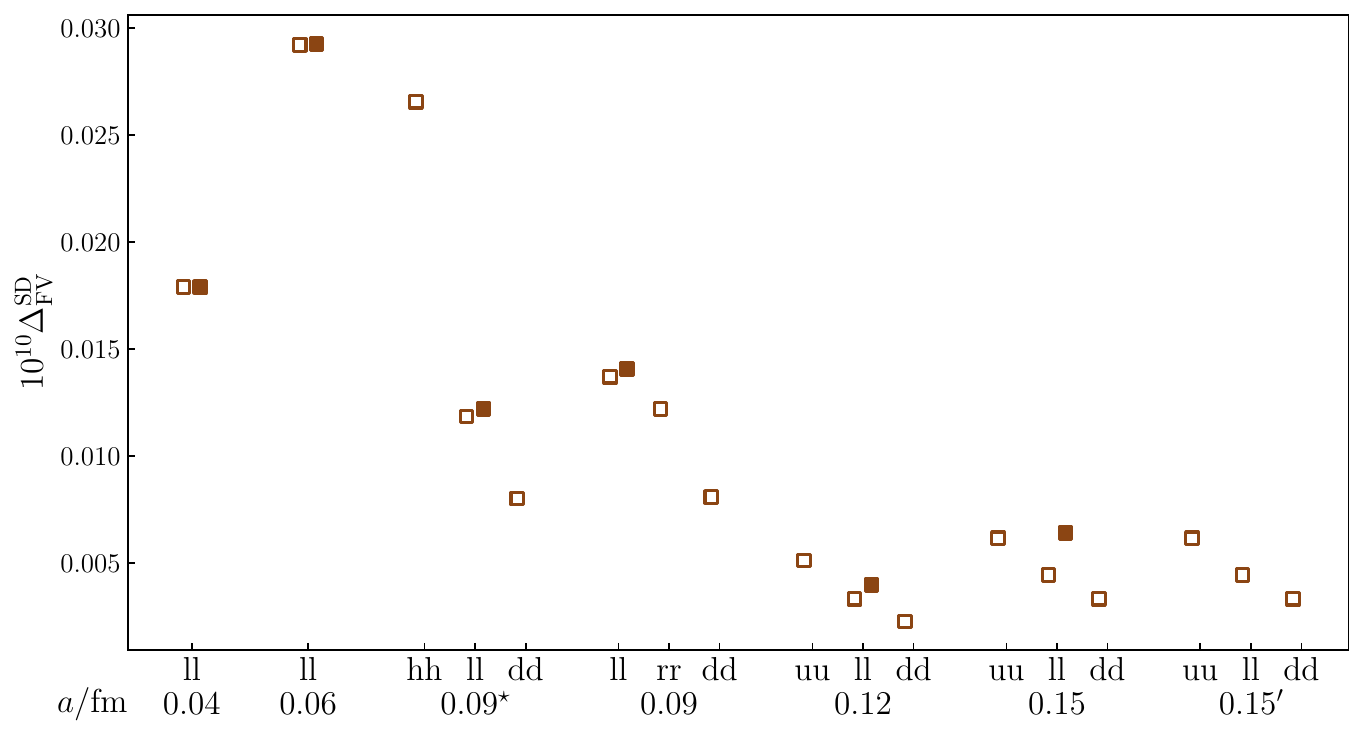}
\vspace{-5mm}
\caption{Finite volume corrections in the Hansen-Patella scheme for each ensemble and valence-quark content for light-quark contributions in the short-distance window region.  The unfilled symbols correspond to the local current and the filled to the one-link.}
\label{fig:FVCorrSD}
\end{figure}

The finite-volume and taste-breaking corrections computed for the intermediate-window observables are given in \cref{fig:FVCorr,fig:TBCorr}, respectively; the finite-volume corrections obtained from the HP scheme, used for the connected contributions in the short-distance window, are given in \cref{fig:FVCorrSD}. The corrections to the disconnected contribution are obtained by multiplying their connected counterparts by $-1/10$.  We do not apply MLLGS to the data at 0.15~fm because the heaviest-taste two-pion pair lies above the $\rho$ mass; see the discussion in the MLLGS section of Ref.~\cite{Borsanyi:2020mff}. The pion-mass mistuning corrections are small where applicable  (at the sub-permille in the intermediate window) due to the accurate tuning of the ensemble masses, and hence are not shown.

\Cref{fig:FVCorr,fig:TBCorr} show that the expected dependencies on the lattice spacing, lattice volumes, and pion masses are largely followed.  Any slight deviations from the expected behavior are likely the result of error introduced by applying the correction schemes slightly outside their range of validity; the use of several EFT and EFT-inspired schemes accounts for this error in the BMA systematic uncertainty.  Additionally, one of the largest sources of systematic uncertainty for $\amuLW$, $\amuDW$, and $\amuSIBCW$ results from the discrepancy between different finite-volume correction schemes---specifically between CM (and, to a lesser extent, NLO  $\chi$PT) and the other EFT and EFT-inspired approaches. Correlations between observables due to this systematic uncertainty are discussed in \cref{sec:bma_cov}. For contributions from heavier flavors, finite volume effects are expected to be negligible compared with the level of precision of our calculation \cite{ExtendedTwistedMass:2024nyi}. 

\subsection{Continuum extrapolations} \label{sec:contextrap}

For each observable, we consider variations of the continuum fit function and the data included , dropping data at certain lattice spacings when appropriate.\footnote{We do not perform fits where the number of degrees of freedom is negative.} The general form of the continuum fit function employed for almost all observables, except for $\amuSIBC$ (see \cref{eq:sib-aMu-conn-cont}), is
\begin{align}
    a_{\mu}(a, M_A)&= a_{\mu}\left[1 + F^{{a}} (a) + F^M (M_A)\right], \label{eq:generalfitfunc}
\end{align}
where $F^{{a}} (a)$ is adjusted as needed to describe the discretization effects of each observable. The function $F^M (M_A)$ accounts for any residual mass-mistuning effects in our extrapolations and is given by
\begin{align}
    F^M (M_A) &= C_{\text{sea}} \sum_{A=\pi, K} \frac{M_{\pi,\,\textrm{phys.}}^2}{M_{A, \,\textrm{phys.}}^2} \delta M^2_A \ , \quad \quad 
    \delta M^2_A = \frac{M_{A, \,\textrm{phys.}}^2-M_{A, \,\textrm{latt.}}^2}{2B_0 \Lambda},\label{eq:mfunc}
\end{align}
unless otherwise specified in the individual analysis subsections of \cref{sec:windowobs}. 
Here $B_0$ is the GMOR constant of proportionality, obtained as $B_0 = \Sigma_0 / F_0^2$ using the values from Refs.~\cite{FlavourLatticeAveragingGroupFLAG:2021npn,MILC:2009ltw,MILC:2010hzw} for SU(3) chiral perturbation theory, and $\Lambda$ is defined below. The parameterization in terms of the meson masses makes it straightforward to match to the isospin-symmetric and isospin-broken QCD schemes defined in \cref{subsec:inputs}.
We impose a stabilizing Gaussian prior of $C_{\text{sea}}=0(2)$. We note that our continuum extrapolations are insensitive to this prior's width as we find that our results (central values and errors) for all observables presented in this paper are essentially unchanged when we increase the width by a factor of ten.

For each $\amuHVP$ observable, we use an empirical Bayes procedure to obtain guidance for the important terms in the continuum fit functions. This procedure was outlined at the end of Sec.~III\,D of Ref.~\cite{FermilabLatticeHPQCD:2023jof}.
Here, we use the following functional form for $F^{{a}} (a)$ in our empirical Bayes fits: 
\begin{align}
F^{{a}}_{\rm emp} (a) &= 
c_{1l} \,x \,\log(a\Lambda) + 
 \sum_{i=1}^4 \sum_{j=0}^4 c_{ij} x^i  \alpha_s^j , \quad \quad x \equiv (a\Lambda)^2. 
\label{eq:GPLfunc}
\end{align}
The scale $\Lambda$ is chosen to maximize the Gaussian Bayes Factor (see Eq.~(28) of Ref.~\cite{Lepage:2001ym}),  which is proportional to the marginal likelihood (model evidence) coefficients in \cref{eq:GPLfunc}. The coefficients are constrained with Gaussian priors $c_{1l},c_{ij} = 0(1)$. Following Ref.~\cite{Bazavov:2017lyh} we use $\alpha_s = \alpha_V(2/a)$ and take $\alpha_V (n_f=4, \mu=5.0~{\rm GeV}) = 0.2530(38)$ from Ref.~\cite{Chakraborty:2014aca}, where we evolve the coupling using the four-loop beta function~\cite{vanRitbergen:1997va,Czakon:2004bu,Zoller:2016sgq}. \Cref{eq:GPLfunc} accounts for log-enhancement effects in short-distance window observables \cite{DellaMorte:2008xb,Ce:2021xgd,Alexandrou:2022amy,Sommer:2022wac}, by allowing for a leading $a^2$ term instead of the expected $a^2 \alpha_s$ for the local current and for a leading $a^2 \log(a)$ for the one-link current; see \cref{sec:logEffects} for further discussion. 
 
The continuum fit functions used in the analyses of the individual HVP observables are specified in the respective subsections of \cref{sec:windowobs}. They are also informed by these theoretical expectations, while the empirical Bayes analyses provide guidance for the scale $\Lambda$ and the sub-leading discretization terms. In order to regulate the degrees of freedom in the continuum fits, we include a prior on the highest order terms of
\begin{align}\label{eq:emp_bayes_prior}
    C_{a^{k}} = 0(2).
\end{align}
This prior has twice the width of the priors used in the empirical Bayes analysis; hence, it does not significantly influence our continuum limit beyond assisting with stability in some cases.

\subsection{Bayesian Model Averaging} \label{sec:BMA}

In the final step, the uncertainties associated with the systematic corrections, needed to obtain the observables in the continuum and infinite volume limits and at the physical point, are estimated using BMA~\cite{Jay:2020jkz,Neil:2022joj}. In the BMA language, the systematic corrections are referred to as ``models.'' This analysis approach is detailed in Ref.~\cite{FermilabLatticeHPQCD:2023jof}. In the context of lattice field theory calculations, BMA using the Akaike information criterion in a frequentist context was pioneered in Refs.~\cite{BMW:2014pzb,Berkowitz:2017gql}; an early use of BMA for lattice QCD was presented in Ref.~\cite{Chang:2018uxx}. The formulas for the mean and variance from the BMA are given by
\begin{align}
    \left\langle a_{\mu}\right\rangle &= \sum_{n=1}^{N_{M}}\left\langle a_{\mu}\right\rangle_{n} \operatorname{pr}\left(M_{n} \mid D\right), \label{eq:BMAMean} \\
    \sigma_{a_{\mu}}^{2}&= \sum_{n=1}^{N_M} \sigma_{a_{\mu}, n}^{2} \mathrm{pr}\left(M_{n} \mid D\right)+\sum_{n=1}^{N_M}\left\langle a_{\mu}\right\rangle_{n}^{2} \mathrm{pr}\left(M_{n} \mid D\right)-\left\langle a_{\mu}\right\rangle^{2}, \label{eq:BMAVar} 
\end{align}
where $N_M$ is the number of models and the probability of a specific model $M$ given the data $D$ is defined through the Bayesian Akaike information criterion (BAIC) weight~\cite{Neil:2022joj,Neil:2023pgt},
\begin{equation}
    \pr(M \mid D) \equiv \pr(M) \exp \left[-\frac{1}{2}\left(\chi_{\rm data}^{2}\left(\mathbf{a}^{\star}\right)+2 k+2 N_{\mathrm{cut}}\right)\right]. \label{eq:modelProb}
\end{equation}
Here, $\chi_{\rm data}^{2}$ is the standard chi-squared function, {\it not} including the contribution of the priors, and $\mathbf{a}^{\star}$ is the posterior mode ({\it i.e.}, the best-fit point for the vector of fit parameters $\mathbf{a}$ when optimized against the augmented chi-squared function \cite{Lepage:2001ym}).  $N_{\mathrm{cut}}$ is the number of data points cut from a data set---in this case, the number of ensembles omitted from a given extrapolation.  The parameter $k$ is the number of independent parameters in a given fit function. The factor $\pr(M)$ is the prior probability of a given $M$.  

For most observables, we take $\pr(M)=1/N_M$, which corresponds to the minimal \textit{a priori} biasing of some models in favor of others.  For $\amuLW$ and $\amuSIBC$, the flat prior $\pr(M)=1/N_M$ implicitly biases the BMA in favor of certain types of models over others due to combinatorial effects; see \cref{subsec:light_analysis,subsubsec:conn_sib_analysis} for the choices of model priors for these observables. The total uncertainties on model-averaged results are given as one-sigma deviations, as derived from the BMA variance of the mean in \cref{eq:BMAVar}. Our procedure for obtaining complete error budgets from the respective BMA analyses is described in Ref.~\cite{FermilabLatticeHPQCD:2023jof}, with any additional considerations described in the relevant analysis sections below.

The probability weights defined in \cref{eq:modelProb} can be used to assess the relative weight of specific analysis choices in the BMA. Comparison of these weights can identify if a particular correction scheme or fit-function variation is preferred or suppressed by the averaging procedure. This identification is achieved by computing the ``subset probability'' of the model subset $S$ from the relative posterior probability of the variations contained in $S$:
\begin{equation}
    \pr(S | D) = \sum_{M_i \in S} \pr(M_i | D). \label{eqn:ssProb}
\end{equation}
The subset probability encapsulates the relative weight of the models in a given subset compared with the whole model space as informed by the data (see, for example, the pie charts in \cref{fig:BMACompareWL,fig:BMACompareSDL}).

In order to account for correlations in the different contributions in  \cref{eq:amuBreakdown} due to common parameter choices and common ensembles, random-sources, and datasets, we use gaussian error propagation, from the software package \textsc{gvar} \cite{gvarGitHub}.
Specifically, we perform separate model averages for each contribution to $\amu$.  To account for statistical and parametric correlations between each contribution, we perform these model averages in series and compute the covariance via
\begin{align}\label{eq:covStat}
\cov^{\rm stat}[\amu^i,\amu^j] = \sum_{m=1}^{N_{M^i}} \cov_{m}[\amu^i,\ev{\amu^j}] \pr(M_m^{i}|D),
\end{align}
where $\cov_{m}[\amu^i,\ev{\amu^j}]$ are the statistical and parametric covariances between $\ev{a^i}_m$ (the individual model results from $\{M_m^i\}$) and $\ev{\amu^j}$ (the model averaged over $\{M_n^j\}$) and are estimated using \textsc{gvar}.  \Cref{eq:covStat} is derived in \cref{sec:bma_cov} under the assumption that the model spaces for each respective contribution are independent, in which case the covariance due to model spread cancels exactly.  
In specific cases where there is significant systematic correlation due to common model space, namely using the same scheme for correcting for finite-volume effects in light-quark observables, we adopt a conservative approach, assuming 100\% model correlation.  The treatment of this correlation adds the additional term
\begin{equation}\label{eq:covTot}
    \cov[a_{\mu}^i,a_{\mu}^j]= \cov^{\rm stat}[a_{\mu}^i,a_{\mu}^j]+\sigma^{\rm FV}_{a_{\mu}^i}\sigma^{\rm FV}_{a_{\mu}^j},
\end{equation}
where $\cov^{\rm stat}[a_{\mu}^i,a_{\mu}^j]$ is estimated using \cref{eq:covStat} and $\sigma^{\rm FV}_{a_{\mu}^i}$ is the contribution to the systematic uncertainty of $a_{\mu}^i$ that is associated with the spread of the  finite-volume correction.  \Cref{eq:covTot} is used to account for correlations between $\amuLW$, $\amuDW$, and $\amuSIBCW$.  Details and derivations of these covariance formulas are given in \cref{sec:bma_cov}. We cross-check our variances obtained from the
\textsc{gvar} approach against
a global bootstrap based approach.

As a final note, the results for each observable obtained from the empirical Bayes procedure, described in \cref{sec:contextrap} can be compared with what we obtain from the corresponding BMA analyses when varying only the continuum fit functions. We find good consistency in the central values and errors reported from the two approaches, in line with the findings of Ref.~\cite{FermilabLatticeHPQCD:2023jof}.

\section{Window observables} \label{sec:windowobs}

\subsection{Light-quark connected}\label{subsec:light_analysis}

\begin{table*}
\centering
\caption{Vector-current correlator datasets used in the calculation of $\amuL$. The first column lists the approximate lattice spacings in~fm. The ensemble parameters are contained in \cref{table:ensParams}. The second and third (fourth and fifth) columns list the number of configurations and wall-sources analyzed for the local (one-link) current. The final column lists the number of eigenvector pairs per configuration.\vspace{1mm}} 
\label{table:lightData}
\begin{tabularx}{\linewidth}{LCCCCR}
\hline \hline
 & \multicolumn{2}{c}{Local} & \multicolumn{2}{c}{One-link} &  \\
$\approx a/\mathrm{fm}$ & $\nconf$ & $\nsrc$ & $\nconf$ & $\nsrc$ & $\neig$ \\ 
\hline
$0.15$       & 957 & 48 & 957 & 48 & 1000\\
$0.12$       & 1060 & 64 & 1060 & 64 & 2000\\ 
$0.09^\star$ & 993 & 96  & 993 & 96 & 2000\\  
$0.06$       & 1009 & 96 & 900 & 96 & 2000\\
$0.04$       & 313 & 144  & 256 & 144 & 2000\\
\hline \hline
\end{tabularx}
\end{table*}

The light-quark connected contribution to $\amuHVP$ is approximately $ 85\%$ and $ 70\%$ of the total intermediate- and short-distance windows, respectively. Hence, the uncertainty on $\amuLW$ and $\amuLSD$ is a limiting factor on the uncertainty of the complete $\amuW$ and $\amuSD$ observables. In this work, we update our previously published result for $\amuLW$, which has a relative uncertainty of $0.5\%$~\cite{FermilabLatticeHPQCD:2023jof}. This updated analysis includes new correlation function data sets, generated using low-mode-averaging on all ensembles (see \cref{subsec:lattice}). These newly generated correlation functions come with reduced statistical noise at large Euclidean times. However, we find that the new data also have better statistical precision at intermediate distances. 

One of the significant sources of systematic uncertainty in the result of Ref.~\cite{FermilabLatticeHPQCD:2023jof} was the mistuned light-quark mass on the $0.09$~fm ensemble. To remedy this, we have generated correlation function data on the new $0.09^\star$~fm ensemble, which has a better tuned light-quark mass, yielding a pion mass within 2\% of the physical value. As a result, there now are no significant pion-mass corrections needed for the entire light-quark-connected data set, when we apply the lattice corrections. 
Another substantial source of uncertainty in the previous calculation was the continuum extrapolation. Here, in order to constrain the continuum limit better, we include correlation functions with a second vector-current discretization, the one-link (taste-singlet) vector current (\cref{eq:vi:1link}), on all the ensembles. In addition, we include local and one-link data sets on an ensemble with a lattice spacing of $0.04$~fm, the finest to date. We currently have fewer configurations on this ensemble, relative to the other ensembles at coarser lattice spacings. Nevertheless, the statistics is sufficient to significantly improve our continuum limit extrapolations for the two window observables considered here. Visually, the improved light-quark dataset is illustrated in \cref{fig:ensBreakdown} with the specific number of configurations, sources, and eigenvectors used in these datasets given in \cref{table:lightData}.

With these improved data, we follow the analysis strategy outlined above in \cref{sec:analysis}, which is similar to the strategy detailed in Ref.~\cite{FermilabLatticeHPQCD:2023jof}. We renormalize the respective vector-current operators with the $Z^{\textrm{RI-SMOM}}_V$ renormalization factors in \cref{table:zvfactors}. This choice takes advantage of the complete and precise set of local and one-link RI-SMOM renormalization factors that are available from Ref.~\cite{Hatton:2019gha}.

For the intermediate window observable, we create datasets corrected for finite-volume and mass-mistuning effects from the EFT-based correction schemes described in \cref{sec:latticeCorrections}.  As already mentioned above, all the correlation functions included in our analysis are generated on well-tuned ensembles, so that quark-mass mistuning corrections are sub-leading. As a result, we are now able to ground all corrections for quark-mass mistuning directly in EFT. 

With the addition of a second current to our data set, we perform independent continuum limit extrapolations, as well as joint ones, to quantify the systematic uncertainty due to residual discretization effects. In the BMA language, these variations are performed by excluding the data for one of the two currents from the continuum extrapolations. For ensembles where we do not have an equal number of local and one-link configurations (0.06~fm and 0.04~fm), we compute the correlation matrix using the configurations shared between both currents and rescale by the variances on the full datasets to obtain the final covariance matrix. Another consequence of including a second current in our data set is that we no longer include taste-breaking corrected data in our Bayesian model average. We do, however, still verify that datasets corrected for taste-breaking give consistent results in the continuum. As mentioned in the preceding section, we apply finite-volume corrections from only the HP scheme, which are at the permille level, to the short-distance observables. The sub-leading valence-mass dependence in this window is being taken care of by the mass term in the continuum fit function, \cref{eq:mfunc}.

After first verifying that independent continuum extrapolations of the local and one-link data agree, our analysis strategy is to include joint, correlated fits to both currents simultaneously. Because the leading discretization effects in the one-link and local currents differ from each other (see \cref{sec:logEffects}), a combined fit gives us a strong handle on the continuum extrapolation of $\amuL$. In this fit, the continuum extrapolated value of $a_{\mu}^{l l}$ ($\amu$ in \cref{eq:generalfitfunc}) and the mass-mistuning coefficient $C_{\text{sea}}$ in \cref{eq:mfunc} are shared parameters. 

The form of $F^{{a}} (a)$ for each observable is informed by theoretical expectations for the two currents, as well as the empirical Bayes procedure. In particular, the one-link current has leading $a^2$ discretization effects (\cref{sec:logEffects}), so we do not include any leading $\alpha_s$ suppression. The empirical Bayes procedure provides guidance for the highest-order discretization terms that can be resolved by our datasets. For $\amuLW$, this corresponds to an $a^4$ term in the fit function, while for $\amuLSD$ we find that an $a^6$ term is resolvable. In our analysis, we therefore consider continuum limit fit functions that include terms up to $a^6$ and $a^8$ for $\amuLW$ and $\amuLSD$, respectively, in order to test for stability of the extrapolated results.

The functional forms of $F^{{a}} (a)$ in \cref{eq:generalfitfunc} for the local and one-link currents are then given by
\begin{align}
    F^{{a}}_{\text{local}} (a) &= C_{a^{2},n}(a\Lambda)^{2} \alpha_s^{n} + \sum^{4}_{k=2} C_{a^{2k}} (a\Lambda)^{2k}, \quad \text{where } n = \{0\},1,2, \label{eq:discfunclocal}\\
    F^{{a}}_{\text{one-link}} (a) &=\{C_{a^{2}\log(a)} (a\Lambda)^{2}\log(a\Lambda)\} + \sum^{4}_{k=1} C_{a^{2k}} (a\Lambda)^{2k}, \label{eq:discfunconelink}
\end{align}
where the terms representing the effects of log enhancement ({\it i.e.}, the terms within $\{\ldots\}$) are included \emph{only} for extrapolations of the short-distance observable. Applying the empirical Bayes procedure for both currents independently and simultaneously, we find the following consistent results for $\Lambda$,
\begin{align}\label{eq:lambda_conn}
    \Lambda_{ll, \text{SD}} = \Lambda_{ll, \text{W}} = 0.6~{\rm GeV}.
\end{align} 

We label fit functions with $C_{a^6}=0$ as ``quadratic,'' fit functions with $C_{a^8}=0$ as ``cubic,'' and fit functions with all terms listed in $F^{{a}} (a)$ as ``quartic.'' Given the additional data point at 0.04~fm, as well as the second current, our continuum extrapolation fit function variation basis has expanded from Ref.~\cite{FermilabLatticeHPQCD:2023jof}. With the constraint that we do not perform fits with zero degrees of freedom or less, we now include variations where we drop the two coarsest spacings, not just the single coarsest. Due to the lower statistics at 0.04~fm, we also include variations for $\amuLW$ where this ensemble is dropped. Aside from the variations, discussed above, where we drop one of the two currents completely, we restrict our variations to ones in which the ensembles are dropped for both currents simultaneously.\footnote{We have verified that including variations where different data points are dropped for the different currents gives a completely consistent final BMA central value and uncertainty.}

For $\amuLSD$, as indicated in \cref{eq:discfunclocal,eq:discfunconelink}, we include an extra subset of fit variations with leading $a^2$ for the local current and include a leading $a^2 \log(a)$ for the one-link current due to the short-distance log-enhancement effects discussed in \cref{sec:logEffects}. We crosscheck this theoretical expectation in our empirical Bayes analysis. Specifically, for the one-link current, the data prefer a fit function that includes the log-enhanced term. Without it, the fit quality, as measured by the $\chidof$, is poor, and the continuum limits of the one-link and local current data are in tension with each other. In contrast, for the local-current $\amuLSD$ data, the empirical Bayes fits prefer a fit function without the log-enhanced term.

In summary, our Bayesian model average incorporates the following analysis variations:
\begin{itemize}
    \item
    \textbf{Finite-volume and mass-mistuning correction:} For $\amuLW$, variations from the following correction schemes are considered: \chpt\,(NLO and NNLO), CM, and MLLGS.  We always correct both currents using the same scheme. In the case of MLLGS, we do not perform extrapolations with the coarsest, $0.15$~fm, data point as discussed in \cref{sec:latticeCorrections}. For $\amuLSD$, we fix the finite-volume correction scheme to HP and account for mass-mistuning in the continuum fit function.
    \item
    \textbf{Correction region ($\amuLW$ only):} We include a variation on the corrections where they are computed from the range $[0.7, 1]$~fm instead of over the full W window interval. The reasoning for this is discussed in Ref.~\cite{FermilabLatticeHPQCD:2023jof}.
    \item
    \textbf{Continuum fits:} We perform continuum extrapolations using all fit function variations described above, including fits to the three and four finest ensembles, and four coarsest ensembles ($\amuLW$ only). Also included are fits that drop one of the two currents entirely.\footnote{We adjust the model priors, $\pr(M)$, of these variations so that they are equal for each current subset, which accounts for the fact that there are more variations for the local-current fits, due to different powers of $n$ in the leading $a^2 \alpha^n_s$ term.}
\end{itemize}

\begin{figure}
\centering
\includegraphics[scale=0.79]{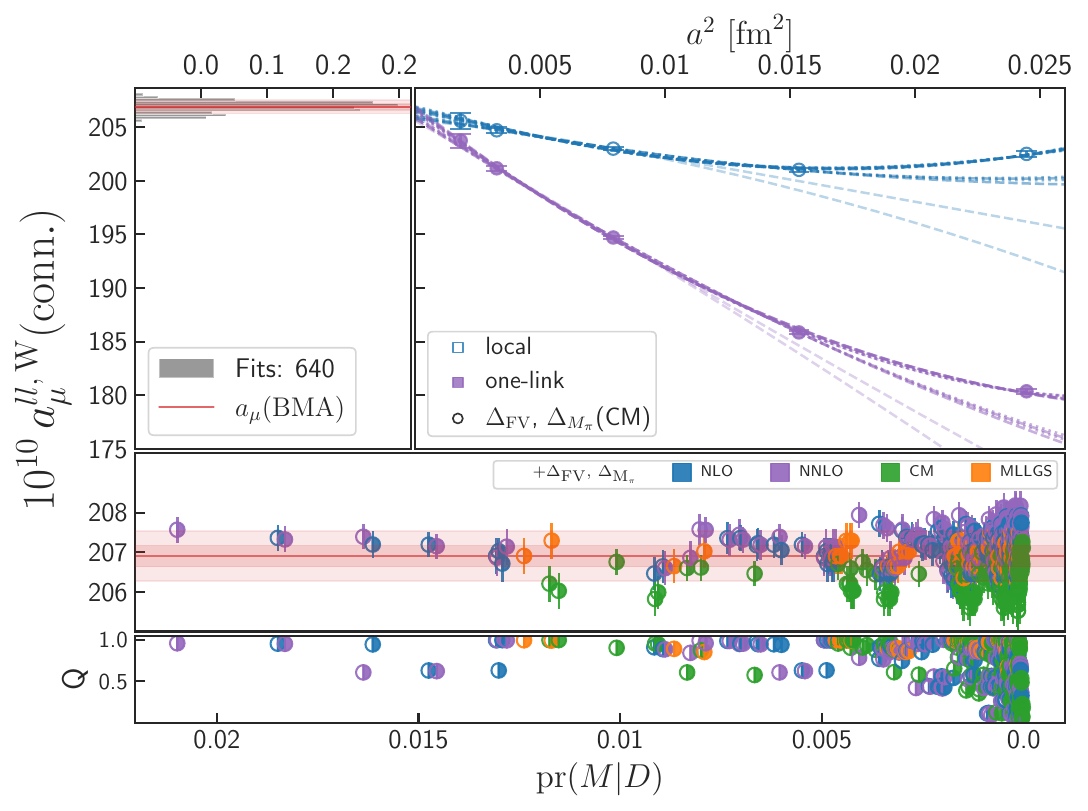}
\vspace{-5mm}
\caption{Results of the BMA procedure applied to $\amuLW$. \panel{Upper left panel}: Histogram of all continuum extrapolations used in the BMA, the inner light-red band corresponds to the first term in \cref{eq:BMAVar}, while the outer is the total error. \panel{Upper right panel}: The subset of data sets and extrapolations corresponding to correcting the local (blue unfilled) and one-link (purple filled) currents with the CM. Different extrapolations correspond to variations of the fit function and ensembles included. \panel{Lower panels}: The best fits according to the model probability, \cref{eq:modelProb}. The middle panel shows the joint-fit results, while the bottom one shows the corresponding $Q$ values \cite{FermilabLattice:2016ipl}. In both panels, the correction schemes employed for $\Delta_{\rm FV}$ and $\Delta_{\rm{M}_{\pi}}$ are indicated by the symbols' color, according to the legend in the middle panel.}
\label{fig:BMAWL}
\end{figure}

\begin{figure}
\centering
\includegraphics[scale=0.75]{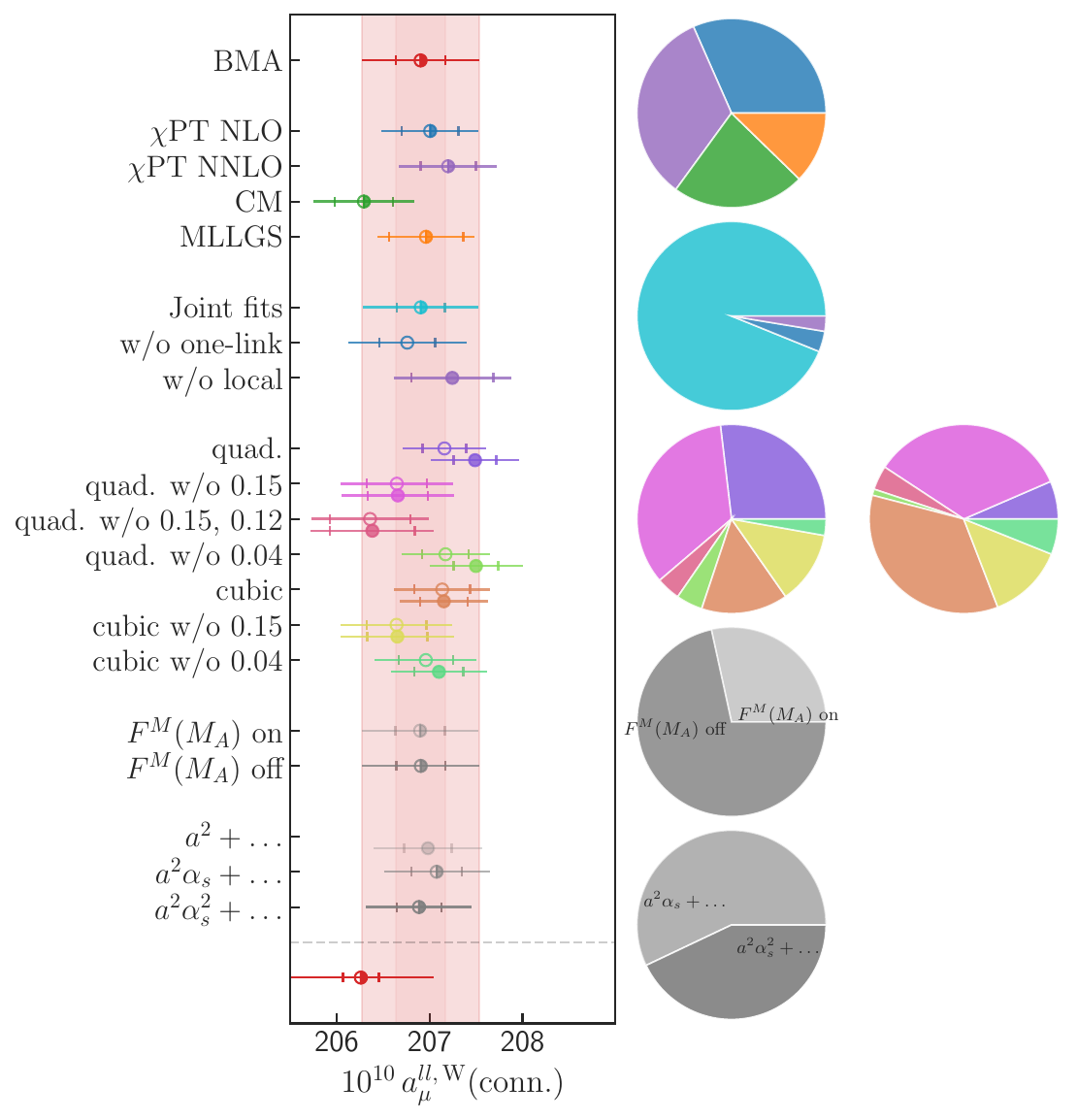}
\vspace{-5mm}
\caption{Breakdown of the results from the BMA applied to $\amuLW$. \panel{Left panel}: From top to bottom, the first, main result (BMA) includes all data sets, schemes, and other variations. The next block of results are the schemes for finite-volume and pion mass corrections. The following three are continuum extrapolations where either neither or one of the two currents is dropped. The next fourteen are subsets with specific continuum fit functions, corresponding to quadratic and cubic, as well as omitting various subsets of the $0.15$~fm, $0.12$~fm or $0.04$~fm ensembles. The unfilled circles correspond to the local current, the filled to the one-link current. The next two correspond to continuum fits with or without the mass-mistuning term. The penultimate three are subsets with differing leading powers of $\alpha_s$ in the fit function where, again, unfilled symbols are the local current and the filled symbols are one-link. Finally, below the dashed line, not included in the final BMA result, is the value one obtains if the model space is doubled by adding the original variations but with the corresponding taste-breaking corrections. The inner error bars on the data points correspond to the first term in \cref{eq:BMAVar}, while the outer is the total error. \panel{Right panels}: Pie charts showing the contributions to the BMA corresponding to the breakdowns in the left panel. The percentages are computed from \cref{eqn:ssProb} for the particular subsets. In the case of the continuum fit function subsets, which are broken up into local and one-link current variations, the left pie-chart corresponds to the local current and the right to the one-link.}
\label{fig:BMACompareWL}
\end{figure}

The result of applying the BMA procedure, with these variations, for $\amuLW$ is shown in \cref{fig:BMAWL}. The top-right panel illustrates the continuum extrapolations using data corrected with CM computed from the full window. The dashed lines indicate the continuum extrapolations for each data set. In total, we include six hundred forty fit results in the Bayesian model average. The histogram of the fit central values is shown in the top left panel of \cref{fig:BMAWL}, where it is overlaid on the BMA result (red line and bands) obtained using \cref{eq:BMAMean,eq:BMAVar}. We stress that this histogram is not the same as the posterior distribution described by \cref{eq:BMAMean,eq:BMAVar}, since it shows only the mean values obtained within each model and does not include the single-model variances.  We also stress that the error band shown does not represent a confidence interval obtained from the histogram, since 100\% of the models considered are summed over to obtain the averaged result. The middle panel shows the results from all the individual fits, ordered by the model probability, in comparison with the BMA result (red line and bands), while the bottom panel gives the associated $Q$ values computed from $\chi_{\rm aug}^{2}$ (see Eq.~B.4 of Ref.~\cite{FermilabLattice:2016ipl}). We find that fits with higher $Q$ values tend to have the larger model probabilities.

Following Ref.~\cite{FermilabLatticeHPQCD:2023jof}, we also perform the Bayesian model averages on specific subsets of the analysis variations. That is, we fix one of the analysis choices but vary the rest as usual. The results of these subset averages for $\amuLW$ are shown in \cref{fig:BMACompareWL} (left). The top data point is the BMA result from \cref{fig:BMAWL}. The four following data points correspond to model subsets for the lattice corrections that are separately limited to each of the four schemes (NLO \chpt, NNLO \chpt, CM, MLLGS). We find tension between the CM scheme and the others, which yields a significant contribution to the systematic uncertainty in our full BMA result. The next three points correspond to model subsets that either do or do not include one of the two currents in the extrapolation, with joint fits including both. We find that these results are in strong agreement with each other, with a gain in precision from performing the joint fits. The following fourteen data points are subsets of the continuum extrapolation where the highest order discretization term is varied, and specific lattice spacings are dropped. We specify these subsets for the local current with unfilled symbols and the one-link current with filled symbols. Here, dropping our two coarsest ensembles or the finest results in the most variation. The next two data-points break the fit function basis into two subsets based on if the mass-mistuning term is included or not. The results here are consistent for these variations. Finally, we split the fit function basis into fits with different leading powers of $\alpha_s$. We find the BMA does not significantly discern among the powers of $\alpha_s$ in the leading term for the local current (unfilled circles). Below the dashed line we include the result obtained by performing all the variations described above and, in addition, including corresponding variations for each scheme including taste-breaking corrections, {\it i.e.}, increasing the size of the model space by a factor of two. We find this result is consistent with our final BMA result.

\begin{figure}
\centering
\includegraphics[scale=0.9]{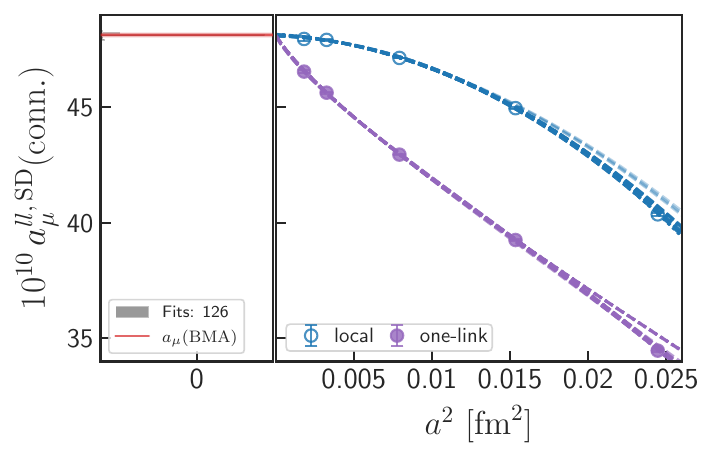}
\vspace{-5mm}
\caption{Results of the BMA procedure applied to $\amuLSD$. \panel{Left panel}: Histogram of all continuum extrapolations used in the BMA, the light-red band is the BMA result. \panel{Right panel}: The subset of data sets and extrapolations corresponding to the two different currents.}
\label{fig:BMASDL}
\end{figure}

The relative probabilities of these subsets, as defined in \cref{eqn:ssProb}, are shown in \cref{fig:BMACompareWL} (right). The top pie chart shows that the correction schemes have roughly equal weight in the BMA, with MLLGS being smaller than the others. This is due to the fact that the number of models with MLLGS corrections is smaller as we do not include variations where the coarsest data point is included, as discussed above. The second pie chart from the top shows the subset probabilities for performing either joint extrapolations or dropping one of the two currents. We see a strong preference for performing the joint extrapolations, largely due to the penalty incurred from dropping the five data points, which corresponds to the full current that is dropped. The third and fourth pie charts, on the same row, indicate that quadratic continuum fit functions are preferred for the local current whereas cubic fit functions are preferred by the one-link current, likely due to the significantly larger discretization effects in the one-link current. In the following pie charts, we observe a preference for continuum fits where the mass-mistuning term is switched off, again due to the penalty incurred in the BAIC for the additional fit parameter. Finally, in the last pie chart, we see a slight preference for a leading power of $\alpha_s$ for the local current.

\begin{figure}
\centering
\includegraphics[scale=0.85]{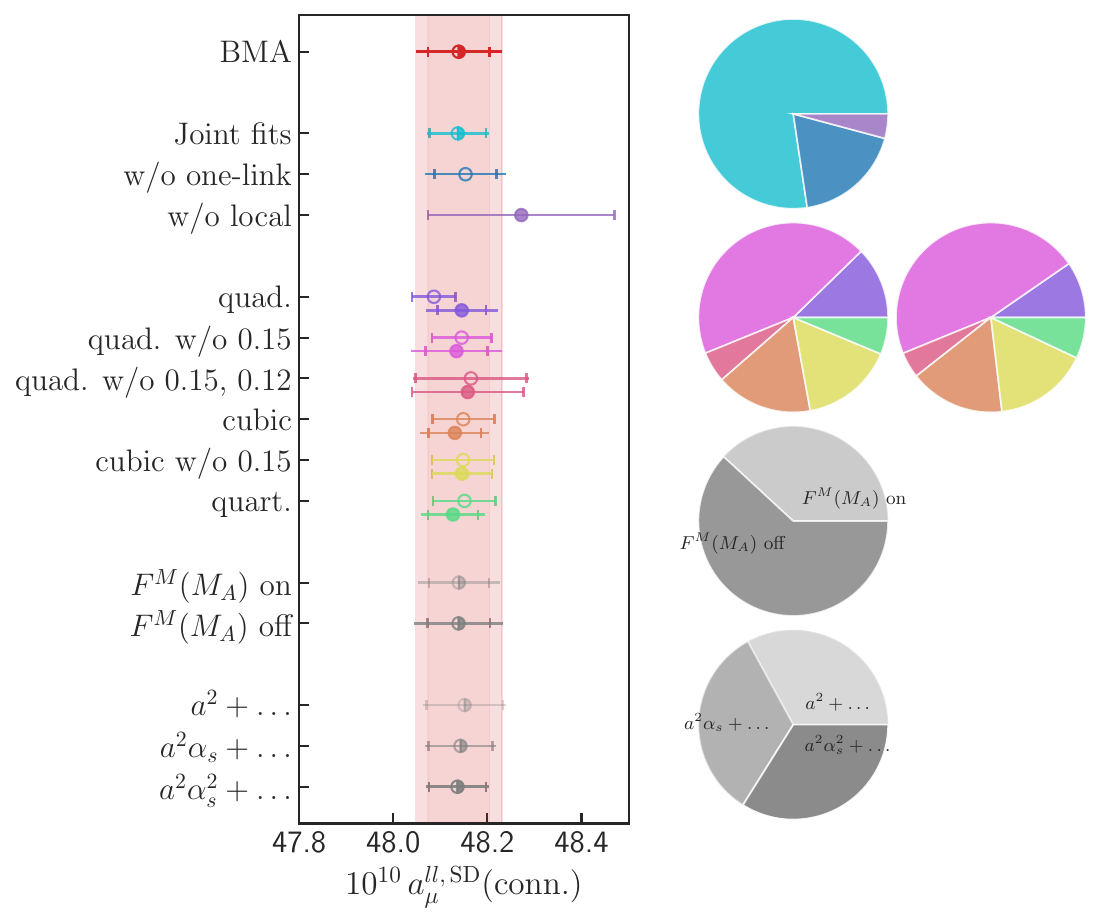}
\vspace{-5mm}
\caption{Breakdown of the results from the BMA applied to $\amuLSD$ (see caption of \cref{fig:BMACompareWL} for a description of the figure).}
\label{fig:BMACompareSDL}
\end{figure}

The result of applying the BMA to $\amuLSD$ is shown in \cref{fig:BMASDL}. Here, we plot all the joint continuum extrapolations to the two current dataset. In total, there are one hundred twenty-six models in our BMA corresponding solely to continuum fit function variations. As expected, the short-distance window observable has much larger discretization effects than $\amuLW$. In particular, the one-link current has a leading $a^2$ discretization effect, which leads to a logarithmic enhancement, as discussed in \cref{sec:logEffects}. Due to these large discretization effects, some fit functions without higher order terms are heavily suppressed in the Bayesian model average. This can be seen in the second and third pie-charts in \cref{fig:BMACompareSDL} where the quadratic fit function is insufficient in capturing the curvature over all five ensembles. In this window, we find significant improvement when performing joint fits as opposed to fits where one of the two currents is dropped. The differing leading discretization effects seem to constrain the continuum result significantly. On the other hand, the fits to just the one-link current result in a very large uncertainty due to the leading $a^2 \log(a)$ dependence. All other fit function variations are reasonably consistent with each other. However, we do observe a large increase in uncertainties when dropping the two coarsest spacings. Finally, we find that the BMA shows no preference for the power of $\alpha_s$ in the leading term of the local current fit function. Other than slightly larger errors in the leading $a^2$ fits, we see almost exactly equal results and subset probabilities.

Our final determinations of $\amuLSD$ and $\amuLW$ are given in \cref{sec:results}, specifically \cref{eqn:amuLWRes,eqn:amuLSDRes}, with complete error budgets given in \cref{table:WIndividualUncertainty,table:SDIndividualUncertainty}. We compare these results with previous determinations in \cref{fig:WIndividualCompare,fig:SDIndividualCompare}, respectively.

\subsection{Strange and charm}\label{subsec:sc_analysis}

\begin{table*}
\centering
\caption{Vector-current correlator datasets used in the calculation of the strange- and charm-quark connected  contributions to $\amuSD$ and $\amuW$. The first column lists the approximate lattice spacings in~fm. The parameters for these ensembles are contained in \cref{table:ensParams}. The second column lists the number of independent configurations analyzed after binning, as discussed in \cref{subsec:lattice}. The third column lists the number of random-wall source solves per configuration.\vspace{1mm}}
\label{table:strangeCharmData}
\begin{tabular}{lcc}
\hline \hline
$\approx a/\mathrm{fm}$ & $\nconf$ & $\nsrc$ \\ 
\hline
$0.15$       & 1001 & 48 \\ 
$0.12$       & 298 & 64 \\ 
$0.09^\star$ & 252 & 48 \\  
$0.06$       & 158 & 24 \\
\hline \hline
\end{tabular}
\end{table*}

As discussed in \cref{subsec:window}, the strange- and charm-quark connected contributions to $\amuHVP$ are the next-largest after the light-quark connected in the short- and intermediate-distance windows. Specifically, as can be seen in \cref{fig:window06Data}, the charm contribution is the second largest to $\amuSD$ while the strange contribution is the second largest to $\amuW$. In fact, almost all the charm contribution is contained in the SD window region, with the W window region containing only a contribution from the long-distance tail of the charm-quark vector-current correlation function. Fortunately, the strange and charm correlation functions do not suffer from the same signal-to-noise issues in the tail as the light-quark contribution does. Additionally, the effects of finite volume and taste-breaking are negligible at the level of precision we require. Hence, the analysis of these two contributions is much simplified compared with observables that contain light quarks.

Our strange and charm datasets are obtained using the local current and have equivalent statistics, with the specific numbers of configurations and sources used in the analysis given in \cref{table:strangeCharmData}. All ensembles have tuned masses within 4\% of the physical strange and charm mesons in \cref{subsec:inputs}, which can be seen from \cref{table:mesonMasses}. The strange and charm local vector currents are renormalized with both sets of factors ($Z^{\textrm{FF}}_{V, \text{ local}}$,  $Z^{\textrm{RI-SMOM}}_{V, \text{ local}}$) given in \cref{table:zvfactors}. The two renormalizations have different discretization errors but should yield consistent results in the continuum limit. Because discretization errors can be a more significant source of uncertainty for the heavier flavors, we use both renormalization schemes in our strange and charm analysis to improve control over the continuum limit extrapolation.
The observables $\amuSSD$, $\amuSW$, $\amuCSD$, and $\amuCW$ are computed on each ensemble. We perform continuum extrapolations with the local continuum fit function $F^{{a}}_{\text{local}}$ (\cref{eq:discfunclocal}) in \cref{eq:generalfitfunc}. For the charm extrapolations, we modify the mass mistuning term $F^M(M_A)$ (\cref{eq:mfunc}) with
\begin{align}
    F^{M}_{\text{charm}}(M_A) =  F^M(M_A) + C_{\text{charm}} \frac{M_{D_s, \,\textrm{phys.}}-M_{D_s, \,\textrm{latt.}}}{M_{D_s, \,\textrm{phys.}}}\label{eq:charmMfunc}
\end{align}
to account for mistuning of the charm-quark mass. The values of the $D_s$ mass on each ensemble are given in \cref{table:mesonMasses}.  We use a Gaussian prior of $C_{\text{charm}}=0(1)$, following Ref.~\cite{Hatton:2020qhk}. Mistuning of the strange-quark mass is captured by the existing sea-mass correction term in $F^{M}(M_A)$; as previously stated, we find that our results are insensitive to the stability prior on this term. From our empirical Bayes procedure we find
\begin{align}
    \Lambda_{ss, \text{SD}} &= \Lambda_{ss, \text{W}} = 0.65~{\rm GeV},\\
    \Lambda_{cc, \text{SD}} &= \Lambda_{cc, \text{W}} = 1.3~{\rm GeV}.
\end{align}
We also find that the charm data are sensitive to higher order discretization terms ($a^6$, $a^8$) while strange data are not. As before, we limit our continuum fit variations to ones in which the degrees of freedom are non-negative, using the prior of \cref{eq:emp_bayes_prior} when necessary to regulate the degrees of freedom.

\begin{figure}
\centering
\includegraphics[scale=0.64]{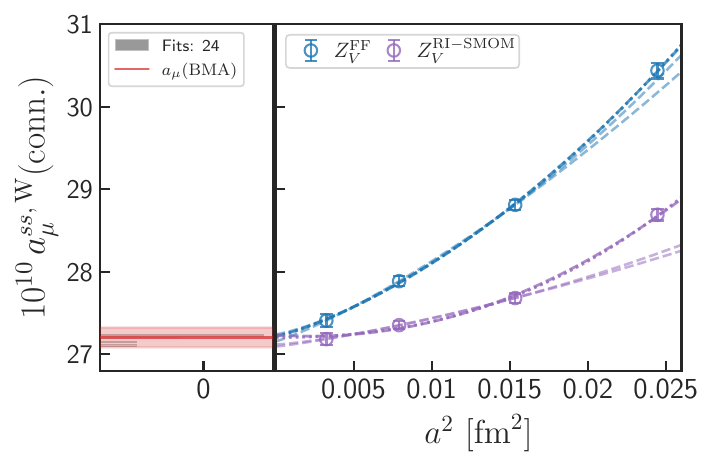}\hspace{4mm}\includegraphics[scale=0.64]{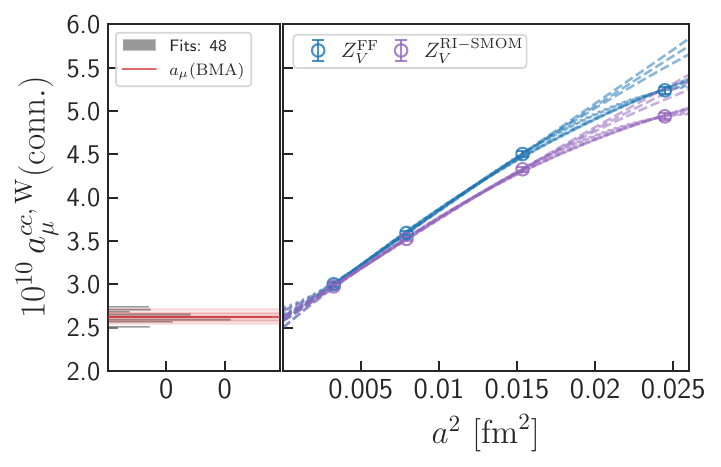}\\
\includegraphics[scale=0.6]{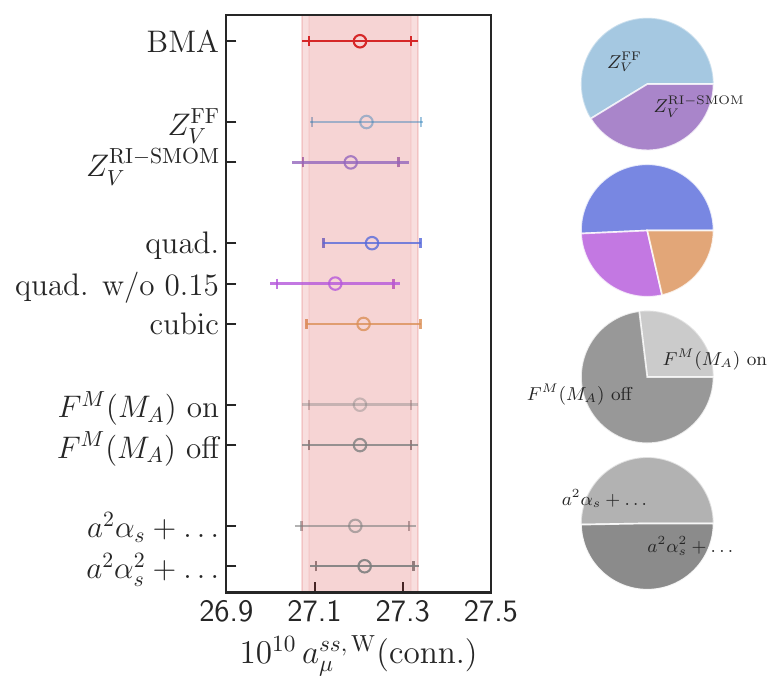}\includegraphics[scale=0.6]{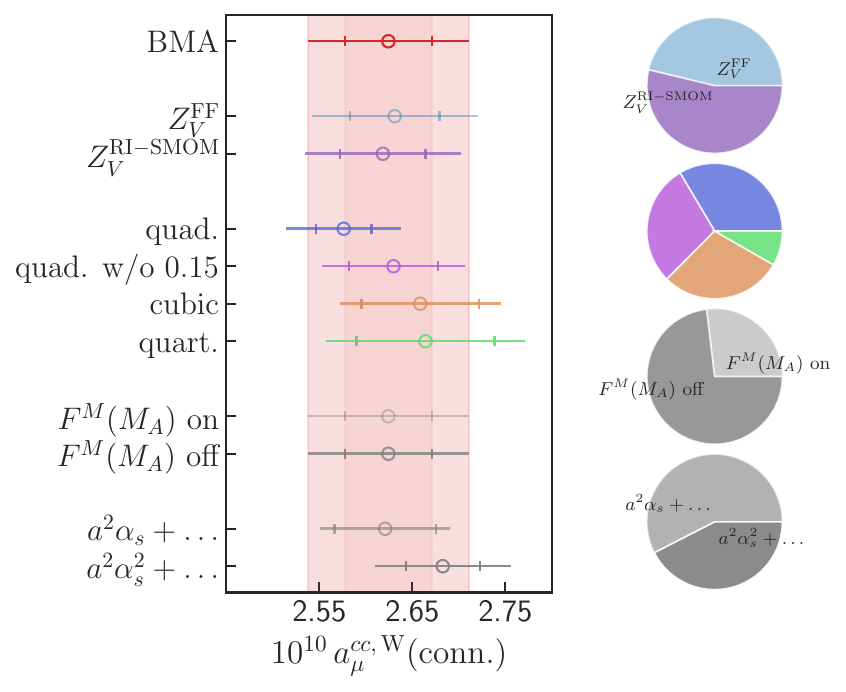}
\vspace{-5mm}
\caption{Results of the BMA procedure applied to $\amuSW$ (left) and $\amuCW$ (right), respectively. \panel{Upper plots, right panels}: All continuum extrapolations used in the BMA for the respective observables. \panel{Upper plots, left panels}: Histogram of all continuum results. The red line and bands are the corresponding BMA result, where the inner light-red band corresponds to the first term in \cref{eq:BMAVar}, while the outer is the total error. \panel{Lower plots, left panels}: Breakdown of BMA result into model subsets. From top to bottom, the main BMA result, containing all variations. After, the two data subsets using the two local-current renormalization factors in \cref{table:zvfactors}. The next group are subsets using specific continuum fit functions: quadratic, quadratic without the $0.15$~fm data point, or cubic. The quadratic data point for $\amuCW$ is outside the range of the x-axis. For this contribution, we also include a quartic variation. The final two data points are subsets with differing leading powers of $\alpha_s$ in the fit function. The inner error bar on the data points corresponds to the first term in \cref{eq:BMAVar}, while the outer is the total error. \panel{Lower plots, right panels}: Pie charts show the relative contributions to the BMA corresponding to the breakdowns in the left panel. The percentages are computed by summing over \cref{eq:modelProb} for the particular subsets, as in \cref{eqn:ssProb}.}
\label{fig:BMACompareWSC}
\end{figure}

The results of these continuum extrapolations are shown in \cref{fig:BMACompareWSC} for $\amuSW$ (top left) and $\amuCW$ (top right). For $\amuSW$, in the right panel, twelve continuum fit variations are performed for the two choices of renormalization factor given in \cref{table:zvfactors}. The histogram of the fit central values is shown in the left panel with the BMA mean and errors, \cref{eq:BMAMean,eq:BMAVar}, given by the red line and inner- (first term in \cref{eq:BMAVar}) and outer-bands (total error). Here, the error is dominated by the first term in \cref{eq:BMAVar}; hence, only the inner band is visible. In the bottom left plot of \cref{fig:BMACompareWSC}, the models are broken up into labeled BMA subsets. We find these subsets are all consistent with our full BMA result (red). For $\amuCW$, \cref{fig:BMACompareWSC} (top right), we have sixteen fit variations (four additional variations from including the $C_{a^8}$ term). We observe that the charm data have additional curvature compared with $\amuSW$ over the range of lattice spacings we compute. And indeed, we find that fit variations with just quadratic terms are not enough to capture this curvature. This can be seen in \cref{fig:BMACompareWSC} (bottom right), where the quadratic continuum extrapolations subset value appears off the plot. From the corresponding pie chart on the right, we observe that BMA suppresses the contribution from this subset, indicating very poor fit quality. Aside from this, all other model variation subsets are in good agreement with our BMA result. The consistency of the quartic subset of fits with everything else indicates that we have stability with respect to adding higher order terms. 
\begin{figure}
\centering
\includegraphics[scale=0.64]{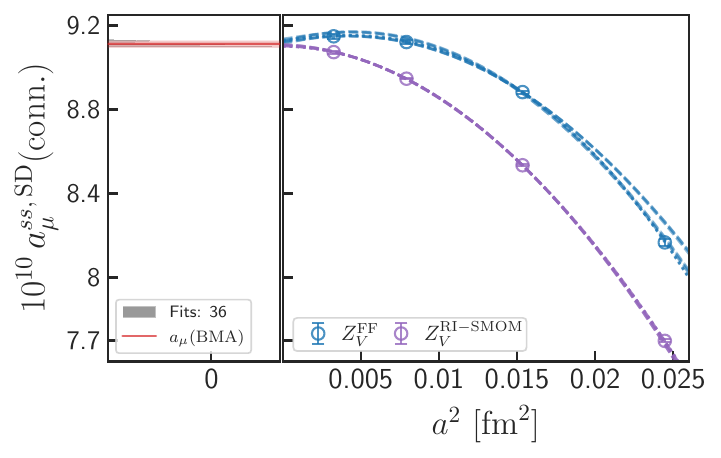}\hspace{4mm}\includegraphics[scale=0.64]{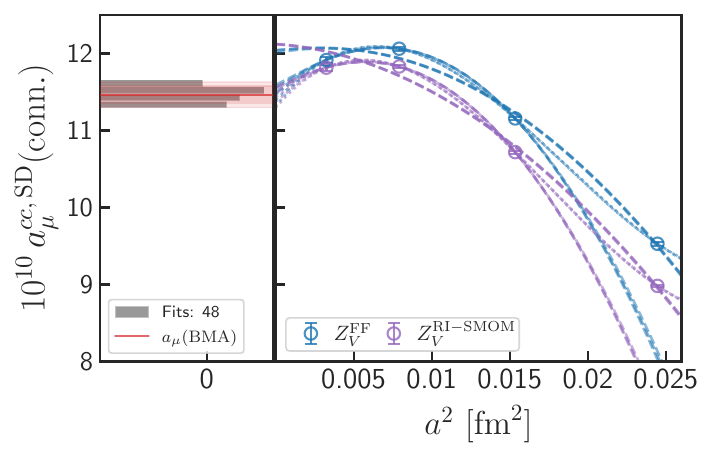}\\
\includegraphics[scale=0.6]{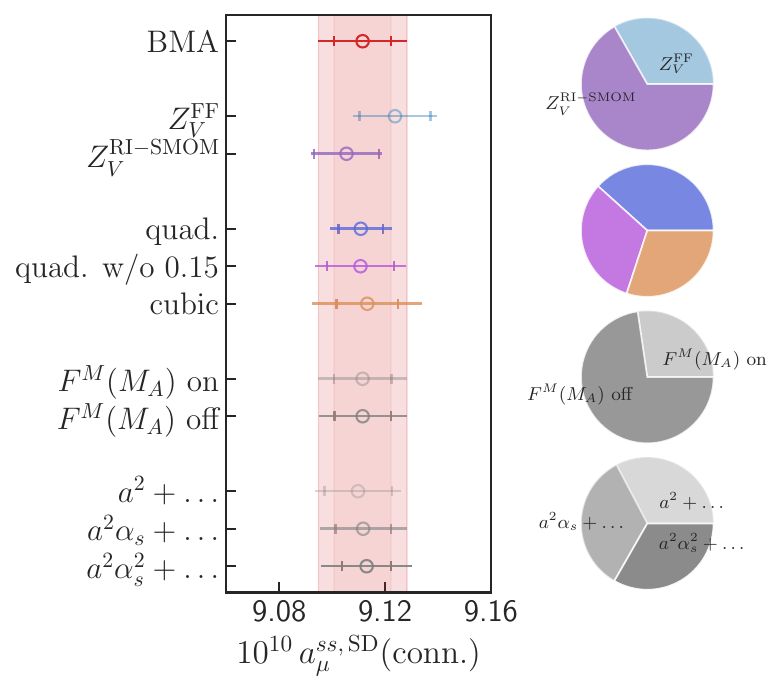}\includegraphics[scale=0.6]{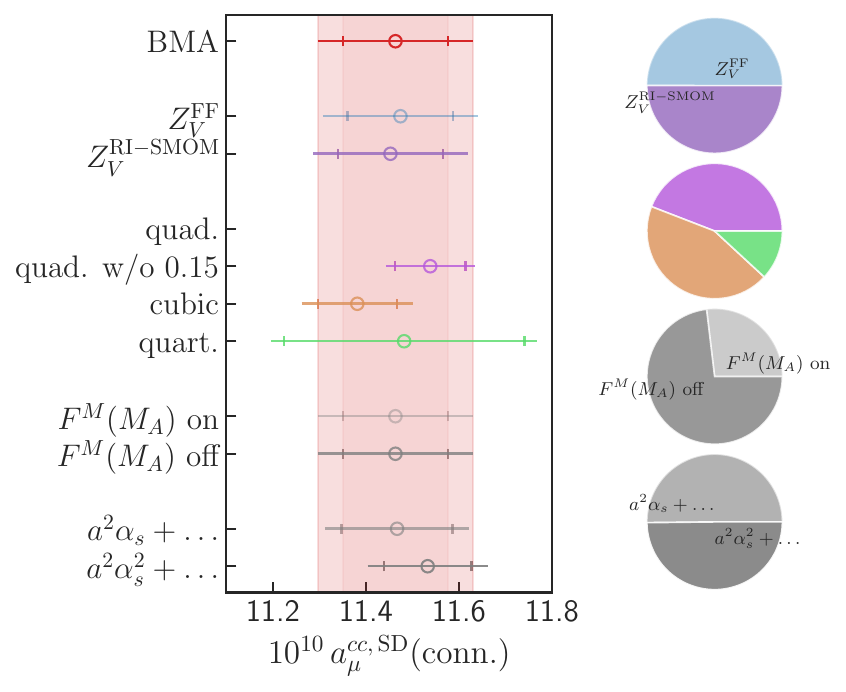}
\vspace{-5mm}
\caption{Results of the BMA procedure applied to $\amuSSD$ and $\amuCSD$, respectively. See caption of \cref{fig:BMACompareWSC} for a description of each plot.}
\label{fig:BMACompareSDSC}
\end{figure}

Continuum extrapolation results for $\amuSSD$ and $\amuCSD$ are shown in \cref{fig:BMACompareSDSC}, (top left) and (top right), respectively. The total fit variations are increased for both strange and charm over the $\amuW$ case due to the inclusion of variations with a leading $a^2$ term, as in the case of the light quark, to test for any log-enhancement effects. As before, we find additional curvature in the charm contribution over the strange. In \cref{fig:BMACompareSDSC} (bottom), the breakdown of these extrapolations into labeled BMA subsets is given. We find that for $\amuSSD$ (left) we have consistent results for all model subset variations, with the quadratic fits to the full dataset being preferred by the BMA. We also see no preference for the power of $\alpha_s$. For $\amuCSD$ (right), as in the case of $\amuCW$, we find that quadratic fits are insufficient to capture the curvature of the full dataset. Outside this particular set of fits, we observe some spread in the remaining variations, though all are statistically consistent with each other and encapsulated by the BMA uncertainty.

Our final determinations of $\amuSSD$, $\amuCSD$, $\amuSW$ and $\amuCW$ are given in \cref{sec:results}, specifically \cref{eqn:amuSWRes,eqn:amuCWRes,eqn:amuSSDRes,eqn:amuCSDRes}, with complete error budgets given in \cref{table:WIndividualUncertainty,table:SDIndividualUncertainty}. We compare these results with previous determinations in \cref{fig:WIndividualCompare,fig:SDIndividualCompare}, respectively. 

\subsection{Disconnected}\label{subsec:disc_analysis}

\begin{table*}
\centering
\caption{One-link current correlator datasets used in the calculation of the disconnected contribution to $\amuSD$ and $\amuW$. The first column lists the approximate lattice spacings in~fm. The parameters for these ensembles are contained in \cref{table:ensParams}. The disconnected correlator data for the coarsest two ensembles are the result of combining two data sets, each analyzed at the outset by separate groups of authors. 
The second column lists the number of independent configurations analyzed by each group, respectively. The third column lists the number of volume random sources per configuration, while the fourth column gives the number of eigenvectors. One data set employs eigenvectors; the other, indicated by the --, does not.}
\vspace{1mm}
\label{table:discData}
\begin{tabularx}{\linewidth}{LLLR}
\hline \hline
$\approx a/\mathrm{fm}$ & $\nconf$ & $\nsrc$ & $\neig$ \\ 
\hline
$0.15$       & 645, 1047 & 256, 256  & 350, -- \\ 
$0.12$       & 150, 562  & 500, 256  & 300, -- \\ 
$0.09$       & 705       & 1000      & 1000  \\  
\hline \hline
\end{tabularx}
\end{table*}

For the disconnected contribution, our dataset spans only our three coarsest lattice spacings (see \cref{fig:ensBreakdown}), which leads to two technical complications. First, MLLGS is not usable at the coarsest spacing (see \cref{sec:latticeCorrections}), so we exclude it from our set of correction schemes. Second, being limited to three spacings constrains the continuum-limit-extrapolation variations, which we usually take to be at least quadratic and hence have three parameters at minimum. A taste-singlet current is necessarily employed, hence we use the one-link discretization, \cref{eq:vi:1link}, along with the renormalization factor $Z_{V,\,\text{one-link}}^{\text{RI-SMOM}}$ from \cref{table:zvfactors}. We also use no mass-mistuning term, as it cannot be resolved with only three data points. Ensembles and corresponding statistics are given in \cref{table:discData}.

\begin{figure}
\centering
\includegraphics[width=0.8\textwidth]{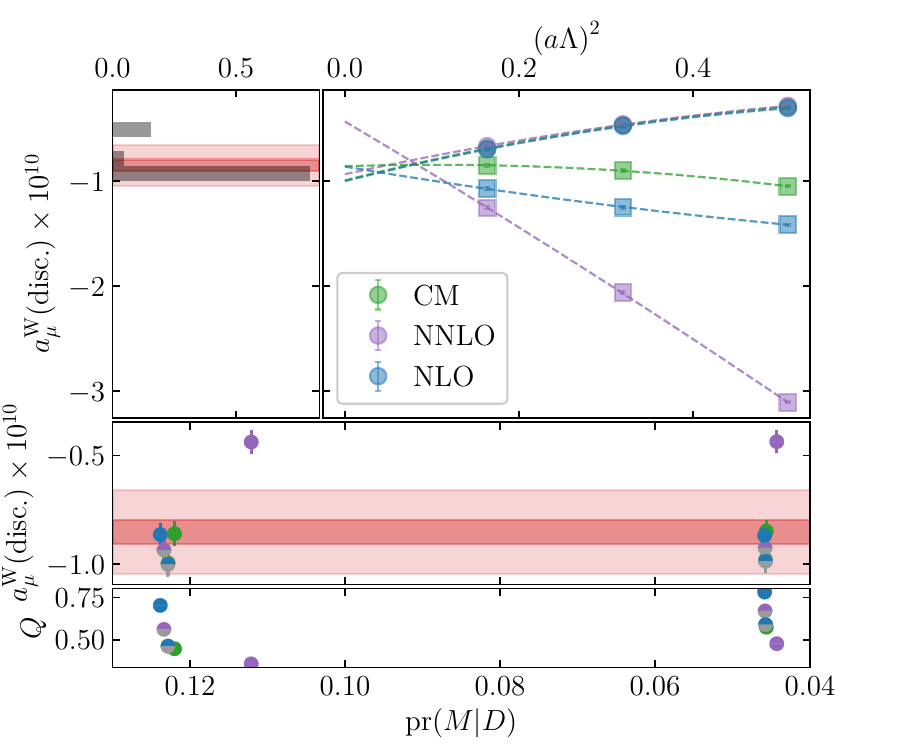}
\caption{Results of the BMA procedure applied to the disconnected contribution in the intermediate window with 12~models in total. The final BMA result is shown in light red, where the inner and outer bands correspond to the error coming from the first term in \cref{eq:BMAVar} and the total error, respectively.
\panel{Upper left panel}: Histogram of $\amuDW$, weighted by $\pr(M|D)$. These results come from the bootstrap, carrying out fits in each sample.
\panel{Upper right panel}: Continuum-limit extrapolations. Data without TB corrections are indicated with circles, while data with TB corrections are indicated with squares. 
\panel{Middle panel}: Results of the continuum fits for individual models going into the BMA, plotted against their model weight. Models without TB corrections are gray in the lower half of their symbols.
\panel{Lower panel}: Goodness of fit $Q$ computed from $\chi_{\rm aug}^{2}$ plotted against the model weight.}
\label{fig:BMAWDisc}
\end{figure}

For $\amuDW$, we perform FV, $M_\pi$, and TB corrections as described in \cref{eq:corrFV,eq:corrMpi,eq:corrTB} of \cref{sec:latticeCorrections}. We consider the CM scheme along with NLO and NNLO \chpt. The empirical Bayes procedure described in \cref{sec:analysis} yields
\begin{align}\label{eq:lambda_disc}
    \Lambda_{\textrm{disc.}, \text{W}} = 0.9~{\rm GeV}.
\end{align}
We observe that our dataset cannot significantly resolve terms beyond the leading $a^2$ for $\amuDW$. Hence, our continuum fits in \cref{eq:generalfitfunc} utilize 
\begin{equation}\label{eq:Fdisc_disconnected}
    F^{{a}}_{\text{disc.}} (a) =  \sum^{m}_{k=1} C_{a^{2k}} (a\Lambda)^{2k},
\end{equation}
where $m=2$ or 3, and we introduce the prior \cref{eq:emp_bayes_prior} on all terms with $k=2$ or 3.

For the short-distance observable, as discussed in \cref{sec:latticeCorrections}, we include the HP FV correction. We find that the data in this window are highly nonmonotonic, which makes it challenging to assign reasonable priors. Hence, we use
\cref{eq:Fdisc_disconnected} with $k=1$ and $k=2$. When $k=1$ we carry out ``solves,'' {\it i.e.}, zero degree of freedom parameterizations, to the finest two and coarsest two points, along with a fit to all three. When $k=2$ we solve for all three. Rather than using BMA, we take the difference between extreme results as a conservative systematic uncertainty, while the spread of the central value over the bootstrap bins is used for the statistical uncertainty. 

\begin{figure}
\centering
\includegraphics[width=0.7\textwidth]{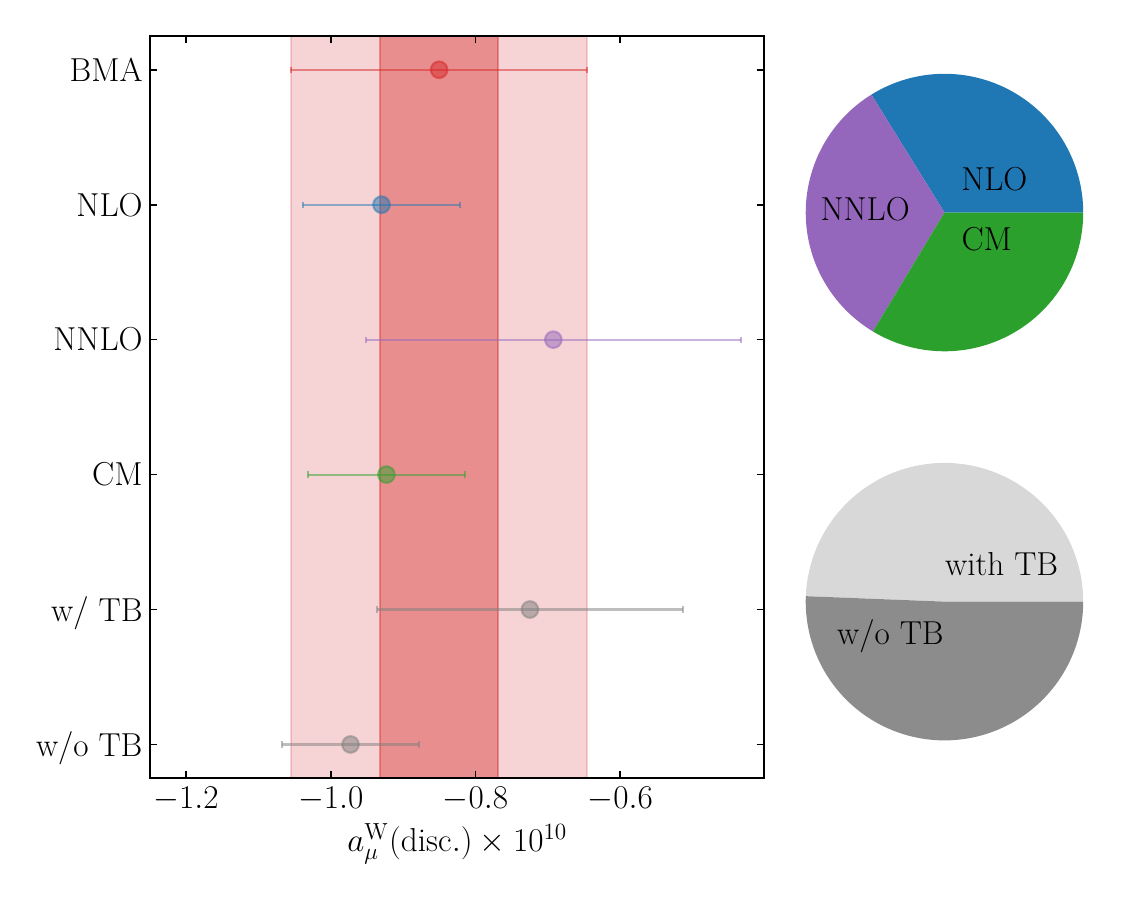}
\caption{Breakdown of the results from the BMA applied to
$\amuDW$. This breakdown is done in the same manner as \cref{fig:BMACompareWL}.}
\label{fig:BMACompareDiscW}
\end{figure}

\begin{figure}
\centering
\includegraphics[width=0.45\textwidth]{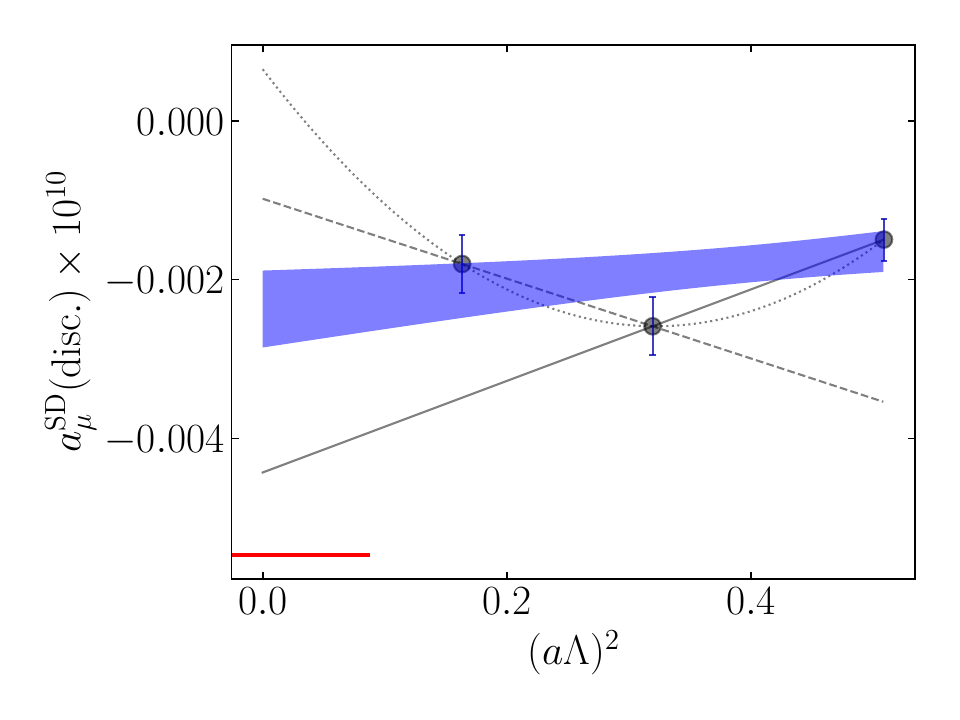}
\includegraphics[width=0.45\textwidth]{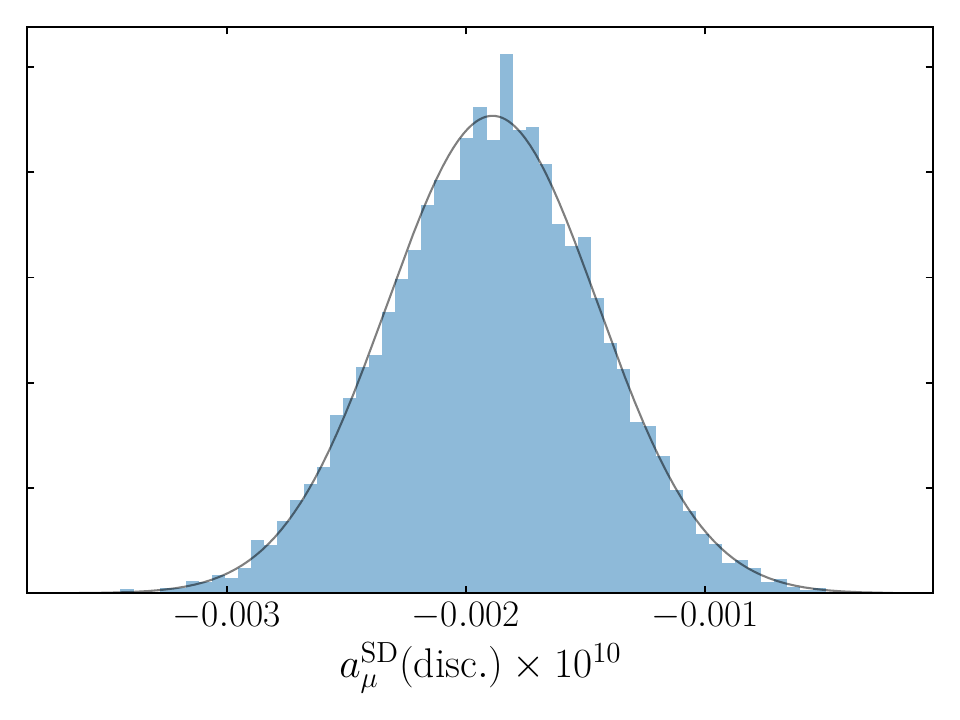}
\caption{Results for $\amuDSD$. \panel{Left panel}: Three solves and
a linear fit in the original data set. The midpoint between the extreme
results in the continuum limit is taken as an estimate of our central value,
and half the spread between the extreme results is taken as an estimate of the systematic error. The red line gives the pQCD prediction,
\cref{eq:amuDpQCD}.
\panel{Right panel}: Histogram of bootstrap results
for the central value. A Gaussian with mean and standard deviation
taken to be the median and half the difference between the limits of the middle 68\% region of the histogram is shown in black for comparison.
}
\label{fig:SDDisc}
\end{figure}

Our BMA procedure applied to $\amuDW$ is shown in \cref{fig:BMAWDisc}.
The BMA runs over the schemes, NLO, NNLO, and CM, and for each scheme, over quadratic and cubic Ans\"atze, with and without taste breaking, which yields a total of 12 models. We see that most extrapolations are in good agreement, except for NNLO with TB. These extrapolations tend to increase the model spread, and correspondingly, we see that our final uncertainty has a considerable systematic component, represented in light red.
Ultimately, the influence of our NNLO points with TB is attenuated
by the fact that they comprise only two of twelve models, represented in the histogram of \cref{fig:BMAWDisc}. A breakdown of the BMA results is shown in \cref{fig:BMACompareDiscW}.

Our procedure for computing $\amuDSD$ is illustrated in \cref{fig:SDDisc}.\footnote{As mentioned earlier, we do not treat the extrapolation of $\amuDSD$ as an expansion about the continuum value. Therefore, we are unable to enforce any sensible priors on the discretization terms, and hence, the fits are insensitive to the choice of $\Lambda$. Nevertheless, we take $\Lambda_{\rm disc.,W}$ from \cref{eq:lambda_disc} as our preferred choice to scale the abscissa in this figure, keeping it consistent with \cref{fig:BMAWDisc}.} The extreme values used to estimate our systematic
error come from the linear solve to the coarsest two spacings and the quadratic solve to all three, which can be seen in the left plot.
The motivation for this conservative estimate lies in the interplay between the SD window function and the behavior of the disconnected correlator. In particular,
as $t/a$ crosses the SD window boundary, the disconnected correlator sharply varies from positive to negative. As a result, due to the ensembles used in this work, there are significant discretization effects.
This is further complicated due to the presence of the oscillating terms in staggered correlation functions; whether there is an ``up-" or ``down-" oscillation at the window boundary varies for each lattice. This is likely the source of the observed nonmonotonicity in $\amuDSD$, and without understanding in detail how this
interplay is resolved at finer spacings, it is difficult to constrain the continuum behavior.

The pQCD prediction of \cref{eq:amuDpQCD} is consistent within
this large systematic. In the right plot, we see the spread of midpoints
coming from the procedure in the left plot carried out in each bootstrap sample. For this observable, the systematic uncertainty dwarfs the statistical uncertainty. Calculations at finer lattice spacings are needed to clarify the continuum-limit behavior of this observable.

Our final determinations of $\amuDSD$ and $\amuDW$ are given in \cref{sec:results}, specifically \cref{eqn:amuDWRes,eqn:amuDSDRes}, with complete error budgets given in \cref{table:WIndividualUncertainty,table:SDIndividualUncertainty}. We compare these results with previous determinations in \cref{fig:WIndividualCompare,fig:SDIndividualCompare}, respectively.

\subsection{Strong-isospin breaking}\label{subsec:sib_analysis}

The strong-isospin breaking (SIB) correction (\cref{eq:sib-def}) can be broken up into connected and disconnected contributions:
\begin{align}
    \Delta a_{\mu}^{ud}(\mathrm{SIB})=\Delta a_{\mu}^{ud}(\mathrm{SIB,conn.})+\Delta a_{\mu}^{ud}(\mathrm{SIB,disc.}).
\end{align}
where
\begin{align}
    \Delta a_{\mu}^{ud}(\mathrm{SIB,conn.}) &\equiv a_{\mu}^{ud}(\mathrm{conn.}) - \amuL \label{eq:SIBConn}, \\
    \Delta a_{\mu}^{ud}(\mathrm{SIB,disc.}) &\equiv a_{\mu}^{ud}(\mathrm{disc.}) - \amuD \label{eq:SIBDisc}.
\end{align}
The last terms on the RHS of \cref{eq:SIBConn,eq:SIBDisc} correspond to the isospin-symmetric quantities in \cref{subsec:light_analysis,subsec:disc_analysis}; 
however, we estimate the RHS differences with distinct analyses to maximize the power of the statistical correlations. These two contributions are computed from different datasets, as shown in \cref{fig:ensBreakdown}.  Specific numbers of configurations and sources are given in \cref{table:sibData}.  The connected SIB-correction dataset is generated using the local vector current, whereas the disconnected-contribution dataset uses the one-link (taste-singlet) current. Furthermore,
the connected SIB analysis employs correlator data generated at 0.12~fm and 0.15~fm using TSM to maintain correlations between the different valence quark flavors.  For the 0.15~fm disconnected dataset, where we do not have an equal number of configurations between $ll$ and $ud$ datasets, we compute the correlation matrix using the configurations shared between both datasets and rescale by the variances on the full datasets to obtain the final covariance matrix.  The local (one-link) vector currents are renormalized with $Z^{\textrm{FF}}_{V, \text{ local}}$ ($Z^{\textrm{RI-SMOM}}_{V, \text{ one-link}}$) given in \cref{table:zvfactors}.  The differences between the connected and disconnected SIB datasets lead to different analysis strategies for each observable.  Here, we describe each of these analyses for both $\amuW$ and $\amuSD$.  The SIB correction is a low-energy effect with about 90\% coming from $t\ge1$~fm, 10\% from the W window region, and almost nothing from the SD window region.

Our final determinations of $\Delta a_{\mu}^{ud,\mathrm{SD}}(\mathrm{SIB,conn.})$, $\Delta a_{\mu}^{ud,\mathrm{SD}}(\mathrm{SIB,disc.})$, $\Delta a_{\mu}^{ud,\mathrm{W}}(\mathrm{SIB,conn.})$, and $\Delta a_{\mu}^{ud,\mathrm{W}}(\mathrm{SIB,disc.})$ are given in \cref{sec:results}---specifically \cref{eqn:amuSIBCWRes,eqn:amuSIBCSDRes,eqn:amuSIBDWRes,eqn:amuSIBDSDRes}, respectively---with complete error budgets for the $\amuW$ and $\amuSD$ observables in \cref{table:WIndividualUncertainty,table:SDIndividualUncertainty}, respectively.  A comparison of $\Delta a_{\mu}^{ud,\mathrm{W}}(\mathrm{SIB,conn.})$ and $\Delta a_{\mu}^{ud,\mathrm{W}}(\mathrm{SIB,disc.})$ with previous results from other groups is given in \cref{fig:WSIBCompare}.

\begin{table*}
\centering
\caption{Vector-current correlator datasets used in the calculation of $\Delta a_{\mu}^{ud}(\mathrm{SIB})$. The first column lists the approximate lattice spacings in~fm. The ensemble parameters are contained in \cref{table:ensParams}.  The second through fourth (sixth through eighth) columns list the valence quark flavors, the number of configurations and wall-sources (volume-sources) analyzed for the local (one-link) current used in the $\Delta a_{\mu}^{ud}(\mathrm{SIB,conn.})$ ($\Delta a_{\mu}^{ud}(\mathrm{SIB,disc.})$) analysis, respectively.  The fifth column lists the number of eigenvector pairs per configuration where applicable for the local current.  Here, $\dagger$ indicates data generated with the truncated solver method and without eigenvectors.}
\label{table:sibData}
\begin{tabularx}{\linewidth}{LCCCCCCCCC}
\hline \hline
 && \multicolumn{4}{c}{Connected (local)} && \multicolumn{3}{c}{Disconnected (one-link)} \\
$\approx a/\mathrm{fm}$ &$\quad$& $qq$ & $\nconf$ &$N_{\text {src}}^{\text{wall}}$ & $\neig$ &$\quad$& $qq$ & $\nconf$ & $N_{\text {src}}^{\text{volume}}$ \\ 
\hline
$0.15^\prime$ && $uu/ll/dd$ & 4949  & 48$^\dagger$ & -- &  & --  & -- &--\\
\multirow{2}{*}{$0.15$} && \multirow{2}{*}{$uu/ll/dd$} & \multirow{2}{*}{10019} & \multirow{2}{*}{48$^\dagger$} & \multirow{2}{*}{--} && $ud$ & 1047 & \multirow{2}{*}{256$^\dagger$}  \\
&  &  &  &  &&& $ll$ & 1692 &  \\
$0.12$ && $uu/ll/dd$ & 9637  & 64$^\dagger$ & --  && $ud/ll$ & 712  & 500$^\dagger$  \\
$0.09$ && $ll/rr/dd$ & 1000 & 48 & 2000 & & $ud/ll$ & 705  & 1000$^\dagger$  \\
$0.09^\star$& & $hh/ll/dd$ & 993   & 96 & 2000 & & --& --  & -- \\
$0.06$ && $ll$ & 1009 & 96 & 2000  &&-- &--  & -- \\
\hline \hline
\end{tabularx}
\end{table*}

\subsubsection{Connected strong-isospin breaking}\label{subsubsec:conn_sib_analysis}

The connected SIB dataset is obtained using the local current (see \cref{table:sibData}).  These data span four lattice spacings in [0.06, 0.15]~fm across six ensembles and are evaluated at various isospin-symmetric light-pion masses; these masses are given in \cref{table:mesonMasses}.  The data on the 0.15~fm, 0.15$^\prime$~fm, and 0.12~fm ensembles are tuned near the physical $uu$, $ll$, and $dd$ values, as defined in \cref{subsec:inputs}.  The 0.09~fm ensemble includes the mistuned (unitary) and retuned light mass values discussed in Ref.~\cite{FermilabLatticeHPQCD:2023jof} (labeled $ll$ and $rr$, respectively) as well as a physical $dd$ point.  The 0.09$^\star$~fm ensemble includes nearly-physical $ll$ and $dd$ values and a point at the harmonic-mean-square mass $hh$ defined in \cref{subsec:lattice}. 
The unphysical-mass points on the 0.09~fm ensembles help constrain the mass dependence of the chiral-continuum extrapolation.  The 0.06~fm ensemble is evaluated only at a nearly physical $ll$ point, which helps constrain the lattice-spacing dependence of the chiral-continuum extrapolation.

We compute $a_{\mu}^{l_ql_q}(\mathrm{conn.})$ at each lattice spacing and available valence quark content $qq$ in both the W and SD window regions.  The notation ``$a_{\mu}^{l_ql_q}$" denotes the light-quark contribution $a_{\mu}^{ll}$ ({\it i.e.}, including the light-quark charge factor defined in \cref{subsec:window}) evaluated at the $qq$ meson mass. We then perform a chiral-continuum fit to these data for each window.  The results of the fit are then used to extrapolate to the continuum and the physical $uu$, $ll$, and $dd$ meson masses defined in \cref{subsec:inputs}.  These values are then used to compute the SIB correction according to \cref{eq:SIBConn}, where
\begin{align}
a_\mu^{ud}(\mathrm{conn.}) & = a_\mu^{uu}(\mathrm{conn.}) + a_\mu^{dd}(\mathrm{conn.}),\\
a_\mu^{uu}(\mathrm{conn.})&=\frac{q_u^2}{q_l^2}a_\mu^{l_ul_u}(\mathrm{conn.}),\\
a_\mu^{dd}(\mathrm{conn.})&=\frac{q_d^2}{q_l^2}a_\mu^{l_dl_d}(\mathrm{conn.}).
\end{align}
In \cref{eq:SIBConn}, the value of $a_{\mu}^{ll}(\mathrm{conn.})$ comes from the SIB analysis rather than the value obtained in \cref{subsec:light_analysis} to preserve its correlations with $a_{\mu}^{uu}(\mathrm{conn.})$ and $a_{\mu}^{dd}(\mathrm{conn.})$, which are not directly available from the analysis in \cref{subsec:light_analysis}. 
We choose to do a joint fit to the lattice-spacing and pion-mass dependencies here, as opposed to computing $\Delta a_{\mu}^{ud}(\mathrm{SIB,conn.})$ at finite lattice spacing and taking the continuum limit (cf. \cref{subsubsec:disc_sib_analysis}). This allows us to take advantage of the data available on the finer ensembles 0.06~fm, 0.09~fm, and 0.09$^\star$~fm where the set of all three physical masses \{$uu$, $ll$, $dd$\} is not complete.

The form of the chiral-continuum fit Ans\"atze is a modified version of the one used to perform data-driven $M_{\pi}$ mistuning corrections in our previous intermediate windows calculation \cite{FermilabLatticeHPQCD:2023jof} and is given by
\begin{align}
    a_{\mu}^{l_ql_q}(a,M_{\pi},M_A)=a_{\mu}^{l_ql_q}(a,M_{\pi})[1+F^M(M_A)],
    \label{eq:sib-conn-cont}
\end{align}
where we take
\begin{align}
\label{eq:sib-aMu-conn-cont} a_{\mu}^{l_ql_q}(a,M_{\pi})&=c_{-1}(a)h(M_\pi/\Lambda) + c_0(a)+c_1(a)(M_{\pi}/\Lambda)^{2}, \\
\label{eq:sib-ci-conn-cont} c_i(a)&=\sum_{j=0}^{n_i}c_{ij}\left(\alpha_s^{\delta_{1j}}\right)^\nu(a\Lambda)^{2j},\quad i\in\{-1,0,1\}.
\end{align}
The term $F^M(M_A)$ is given by \cref{eq:mfunc}, and $\Lambda=0.6$ GeV as in \cref{eq:lambda_conn}. The variations of the fit form above included in the BMA are as follows:
\begin{itemize}
\item Powers of $M_{\pi}^2$: The pion mass dependence in \cref{eq:sib-aMu-conn-cont} is varied through the choice of the function $h(x)$, which is taken to be one from the set $\{ 0,x^{-2},\log x^2,x^4\}$ for $\amuW$ observables and $\{0,\log x^2,x^4\}$ for $\amuSD$ observables.  At short distances, the mass dependence is expected to mild; indeed, we find that the term $h(x) = x^{-2}$ doesn't describe our data.  The forms of these variations are inspired by chiral perturbation theory \cite{Golterman:2017njs,Golterman:2017qtu}; see Ref.~\cite{FermilabLatticeHPQCD:2023jof} for further discussion.  In addition to the variations used in Ref.~\cite{FermilabLatticeHPQCD:2023jof}, we include the $h(x)=0$ variation. Other forms of the $M_{\pi}$ dependence are also considered, namely other two-term variations including either $M_{\pi}^{-2}$ and $M_{\pi}^0$ terms or $\log M_{\pi}^2$ and $M_{\pi}^0$ terms, {\it i.e.}, with $h(x)$ in $\{x^{-2},\log x^2\}$ but no $M_{\pi}^2$ ($i=1$) term.  These variations have negligible weight in the BMA due to a poor goodness of fit and thus are not included in the final analysis.

\item Order in $a^2$: The $a^2$ dependence is controlled by the parameters $\{n_i\}$, which are the upper limits of the sums in \cref{eq:sib-ci-conn-cont}.  We take $n_0=0,1,2,3$ and $n_{i\neq0}=0,1,2$ with the additional constraints that $\min(\{n_i\})\geq 1$, and no fits with a negative number of degrees of freedom are included.\footnote{A naive frequentist degree-of-freedom counting would indicate some fits with sea-mass corrections have a negative number of degrees of freedom; however, the tight prior on $C_{\rm sea}$ (see \cref{subsec:light_analysis}) leads to a positive number of Bayesian degrees of freedom for these fits.}

\item $\alpha_s$ dependence of the leading order $a^2$ term: The dependence on the strong coupling
$\alpha_s$ is controlled by the parameter $\nu$ in \cref{eq:sib-ci-conn-cont}, which sets the power of the $\alpha_s$ prefactor in the $a^2$ terms in the fit.  For $\amuSIBW$, $\nu=0,1,2$ variations are included, while for $\amuSIBSD$, only the $\nu=0$ variation is included for reasons discussed in \cref{sec:logEffects}.

\item Sea-mass corrections:  The form of the sea-mass correction $F^M(M_A)$ is the same as in \cref{eq:mfunc}.  For the 0.15$^{\prime}$~fm ensemble, which has distinct up and down sea quark masses, an additional term is included to correct $\Delta M_K^2$ defined in \cref{eq:DeltaK}.  Specifically,
\begin{align}
F^M_{0.15^{\prime}\text{ fm}} (M_A) &= C_{\text{sea}} \sum_{A=\pi, K, \Delta K} \frac{M_{\pi,\,\textrm{phys.}}^2}{M_{A, \,\textrm{phys.}}^2} \delta M^2_A \ , \quad \quad \delta M^2_{\Delta K} = \frac{\Delta M_{K, \,\textrm{phys.}}^2-\Delta M_{K, \,\textrm{latt.}}^2}{2B_0 \Lambda}.
\end{align}
Variations with and without the sea-mass correction term are included.

\item Finite-volume corrections: For $\amuW$, four of the finite-volume correction schemes introduced in \cref{sec:analysis} are included, namely $\chi$PT (NLO and NNLO), CM, and HP schemes; note that pion-mass mistunings are corrected through the data-driven approach described above.  For reasons discussed in \cref{sec:latticeCorrections}, MLLGS is not usable at the 0.15~fm points.
This means that all chiral-continuum fit variations in the MLLGS scheme would be based on only two lattice spacings with which to resolve the mass dependence ($[0.09, 0.12]$ fm); therefore, we do not use MLLGS for $\amuW$.
For $\amuSD$, finite-volume corrections obtained in only the HP scheme are included for reasons discussed in \cref{sec:latticeCorrections}.  Unlike many of the other observables, taste-breaking corrections in part cancel when computing $\Delta a_{\mu}^{ud}(\mathrm{SIB})$ and hence have virtually no impact on the continuum limit.  Thus, no variations that include taste-breaking corrections are considered here.

\item Data subsets: In addition to using the full range of lattice spacings over $[0.06,0.15]$~fm, we also include variations that exclude the coarsest ensembles at 0.15~fm and the finest ensemble at 0.06~fm.  The former variation is considered since it reduces discretization effects in the chiral-continuum extrapolation.  The latter is considered since the 0.06~fm ensemble includes only the $ll$ point and thus does not help constrain the $M_{\pi}$ dependence in the chiral-continuum extrapolation; in practice, the presence of the 0.06~fm point negligibly affects the continuum value of $\Delta a_{\mu}^{ud}(\mathrm{SIB,conn.})$.  In fits that cut away data, the maximum order of the $a^2$ dependence is reduced to avoid fits with a negative number of degrees of freedom.  Hence, the constraints listed above become $n_0=0,1,2$ and $n_{i\neq0}=0,1$ for fits over the range $[0.06,0.12]$~fm and $n_i=0,1,2$ for fits over the range $[0.09,0.15]$~fm.\footnote{The fits on only $[0.06,0.12]$~fm remove $uu$, $ll$, and $dd$ data at 0.15~fm, so the maximum $n_{i\neq0}$ is reduced.  The fits on only $[0.09,0.15]$~fm remove $ll$ data and thus do not have a reduced range of $n_{i\neq0}$ values.}
\end{itemize}

We impose the Gaussian prior constraint discussed in \cref{sec:contextrap} on the sea-mass correction; all other parameters are unconstrained. Because of combinatorial effects introduced by the degree-of-freedom counting discussed above, there would be \textit{a priori} model biasing introduced by a naive, flat model-prior probability $\pr(M)=1/N_M$ in \cref{eq:modelProb}.  For example, there are more possible fit variations for each of $h(x)$ in $\{x^{-2},\log x^2,x^4\}$ than those with $h(x)=0$ because, in the latter case, there are no additional variations from the $c_{-1}(a)$ polynomial.  This discrepancy biases the final BMA result by disfavoring the $h(x)=0$ variations.  To account for such biases, the model-prior probabilities are chosen to approximately eliminate this bias over the data subset and $M_{\pi}$ dependence variations.  Specifically, denoting a model fit to data subset $s$ with $M_{\pi}$ dependence $h$ as $M_{s,h}$, we take
\begin{align}
\pr(M_{s,h})=\frac{1}{N_{\mathrm{subsets}}}\frac{1}{N_{M_{\pi}}}\frac{1}{N_s}\frac{1}{N_h},
\end{align}
where $N_{\mathrm{subsets}}$ is the number of data subset variations in the model space (three here), $N_{M_{\pi}}$ is the number of different $M_{\pi}$ variations (four for $\amuW$, three for $\amuSD$), $N_s$ is the number of models fit to the data subset $s$, and $N_h$ is the number of models with the $M_{\pi}$ dependence~$h$.

\begin{figure}
\centering
\includegraphics[width=0.7\textwidth]{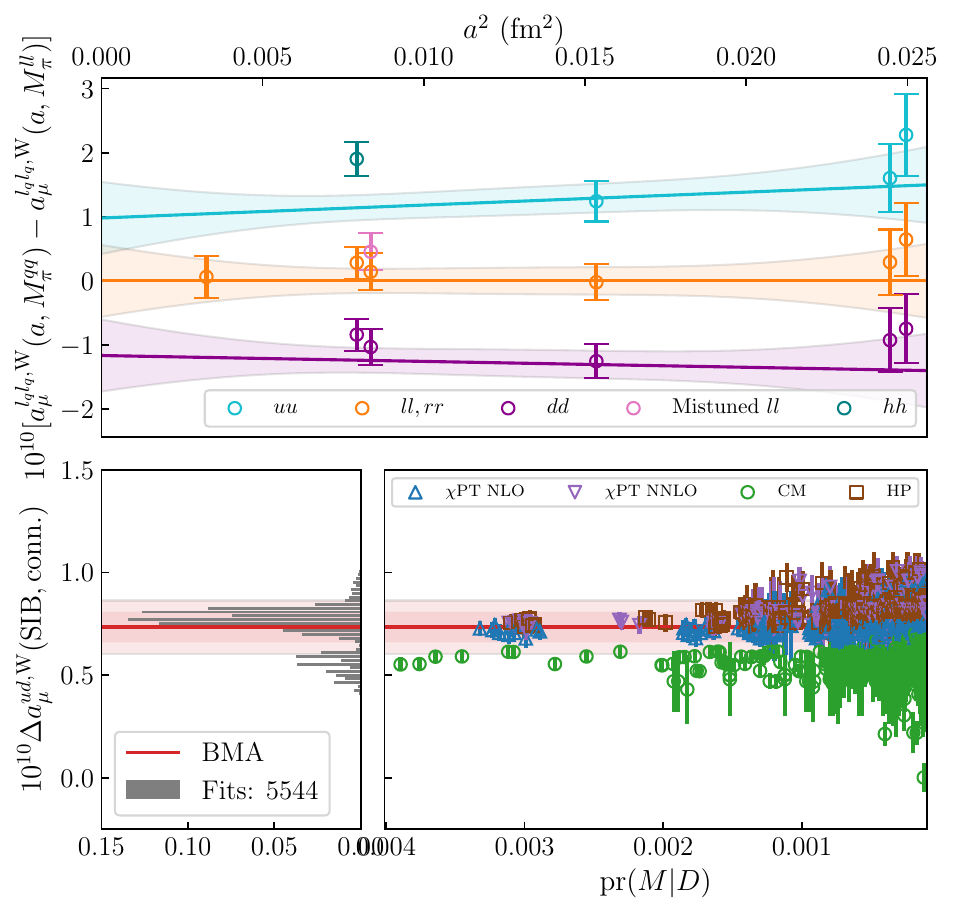}
\vspace{-5mm}
\caption{Results of the BMA procedure applied to $\Delta a_{\mu}^{ud,\mathrm{W}}(\mathrm{SIB,conn.})$.
\panel{Upper panel}: The chiral-continuum fit with the maximum model probability, as well as the corresponding data, relative to the central value of the extrapolation curve evaluated at the physical $ll$ point.  The curves are extrapolations evaluated at physical pion masses for $a_{\mu}^{uu}(\mathrm{conn.})$, $a_{\mu}^{ll}(\mathrm{conn.})$, and $a_{\mu}^{dd}(\mathrm{conn.})$ (light-cyan, -orange, and -purple bands, respectively); the data match this coloring except for the mistuned $ll$ point at 0.09~fm (pink) and the $hh$ point at 0.09$^\star$~fm (teal). Data from the 0.09~fm and $0.15^{\prime}$~fm ensembles include a small, rightward offset for visual clarity.  The symbol shapes indicate that the data are corrected with the CM finite-volume scheme (see description of the lower right panel).
\panel{Lower left panel}: Model probability-weighted histogram of all continuum $\Delta a_{\mu}^{ud,\mathrm{W}}(\mathrm{SIB,conn.})$ extrapolations used in the BMA.
\panel{Lower right panel}: Individual model predictions of $\Delta a_{\mu}^{ud,\mathrm{W}}(\mathrm{SIB,conn.})$ versus the corresponding model probability defined in \cref{eq:modelProb}.  Note that the model probabilities decrease in going from left to right.  The colors and symbols indicate the finite-volume scheme of each fit.  Models with $\pr(M|D)<10^{-4}$ are not shown.
In the lower panels, the BMA result is shown in light-red, where the inner and outer bands correspond to the error coming from the first term in \cref{eq:BMAVar} and the total error, respectively.}
\label{fig:BMASIBConnW}
\end{figure}

The results of the BMA procedure with the variations to $\Delta a_{\mu}^{ud,\mathrm{W}}(\mathrm{SIB,conn.})$ listed above are shown in \cref{fig:BMASIBConnW}.  The top panel shows the best-fit chiral-continuum extrapolations evaluated at the physical $M_{\pi}^{uu},M_{\pi}^{ll},M_{\pi}^{dd}$ masses versus the lattice spacing; extrapolation curves are shown relative to the center value of the physical $ll$ curve $a_{\mu}^{l_ql_q}(a,M_{\pi}^{ll})$.  The bottom panels summarize the results of all fits as a weighted histogram (left panel) and versus their respective model probabilities (right panel) compared with the BMA result (red line and bands).

\begin{figure}
\centering
\includegraphics[width=0.6\textwidth]{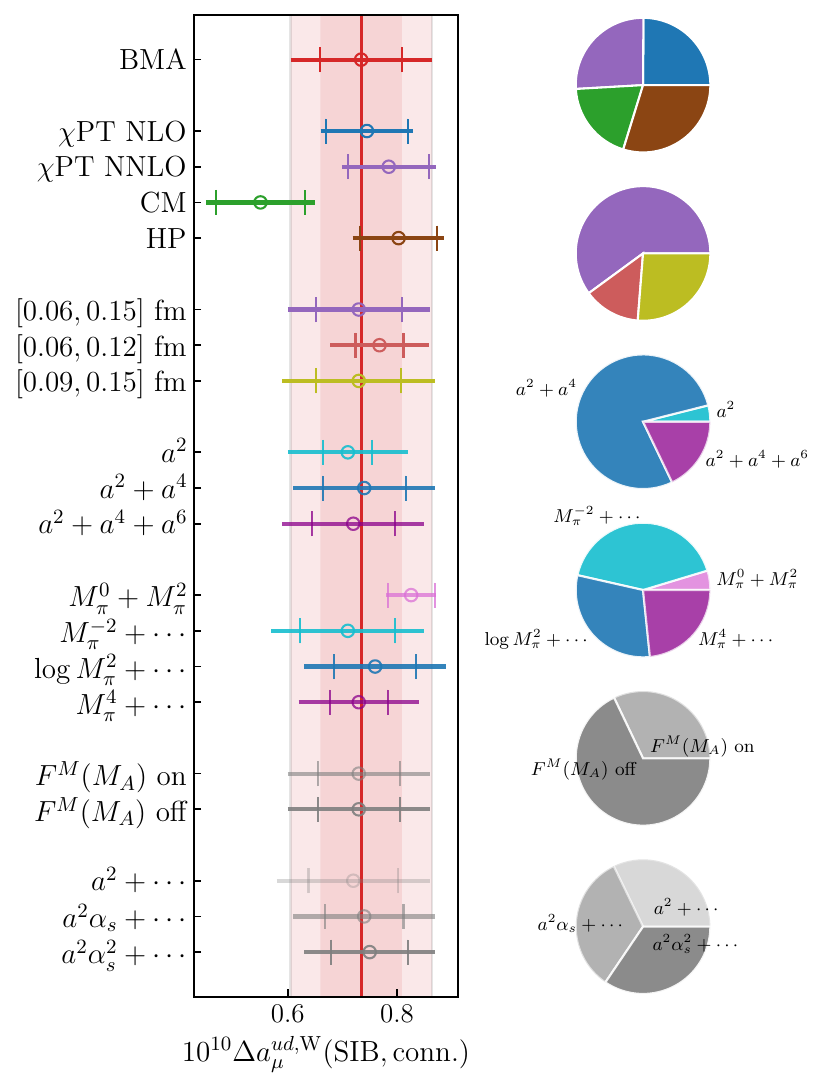}
\vspace{-5mm}
\caption{Breakdown of the BMA procedure applied to $\Delta a_{\mu}^{ud,\mathrm{W}}(\mathrm{SIB,conn.})$.
\panel{Left panel}: From top to bottom, the first point is the full BMA result, including all model variations.  The next block of four points show the result of varying the finite-volume correction schemes.  The next three points show the result of varying the data subsets.  The next two blocks of results come from variations in the chiral-continuum extrapolation, namely the lattice-spacing and the pion-mass dependencies, respectively.  In the former case, the set of models is partitioned by the highest-order term in the extrapolation; in the latter case, by the variations described in the text. The next two show results of fit variations that include or do not include the sea-mass correction $F^M(M_{A})$.  The final three show the result of varying powers of $\alpha_{s}$ in the $a^2$ term of the fit function.  The inner error bars on the data points correspond to the first term in \cref{eq:BMAVar}, while the outer are the total errors.
\panel{Right panels}: Pie charts showing the contributions to the BMA corresponding to the partitions in the left panel.  The percentages are computed according to \cref{eqn:ssProb}.}
\label{fig:BMACompareSIBConnW}
\end{figure}

Similarly to previous sections, we perform Bayesian model averages over subsets of the model space, holding a single model choice fixed while varying the others.  The results of this procedure for $\Delta a_{\mu}^{ud,\mathrm{W}}(\mathrm{SIB,conn.})$ are shown in \cref{fig:BMACompareSIBConnW}.
The individual model subset averages are shown on the left and the relative probabilities of these subsets, as defined in \cref{eqn:ssProb}, are shown on the right.  The full BMA result is shown on the top of the left panels and in the corresponding red line and bands.  The four following points on the left and top pie chart on the right correspond to the four finite-volume schemes (NLO and NNLO $\chi$PT, CM, and HP). The next three points and corresponding pie chart shows the three data subset variations.  We see that the results for $\Delta a_{\mu}^{ud,\mathrm{W}}(\mathrm{SIB,conn.})$ are fairly insensitive to data cuts, with the full data set being favored by the BMA due the $N_{\mathrm{cut}}$ penalty in \cref{eq:modelProb}.  The next seven points and two pie charts correspond to variations in the lattice-spacing dependence (first three points and pie chart) and $M_{\pi}$ dependence (next four points and pie chart) in the chiral-continuum fit Ans\"atze.  The next two points and corresponding pie chart show the effect of including the sea-mass correction or not.  Again, this correction has a small impact on the final BMA results.  The last three points and corresponding pie chart shows the $\alpha_s$ dependence variations.  While the results vary slightly with different powers of $\alpha_s$, none of these variants is favored over the others by the BMA.

As expected from \cref{fig:FVCorr} and shown in \cref{fig:BMASIBConnW,fig:BMACompareSIBConnW}, the tension between the CM scheme and the other schemes informs the estimate of the corresponding systematic error for this observable.  
We emphasize that the systematic error estimate from the BMA already properly accounts for the spread due to the outliers.

\begin{figure}
\centering
\includegraphics[width=0.5\textwidth]{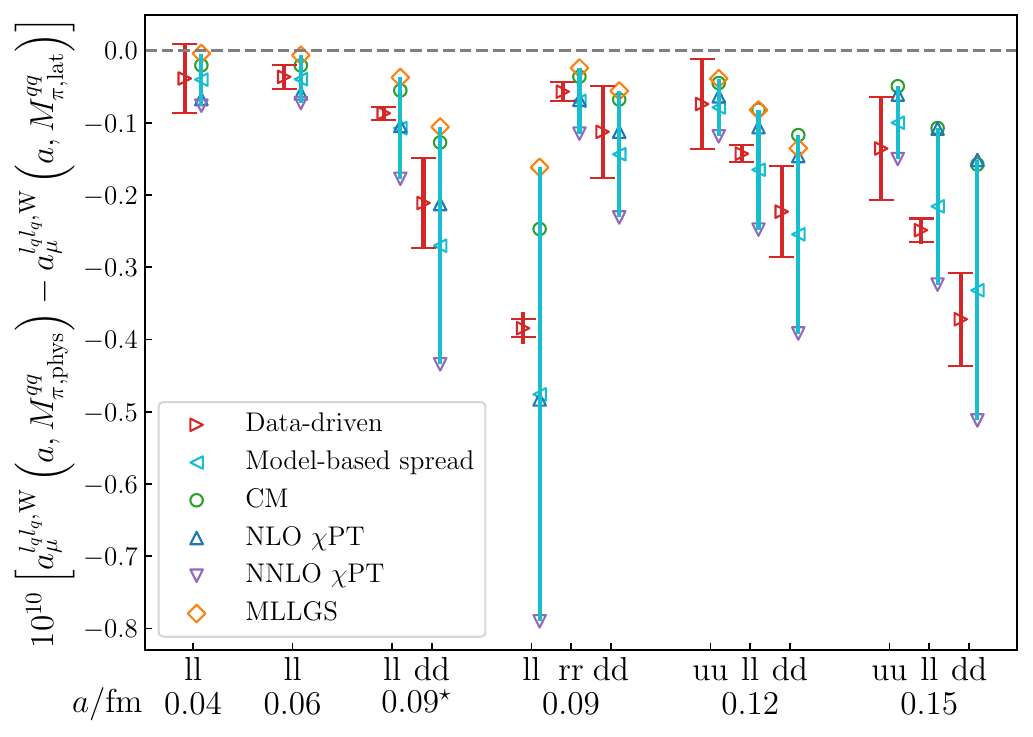}
\vspace{-5mm}
    \caption{Comparison of various EFT and EFT-inspired schemes and the data-driven corrections with the light-quark connected $a_\mu^{\mathrm{W}}$ resulting from adjustments to the value of $M_{\pi}$ for each ensemble and valence quark content.  Individual EFT and EFT-inspired schemes corresponding to the local current are denoted by open symbols; the cyan error bars denote the full spread of the scheme-dependent results; the red error bars are the data-driven determinations from the BMA.  The scheme-dependent results are shown with a small, rightward offset for visual clarity.}
\label{fig:MpiCorr}
\end{figure}

As a cross-check, we can compare the EFT-based corrections for $\Delta_{M_{\pi}}$ defined in \cref{sec:latticeCorrections} with a data-driven estimate that can be obtained from the $\amuSIBCW$ BMA.  Specifically, we perform a Bayesian model average for the difference in $\amuW$ between the fit results at the physical pion masses and measured lattice pion masses for each ensemble and pion flavor $qq$ using the same model space described above. We do not perform this comparison at $hh$ on $0.09^{\star}$~fm where the mass shift is too large to be reliably estimated by the EFT and EFT-inspired schemes or at 0.15$^{\prime}$~fm where the corrected results are nearly identical to those for the 0.15~fm ensemble, which dominates the fits statistically at this lattice spacing.  The results of this comparison are given in \cref{fig:MpiCorr}, which shows that the EFT-inspired corrections and data-driven results are in agreement.

\begin{figure}
\centering
\includegraphics[width=0.7\textwidth]{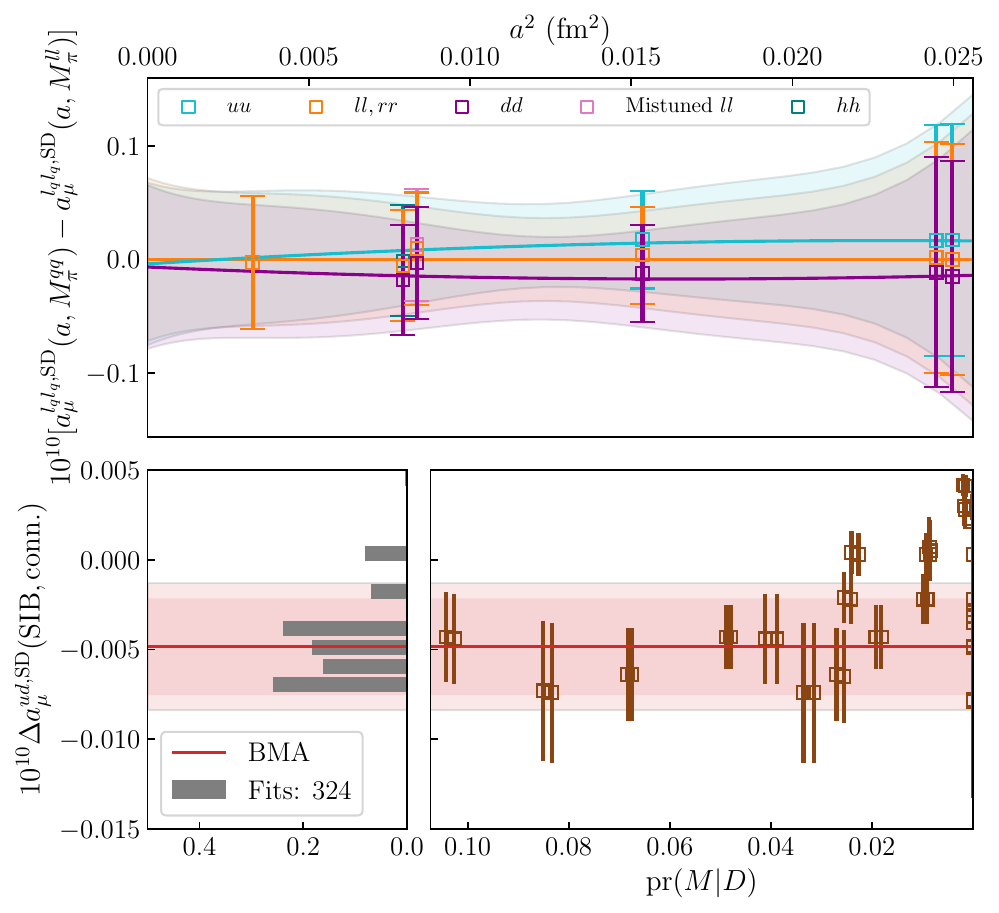}
\vspace{-5mm}
\caption{Results of the BMA procedure applied to $\Delta a_{\mu}^{ud,\mathrm{SD}}(\mathrm{SIB,conn.})$.  See caption of \cref{fig:BMASIBConnW} for a description of the figure. The curves shown in the upper panel correspond to the best fit that uses all of the data on [0.06,0.15]~fm---which is the second best fit in this case---since the best fit excludes the 0.06~fm ensemble. Note that only the HP scheme is used for finite-volume corrections to $\Delta a_{\mu}^{ud,\mathrm{SD}}(\mathrm{SIB,conn.})$.}
\label{fig:BMASIBConnSD}
\end{figure}

\begin{figure}
\centering
\includegraphics[width=0.6\textwidth]{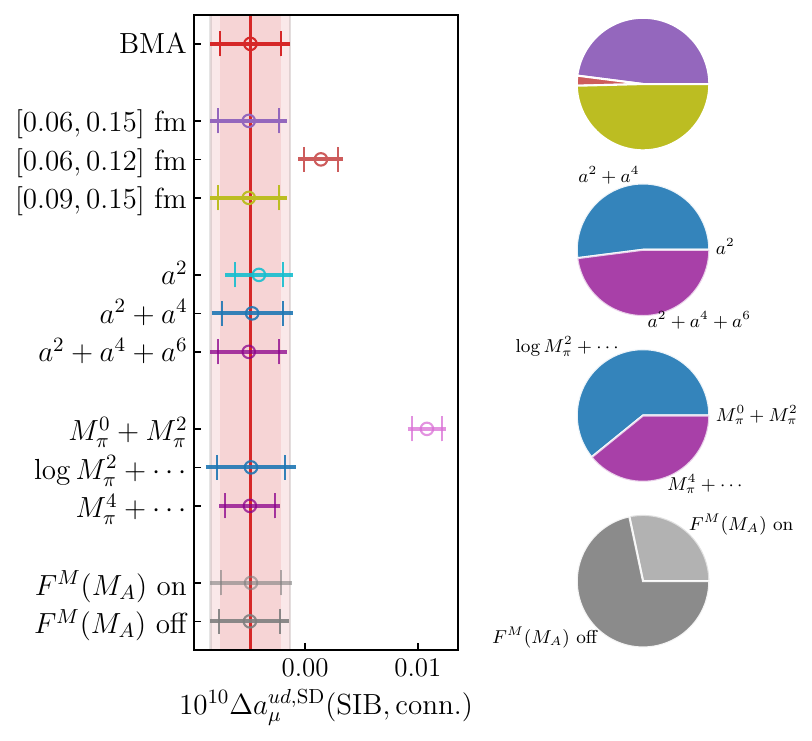}
\vspace{-5mm}
\caption{Breakdown of the BMA procedure applied to $\Delta a_{\mu}^{ud,\mathrm{SD}}(\mathrm{SIB,conn.})$.  See caption of \cref{fig:BMACompareSIBConnW} for a description of the figure.}
\label{fig:BMACompareSIBConnSD}
\end{figure}

The results of the BMA over all fit variations to  $\amuSIBCSD$ are shown in \cref{fig:BMASIBConnSD} with the model subspace averaging procedure, which is a subset of that discussed above for $\amuSIBCW$, shown in \cref{fig:BMACompareSIBConnSD}.  As expected, the SIB correction in the SD window region is much smaller than in the W window region due to the significantly weaker mass dependence at short distances.

\subsubsection{Disconnected strong-isospin breaking}\label{subsubsec:disc_sib_analysis}

The disconnected SIB dataset is obtained using the one-link current (see \cref{table:sibData}).  These data span three lattice spacings---0.09~fm, 0.12~fm, and 0.15~fm---and are evaluated for light quarks $l$ and $s$ in the isospin-symmetric limit and for light quarks $u$, $d$, and $s$ with isospin breaking. The quark masses are those used to produce \cref{table:mesonMasses}.

As in \cref{eq:SIBDisc}, we compute $\Delta a_{\mu}^{ud}(\mathrm{SIB,disc.})$ at each lattice spacing for both the W and SD window regions. At finite lattice spacing, scheme-dependent finite-volume and taste-breaking corrections to $\Delta a_{\mu}^{ud}(\mathrm{SIB,disc.})$ cancel at leading order in the light quark mass splitting $m_d-m_u$ and hence are beyond the present scope of leading-order strong-isospin breaking effects. We perform a continuum fit to these data in for each window observable. The continuum fit Ans\"atze are of the same form as \cref{eq:generalfitfunc}, which is analogous to \cref{eq:sib-conn-cont} with the $M_{\pi}$ dependence removed.  Here, $F^{a}$ is given by
\begin{align}
\label{eq:sib-Fdisc-disc-cont} F^{{a}}_{\rm SIB}(a)&=\sum_{j=1}^{n}c_{j}\left(\alpha_s^{\delta_{1j}}\right)^\nu(a\Lambda)^{2j},
\end{align}
$F^M(M_A)$ is given by \cref{eq:mfunc}, and $\Lambda=0.9$ GeV as in \cref{eq:lambda_disc}.  Included in the BMA are the following  variations of the fit form:
\begin{itemize}
\item Order in $a^2$: The $a^2$ dependence is controlled by the parameter $n$, which is the upper limit of the sum in \cref{eq:sib-Fdisc-disc-cont}.  We take $n=1,2$ with the additional constraint that no fits with a negative number of degrees of freedom are included.  For $\amuSD$, we find that log enhancement of the one-link current (see \cref{sec:logEffects}) leads to a functional dependence of the form $a^2\log a$ in the data.  Thus for $\amuSD$, we also include variations of the form $a^2\log a$, $a^2\log a+a^2$, and $a^2\log a+a^4$.

\item $\alpha_s$ dependence of the leading order $a^2$ term: The $\alpha_s$-dependence variations of other observables using one-link current data are determined using a combination of the results from \cref{sec:logEffects} and empirical Bayes analyses; however, since we obtain $\amuSIBD$ by extrapolating the difference of two values of the disconnected contribution to $\amu$, these arguments cannot be applied directly.  Instead, we determine the $\alpha_s$-dependence variations for $\amuSIBD$ empirically based on results from the BMA, selecting those variations with favorable model weights.  We find a reasonable yet conservative set of variations are $\nu=0,1,2$ for $\amuW$ and $\nu=0$ only for $\amuSD$.

\item Mass-mistuning corrections:  The form of the mass-mistuning correction $F^M(M_A)$ is the same as from \cref{subsec:light_analysis}.  Variations with and without the mass correction term are included.

\item Data subsets: In addition to using the full range of lattice spacings over $[0.09,0.15]$~fm, we also include variations that exclude the coarsest ensemble at 0.15~fm and the finest ensemble at 0.09~fm.  The former is included to reduce discretization effects in the chiral-continuum extrapolation.  The latter is included to inflate the systematic error, which is similar to what is done in the $a_{\mu}^{\mathrm{SD}}(\mathrm{disc.})$ analysis; this error inflation reflects the uncertainty associated with the limited amount of disconnected data available.  Similarly to \cref{subsubsec:conn_sib_analysis}, fits in which data have been cut away do not include the $n=2$ variation.
\end{itemize}
The parameter priors are analogous to those discussed in \cref{subsubsec:conn_sib_analysis}; the bias introduced by using flat model priors $\pr(M)=1/{N_M}$ is small for these observables.

\begin{figure}
\centering
\includegraphics[width=0.7\textwidth]{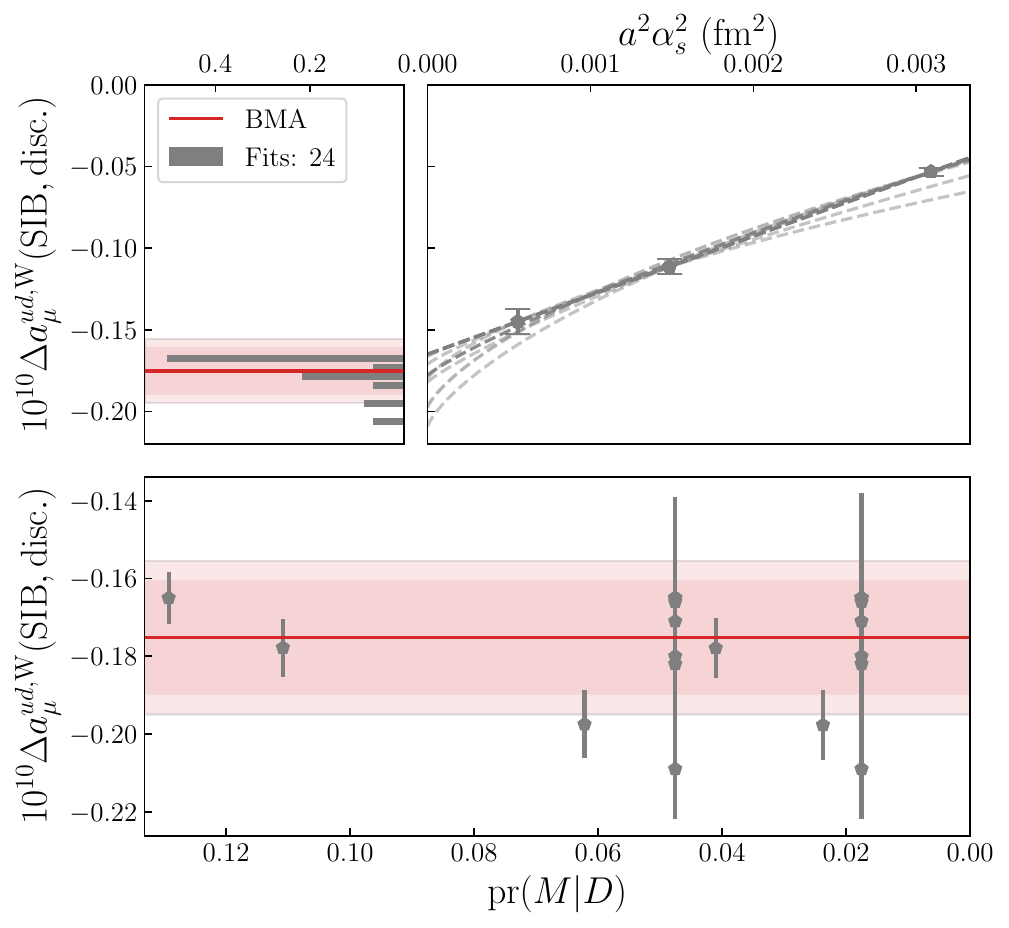}
\vspace{-5mm}
\caption{Results of the BMA procedure applied to $\Delta a_{\mu}^{ud,\mathrm{W}}(\mathrm{SIB,disc.})$.
\panel{Upper left panel}: Model probability-weighted histogram of all continuum $\Delta a_{\mu}^{ud,\mathrm{W}}(\mathrm{SIB,disc.})$ extrapolations used in the BMA.
\panel{Upper right panel}: Extrapolation curves and corresponding data for all model variations included, with opacities proportional to the respective model probabilities.  The shown data points are corrected for mass-mistuning where applicable, which is a small effect.
\panel{Lower panel}: Individual model predictions of $\Delta a_{\mu}^{ud,\mathrm{W}}(\mathrm{SIB,disc.})$ versus the corresponding model probability defined in \cref{eq:modelProb}.  Note that the model probabilities decrease in going from left to right.  In the upper left and lower panels, the BMA result is shown in light-red, where the inner and outer bands correspond to the error coming from the first term in \cref{eq:BMAVar} and the total error, respectively.}
\label{fig:BMASIBDiscW}
\end{figure}

The results of the BMA procedure with the variations to $\Delta a_{\mu}^{ud,\mathrm{W}}(\mathrm{SIB,disc.})$ listed above are shown in \cref{fig:BMASIBDiscW}.  The top panels show the results of all continuum extrapolations, with a weighted histogram of the continuum results in the left panel and all the extrapolation fits versus the lattice spacing in the right panel.  The opacities of the continuum extrapolations are proportional to the respective model weights; for fits with mass-mistuning corrections, the mass-corrected data points are also shown but are difficult to see due to the smallness of these corrections and the light opacity of these variations.  The bottom panel shows the continuum results of the fits versus their respective model probabilities.  In the upper left and bottom panels, the BMA result is shown by the red line and bands.  The final BMA error of $\Delta a_{\mu}^{ud,\mathrm{W}}(\mathrm{SIB,disc.})$ is dominated by the first term in \cref{eq:BMAVar} as shown by the inner red band.

\begin{figure}
\centering
\includegraphics[width=0.5\textwidth]{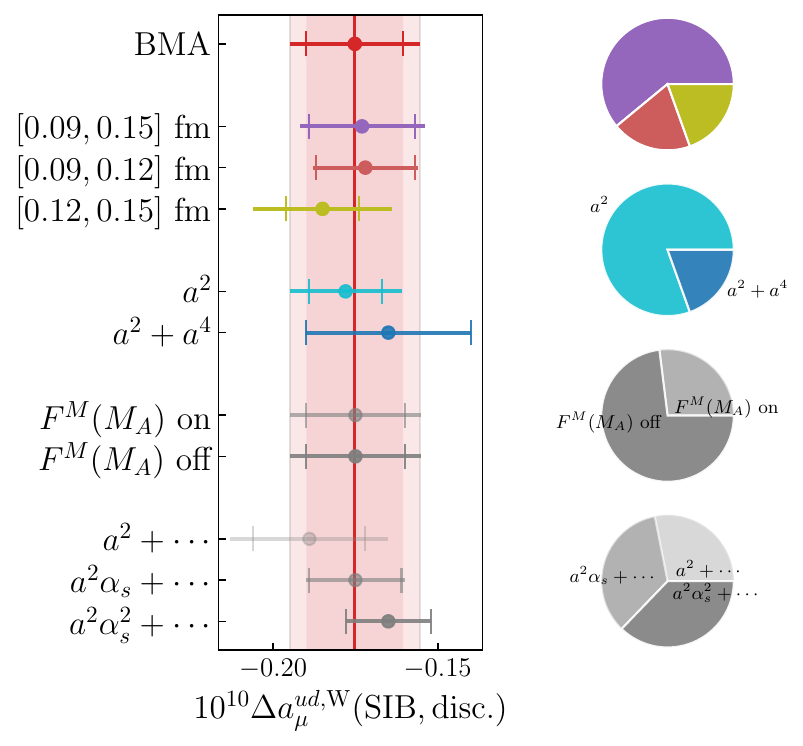}
\vspace{-5mm}
\caption{Breakdown of the BMA procedure applied to $\Delta a_{\mu}^{ud,\mathrm{W}}(\mathrm{SIB,disc.})$.
\panel{Left panel}: From top to bottom, the first point is the full BMA result including all model variations.  The next block of three points show the effect of  varying the data subset.  The next two points are from varying the lattice spacing dependence in the continuum extrapolation, as distinguished by the highest order term in the extrapolation.  The next two come from fit variations that include or exclude the mass-mistuning correction $F^M(M_{A})$, respectively. The final three show the effect of varying the powers of $\alpha_{s}$ in the $a^2$ term of the fit function.  The inner error bars on the data points correspond to the first term in \cref{eq:BMAVar}, while the outer are the total errors.
\panel{Right panels}: Pie charts showing the contributions to the BMA corresponding to the breakdowns in the left panel.  The percentages are computed according to \cref{eqn:ssProb}.}
\label{fig:BMACompareSIBDiscW}
\end{figure}

\begin{figure}
\centering
\includegraphics[width=0.6\textwidth]{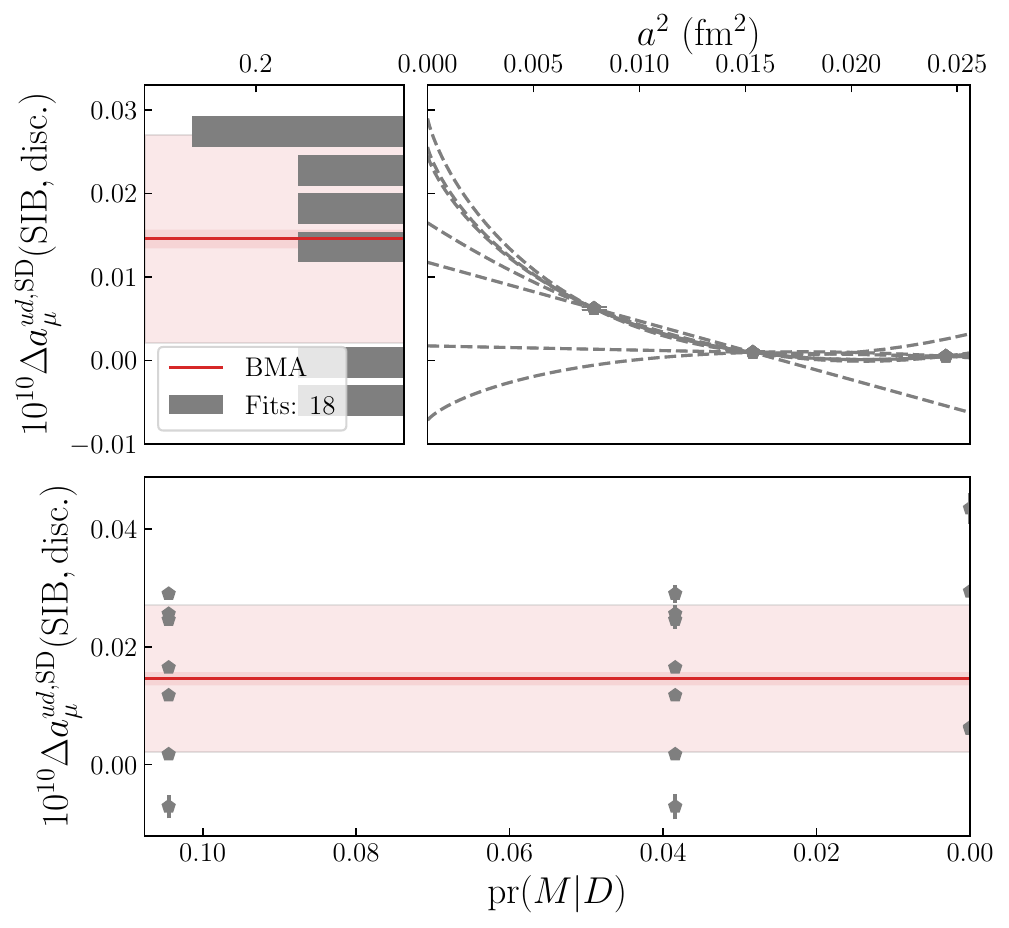}
\vspace{-5mm}
\caption{Results of the BMA procedure applied to $\Delta a_{\mu}^{ud,\mathrm{SD}}(\mathrm{SIB,disc.})$.  See caption of \cref{fig:BMASIBDiscW} for a description of the figure.}
\label{fig:BMASIBDiscSD}
\end{figure}

\begin{figure}
\centering
\includegraphics[width=0.65\textwidth]{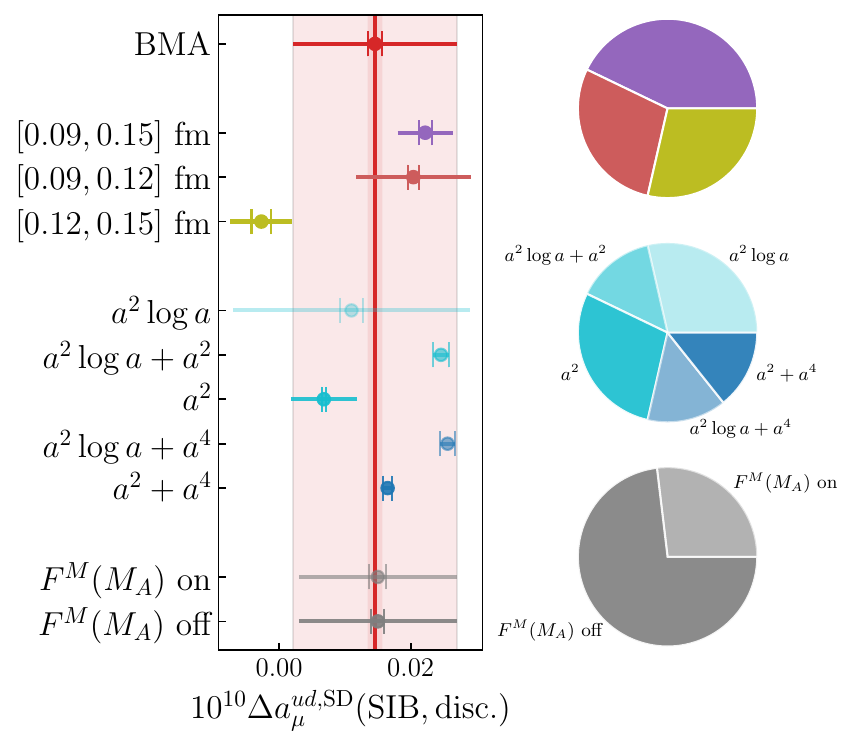}
\vspace{-5mm}
\caption{Breakdown of the BMA procedure applied to $\Delta a_{\mu}^{ud,\mathrm{SD}}(\mathrm{SIB,disc.})$.  See caption of \cref{fig:BMACompareSIBDiscW} for a description of the figure.}
\label{fig:BMACompareSIBDiscSD}
\end{figure}

Again, we perform Bayesian model averages over subsets of the model space where a single model choice has been held fixed.  The results of this procedure for $\Delta a_{\mu}^{ud,\mathrm{W}}(\mathrm{SIB,disc.})$ are shown in \cref{fig:BMACompareSIBDiscW}.  The individual model subset averages are shown on the left and the relative probabilities of these subsets, as defined in \cref{eqn:ssProb}, are shown on the right.  The full BMA result is shown on the top of the left panels and in the corresponding red line and bands.  The following three points and corresponding pie chart show the three data subset variations.  We see that the full data set is favored by the BMA due the $N_{\mathrm{cut}}$ penalty in \cref{eq:modelProb}.  The next two points and pie chart correspond to the different lattice spacing dependence.  The next two points and corresponding pie chart show the effect of including the mass-mistuning correction or not, which behaves similarly to those in \cref{fig:BMACompareSIBConnW}.  The last three points and corresponding pie chart show the $\alpha_s$ dependence variations.  As for $\Delta a_{\mu}^{ud,\mathrm{W}}(\mathrm{SIB,conn.})$, none of these variants are significantly favored over the others by the BMA.

The results of the BMA over all fit variations to $\Delta a_{\mu}^{ud,\mathrm{SD}}(\mathrm{SIB,disc.})$ are shown in \cref{fig:BMASIBDiscSD} with the model subspace averaging procedure shown in \cref{fig:BMACompareSIBDiscSD}.  The set of variations in \cref{fig:BMACompareSIBDiscSD} is a subset of those in \cref{fig:BMACompareSIBDiscW} for $\amuSIBDW$ except for the lattice spacing dependence variations, which now also include the fits with $a^2\log a$ terms.  Again, the SIB correction in the SD window region is much smaller than the corresponding one in the W window region due to the weaker mass dependence short distances.  For this observable, the systematic error is significantly increased by the coarse data subset $[0.12,0.15]$~fm.  Even with this error inflation procedure, the resultant uncertainty on $\Delta a_{\mu}^{ud,\mathrm{SD}}(\mathrm{SIB,disc.})$ is small compared with that of other observables.

\vspace{10mm}

\subsection{Perturbative QCD estimates of short-distance observables}
\label{subsec:pQCD_analysis}

As a consistency check on our short-distance window results, we compare our lattice determinations with predictions from perturbative QCD (pQCD). Specifically, we use the $\mathcal{O}(\alpha_s^4)$ determination of the $R$~ratio, $R(s)$, from the software package \texttt{rhad}~\cite{Harlander:2002ur}. Results produced from this code have been used to supplement other short-distance (or high-energy) lattice determinations of $\amuHVP$ \cite{Budapest-Marseille-Wuppertal:2017okr,Alexandrou:2022amy,Boccaletti:2024guq}. pQCD results for $R(s)$ can be converted to the Euclidean time domain using a Laplace transform,
\begin{equation}
    C(t) = \frac{1}{12\pi^2} \int_0^\infty \dd{E} E^2 R(E^2) \e^{-Et}. \label{eqn:CtPert}
\end{equation}
Alternatively, $C(t)$ is the Fourier transform of the vector correlator in momentum space~\cite{Chetyrkin:2010dx}; see Ref.~\cite{Blum:2023qou} for a comparison with lattice-QCD calculations. With either approach, the resulting $C(t)$ can then be used to compute window observables via \cref{eq:amuTintWin}.

Comparisons of pQCD calculations with hadronic results (from lattice or $e^+e^-$ data) rely on quark-hadron duality. The Laplace transform smears out the peaked structure of $R(s)$~\cite{Poggio:1975af,Melnitchouk:2005zr}, so this duality is expected to hold rather accurately for $a_\mu$ and related windows.

For the connected contributions, we pursue an approach employed in Ref.~\cite{Blum:2023qou} for the light-quark case. Specifically, we extrapolate our lattice results to the continuum for a flexible window $\mathcal{W}_2(t,t^\prime,0.4,0.15)$. The contribution from the complementary short-distance window, $\mathcal{W}_1(t, t^\prime, 0.15)$, is then computed using the pQCD determination of $C(t)$ via $R(s)$ from \texttt{rhad}~\cite{Harlander:2002ur}. The sum of the lattice and perturbative results are then computed as a function of $t^\prime$. As in Ref.~\cite{Blum:2023qou}, we monitor the combined result for stability out to the edge of the standard short-distance window, $t^\prime =0.4$~fm. In addition to studying the light-quark contribution in this way, we also examine the cases of strange and charm.

The \texttt{rhad} software package requires as inputs a threshold and quark masses for the loops, which we choose as follows. For the connected light and strange contributions, the lower threshold is taken to be $2\bar{m}_q$, where $\bar{m}_{q}$ is the $\MSbar$ mass at $\mu=\bar{m}_q$, taken from Ref.~\cite{FlavourLatticeAveragingGroupFLAG:2021npn} (based on Refs.~\cite{EuropeanTwistedMass:2014osg,Chakraborty:2014aca,Lytle:2018evc,FermilabLattice:2018est}). The light and strange propagators in the loops are taken to be massless. For the charm quark, we employ both available options in the software, namely either using the pole mass or the running $\MSbar$ mass $\bar{m}_c(\sqrt{s})$. The charm threshold is then taken to be $2m_c$. We also evaluate the light-strange disconnected contribution (called ``singlet'' in \texttt{rhad}) in pQCD. Here, we use massless light quarks and a massive strange quark. The QED contribution is computed for the massless light and strange quarks, and the $\amuSIBSD$ contribution is negligible and, thus, ignored. 

\begin{figure}
\centering
\includegraphics[scale=0.485]{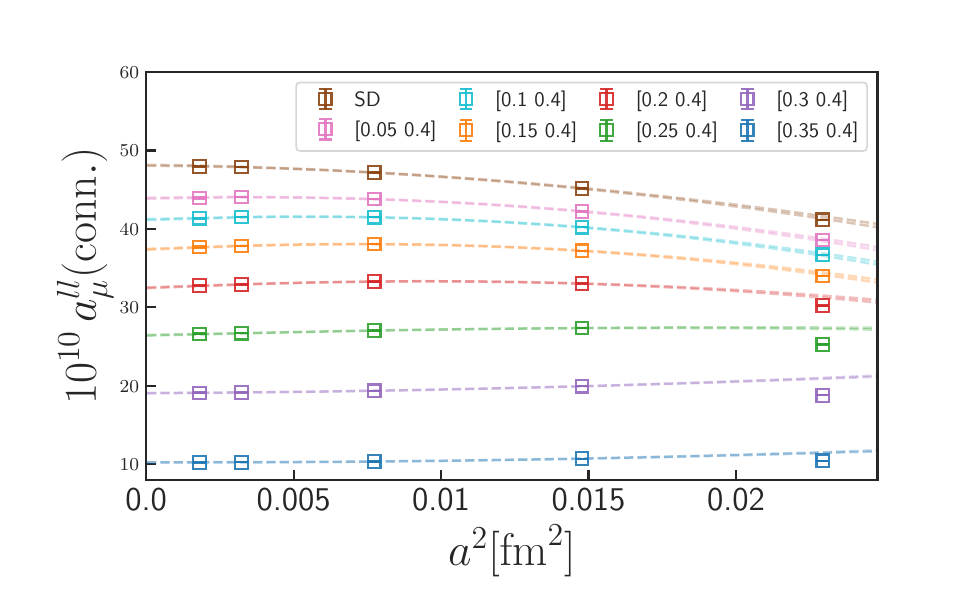}\hspace{4mm}\includegraphics[scale=0.5]{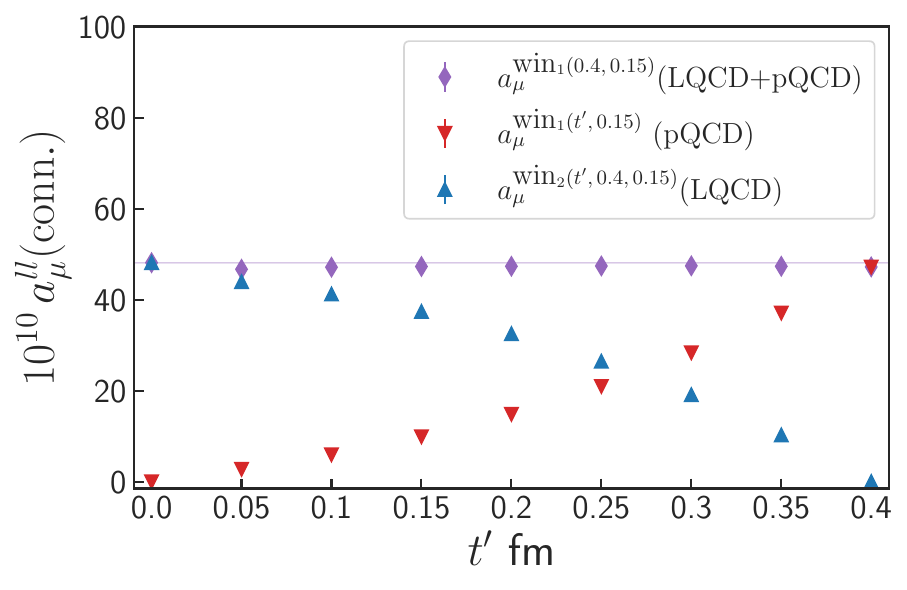}\\
\includegraphics[scale=0.485]{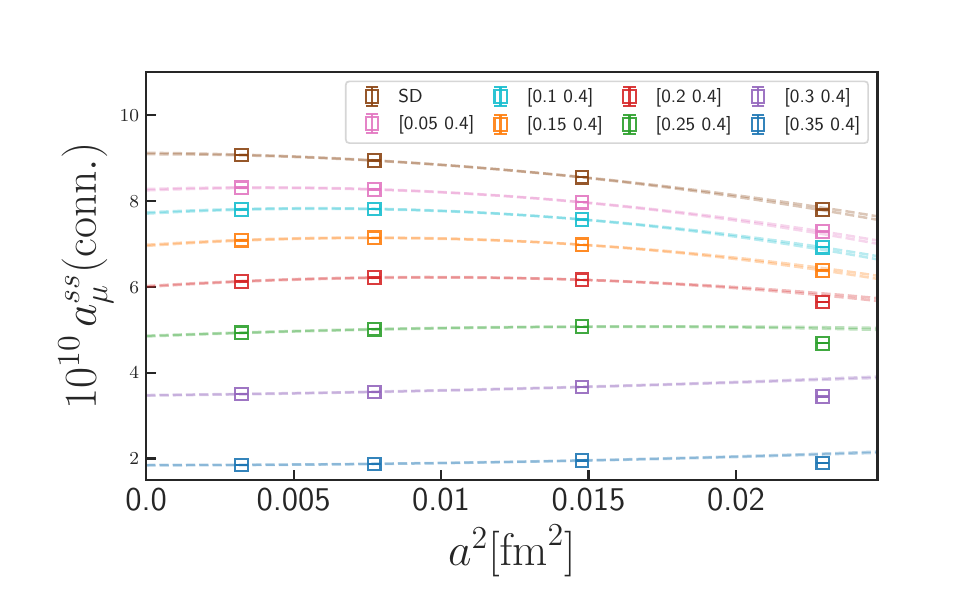}\hspace{5mm}\includegraphics[scale=0.5]{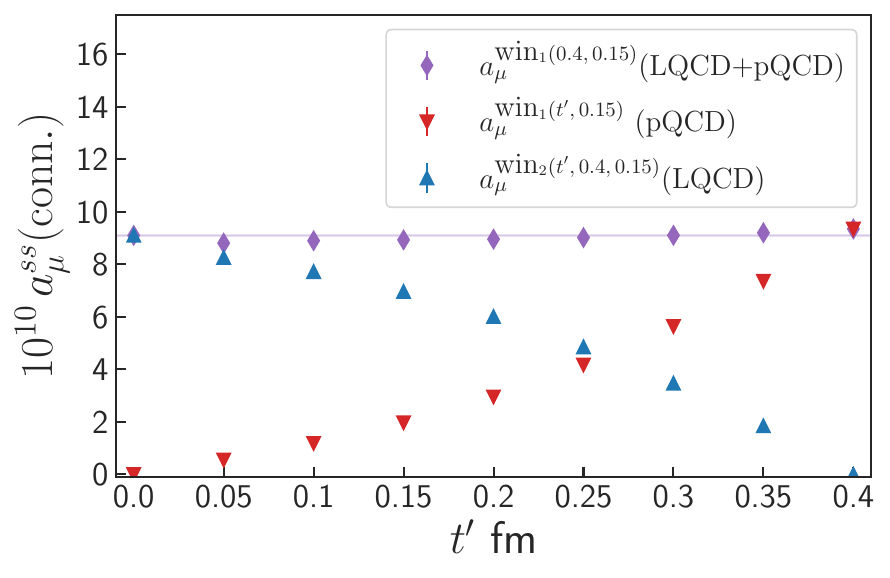}\\
\includegraphics[scale=0.485]{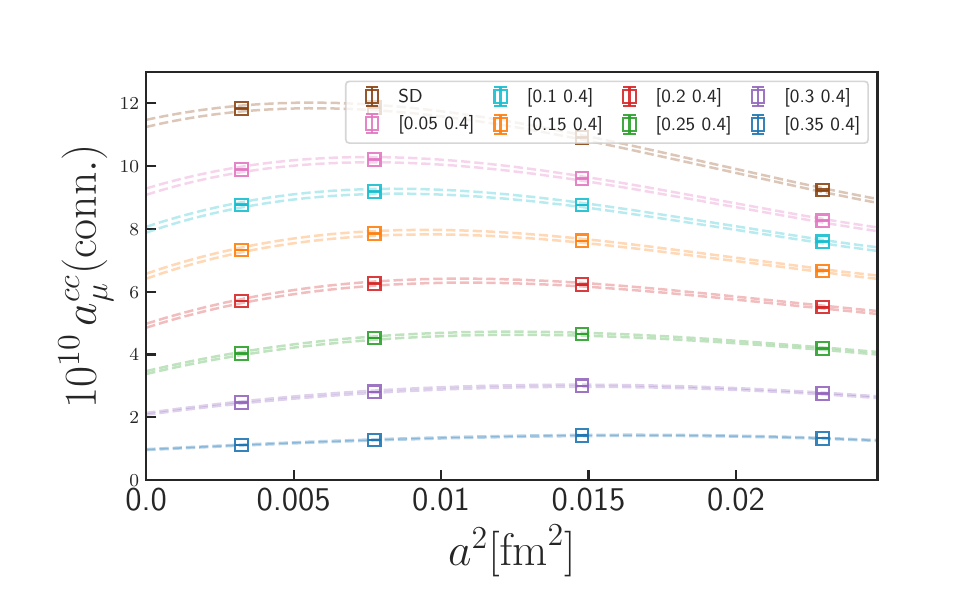}\hspace{4mm}\includegraphics[scale=0.5]{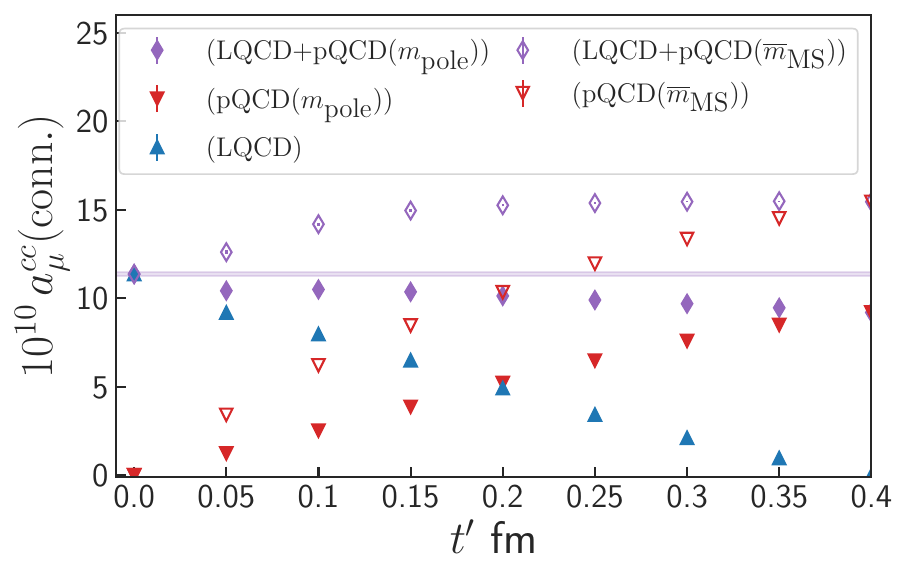}
\caption{Stability analysis for the light- (top), strange- (middle), and charm-quark (bottom) contributions in the short-distance window region. \panel{Left panels}: Results of continuum extrapolations for the short-distance sub-window contributions with the lattice data. \panel{Right panels}: Combinations of lattice and perturbative QCD results from the sub-window regions totaling the full short-distance window region as a function of $t^\prime$.}
\label{fig:pertStab}
\end{figure}

Results for the perturbative replacement procedure described above are shown in \cref{fig:pertStab} for the light, strange, and charm contributions, respectively. For this test, it suffices to use only the local vector current and a single fit to the data. A full BMA as in \cref{subsec:light_analysis,subsec:sc_analysis} would be unnecessarily arduous, so we choose a representative fit from amongst the fits with the highest BMA weights in the full $\amuLSD$, $\amuSSD$, and $\amuCSD$ analysis. We find that, for the light and strange contributions, the massless perturbative result is consistent with the lattice result, and we have stability out to the end of the short-distance window as in Ref.~\cite{Blum:2023qou}. For charm, where the quark-mass scheme is relevant, such stability is not observed. We find that when using the pole mass for the charm quark, the perturbative result is marginally smaller than the lattice result, and we see only partial stability out to $t^\prime=0.1$~fm. Using the $\MSbar$ mass and allowing the mass to run, we obtain a perturbative result that is significantly larger than the lattice result. The authors of Ref.~\cite{Harlander:2002ur} recommend using the pole mass near threshold and switching to the $\MSbar$ mass at larger energies. In theory, tuning can be performed to match the lattice results by playing with a switch-over energy, but we did not find this exercise illuminating.

We note that the Extended Twisted Mass Collaboration (ETMC)~\cite{Alexandrou:2022amy} used the running $\MSbar$ mass approach to obtain a result for the bottom contribution to the short-distance window of
\begin{align}
    a^{bb,\, \textrm{SD}, \,\text{pQCD, ETMC}}_{\mu}(\text{conn.}) =0.32 \times 10^{-10}.
\end{align}
This result is in good agreement with previous lattice determinations of this quantity for the full integrand \cite{Hatton:2021dvg},
\begin{align}
    a^{bb, \,\text{latt, HPQCD}}_{\mu}(\text{conn.}) = 0.300(15) \times 10^{-10},
\end{align}
with the difference between the short-distance and full integrand contribution for bottom expected to be in the third digit. For completeness, we give the pQCD result for this contribution using the pole mass for the bottom quark from Ref.~\cite{Harlander:2002ur}, 
\begin{align}
    a^{bb,\, \textrm{SD} \,\text{pQCD, \textrm{pole mass}}}_{\mu}(\text{conn.}) =0.28 \times 10^{-10}.
\end{align}
As in the case of charm-quark, we observe that the two perturbative determinations bound the lattice result. Additionally, we use the pQCD estimate for bottom contribution to the intermediate window from the following difference, 
\begin{align}
    \amuBW = a^{bb,\,\text{pQCD}}_\mu(\textrm{conn.}) - a^{bb,\,\text{SD},\,\text{pQCD}}_\mu(\textrm{conn.})=0.004  \times 10^{-10}.\label{eqn:amuBottomW}
\end{align}
using the pole mass.

For the disconnected contribution, we obtain a pQCD result of
\begin{align}\label{eq:amuDpQCD}
    a^{\text{pQCD}}_{\mu}(\text{disc.}) = -0.00547 \times 10^{-10}. 
\end{align}
As discussed in \cref{subsec:disc_analysis}, this result does not enter the analysis. We compare it with our result for $\amuDSD$ in \cref{fig:SDDisc}. 

Finally, we compute the $\mathcal{O}(\alpha\alpha_s)$ QED corrections to the massless light and strange contributions in the short-distance window,
\begin{align}
    \amuQEDLSD &= 0.0265 \times 10^{-10}, \label{eqn:QEDLSDPQCD} \\
    \amuQEDSSD &= 0.0016 \times 10^{-10}. \label{eqn:QEDSSDPQCD}
\end{align}
Higher order contributions, including the QED correction to the disconnected diagram, which enter at $\mathcal{O}(\alpha\alpha^3_s)$, are neglected in this estimate \cite{Harlander:2002ur}.

\section{Results}\label{sec:results}

\subsection{Short-distance window observables}\label{subsec:sd_results}

\begin{table}
\centering
\caption{Approximate absolute error budgets (in units of $10^{-10}$) for the uncertainties reported in \cref{eqn:amuLSDRes,eqn:amuSSDRes,eqn:amuCSDRes,eqn:amuDSDRes,eqn:amuSIBCSDRes,eqn:amuSIBDSDRes}.}
\label{table:SDIndividualUncertainty}
\begin{tabularx}{\linewidth}{rCcCcCr}
\hline \hline
Contribution &  stat.    & $a \to 0$  &  $a$  & $Z_V$ & $m_u/m_d$ & Total\\ 
\hline
$\amuLSD$ & 0.011 & 0.06 & 0.028 & 0.063 & --- & 0.092  \\
$\amuSSD$ & 0.003 & 0.01 & 0.005 & 0.012 & --- & 0.017  \\
$\amuCSD$ & 0.0   & 0.14 & 0.09  & 0.02  & --- & 0.17 \\
$\amuDSD$ & 0.0004 &0.0026 &0.0000 &0.0000 &--- &0.0026  \\
$\Delta a_{\mu}^{ud,\mathrm{SD}}(\mathrm{SIB,conn.})$  & 0.0026 & 0.0023 & 0.0004 & 0.0000 & 0.0006 & 0.0035\\
$\Delta a_{\mu}^{ud,\mathrm{SD}}(\mathrm{SIB,disc.})$ & 0.001 & 0.012 & 0.000 & 0.000 & --- & 0.012 \\
\hline \hline
\end{tabularx}
\end{table}

\begin{figure}
    \centering
    \includegraphics[scale=0.9]{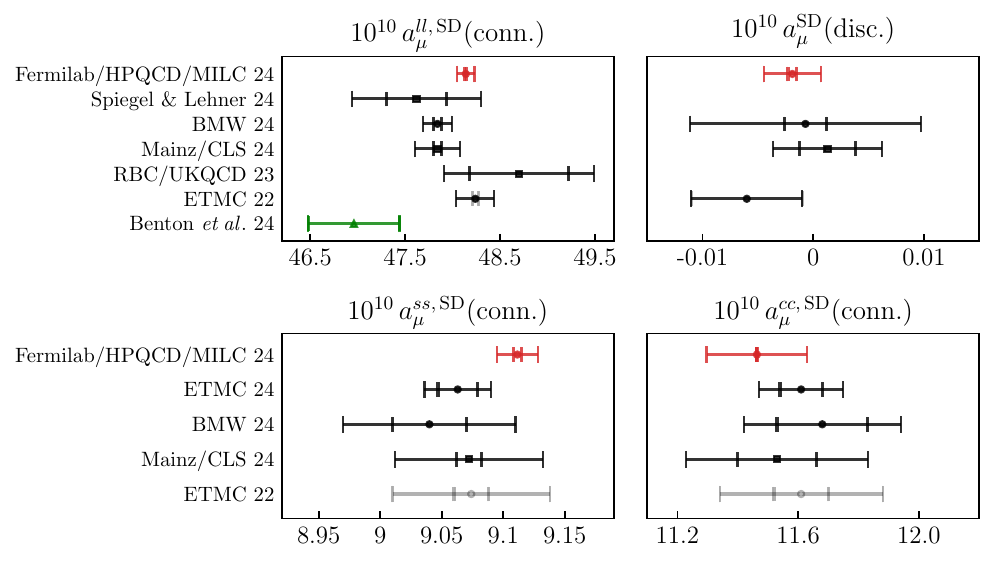}
    \caption{Comparison of our lattice determinations for $\amuLSD$, $\amuSSD$, $\amuDSD$, and $\amuCSD$  (red circles) labeled ``Fermilab/HPQCD/MILC~24'' with $n_f=2+1+1$ (black circles) and $n_f=2+1$ (black squares) lattice-QCD calculations by ETMC~24 \cite{ExtendedTwistedMass:2024nyi}, Spiegel \& Lehner~24 \cite{Spiegel:2024dec}, BMW~24 \cite{Boccaletti:2024guq}, Mainz/CLS~24 \cite{Kuberski:2024bcj}, RBC/UKQCD~23 \cite{Blum:2023qou} and ETMC~22 \cite{Alexandrou:2022amy}. The inner error bar shown for our result is from Monte Carlo statistics.  Also shown is a data-driven evaluation of $\amuLSD$ using $e^+e^-$ cross section data (green triangle) by Benton {\it et al.}~24 \cite{Benton:2024kwp}.}\label{fig:SDIndividualCompare}
\end{figure}

Our final determinations for the individual contributions to $\amuHVP$ in the short-distance window are 
\begin{align}
    \amuLSDRes \label{eqn:amuLSDRes}\,,\\
    \amuSSDRes \label{eqn:amuSSDRes}\,,\\
    \amuCSDRes \label{eqn:amuCSDRes}\,,\\
    \amuDSDRes \label{eqn:amuDSDRes}\,,\\
    \amuSIBCSDRes \label{eqn:amuSIBCSDRes}\,, \\
    \amuSIBDSDRes \label{eqn:amuSIBDSDRes}\,. 
\end{align}
where the uncertainties are statistical (second column of \cref{table:SDIndividualUncertainty}), systematic (third to sixth columns of \cref{table:SDIndividualUncertainty}) and total.
We compare with previous lattice determinations in \cref{fig:SDIndividualCompare}. 
For the light-quark connected contribution, we obtain a result with a relative error of $0.19\%$. The benefit of the HISQ local current mitigating the log-enhancement effects associated with short-distance $\amuHVP$ observables (see \cref{sec:logEffects}) is apparent. This is the most precise determination of this quantity to date and is in reasonable agreement with all previous determinations (left side of \cref{fig:SDIndividualCompare}). Indeed, the strengths of the local current are apparent in the heavier flavor contributions, $\amuSSD$ and  $\amuCSD$, also, with our determinations being the most precise to date. We note that our calculation of $\amuLSD$ differs from the data-driven evaluation of Ref.\cite{Benton:2024kwp} by $2.4\sigma$.

The particulars of the disconnected contribution in the short-distance window are such that the quantity is nearly zero with difficult to determine systematic effects associated with the short-distance window boundary, the behavior of the disconnected correlation function, and staggered oscillating effects. As such, we estimate this quantity with a conservative approach and obtain a result consistent and competitive with previous determinations in both central value and uncertainties.
Our results for the strong-isospin-breaking contribution to $\amuSD$ [\cref{eqn:amuSIBCSDRes,eqn:amuSIBDSDRes}] are both near zero, as expected. We are the first to put forth a lattice calculation of this quantity. 

For the bottom-quark connected and QED contributions to $\amuSD$ we consider existing lattice and phenomenological results in our estimates. In particular, for $\amuBSD$, we take the previous HPQCD determination \cite{Hatton:2021dvg} and subtract from it the estimate of $\amuBW$ in \cref{eqn:amuBottomW}. This yields
\begin{align}
    \amuBSD = 0.296(15), \label{eqn:amuBSDres}
\end{align}
which we include in our complete result assuming 100\% correlation given the shared HISQ ensembles in that work. This is likely a vast overestimate of the true correlation.

For the QED correction to this window observable, we take the estimates of \cref{eqn:QEDLSDPQCD,eqn:QEDSSDPQCD}, to which we assign a $100\%$ error, yielding
\begin{align}
    \amuQEDSDRes . \label{eqn:amuQEDSDres}
\end{align}

\subsection{Intermediate-distance window observables}\label{subsec:w_results}

\begin{table}
\centering
\caption{Approximate absolute error budgets (in units of $10^{-10}$) for the uncertainties reported in \cref{eqn:amuLWRes,eqn:amuSWRes,eqn:amuCWRes,eqn:amuDWRes,eqn:amuSIBCWRes,eqn:amuSIBDWRes}.}
\label{table:WIndividualUncertainty}
\vspace{1mm}
\begin{tabularx}{\linewidth}{rcCcClcr}
\hline \hline
Contribution &  stat.    & $a\to0$, $\Delta_{\textrm{TB}}$   &  $\Delta_{\textrm{FV}}$, $\Delta_{M_{\pi}}$ &    $a$  & $Z_V$ & $m_u/m_d$ & Total\\ 
\hline
$\amuLW$ & 0.14  & 0.37   & 0.34  & 0.31   & 0.18  & --- & 0.63 \\
$\amuSW$ & 0.01  & 0.06   & ---   & 0.10   & 0.04  & --- & 0.13 \\
$\amuCW$ & 0.000 & 0.063  & ---   & 0.059  & 0.005 & --- & 0.087 \\
$\amuDW$ & 0.06 &0.18  &0.03  &0.00 & 0.00&--- &0.20 \\
$\Delta a_{\mu}^{ud,\mathrm{W}}(\mathrm{SIB,conn.})$ & 0.07 & 0.05 & 0.09 & 0.01 & 0.01 & 0.03 & 0.13 \\
$\Delta a_{\mu}^{ud,\mathrm{W}}(\mathrm{SIB,disc.})$ & 0.015 & 0.013 & --- & 0.002 & 0.000 & --- & 0.020 \\
\hline \hline
\end{tabularx}
\end{table}

\begin{figure}
    \centering
    \includegraphics[scale=0.9]{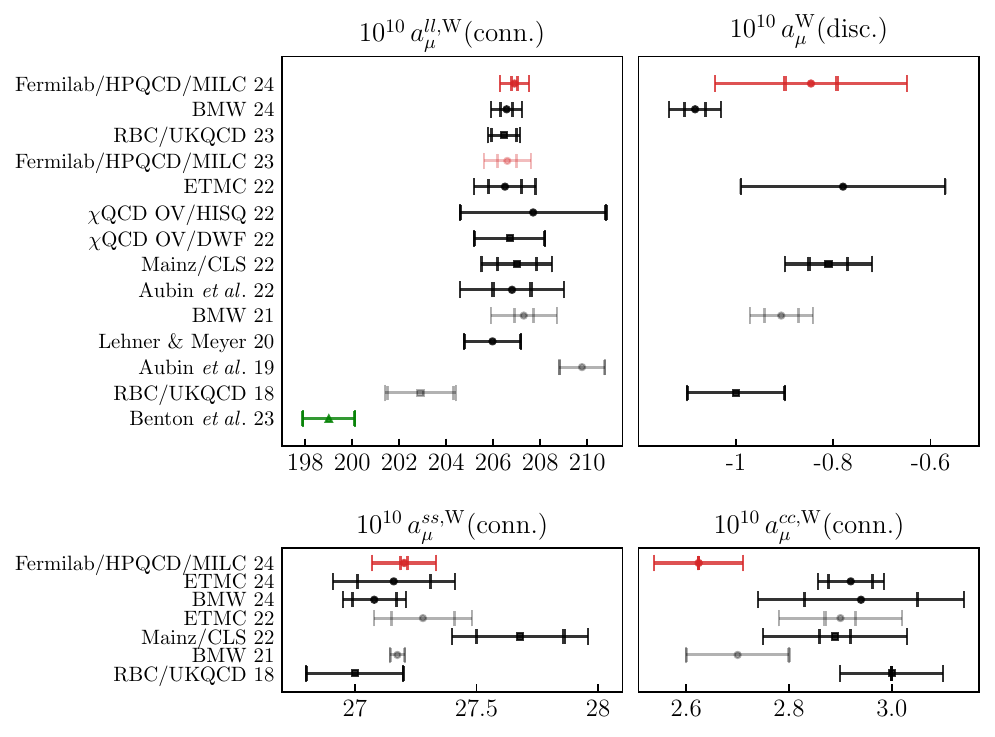}
    \caption{Comparison of our lattice determinations for $\amuLW$, $\amuDW$, $\amuSW$, and $\amuCW$ (red circles) labeled ``Fermilab/HPQCD/MILC~24'' with $n_f=2+1+1$ (black circles) and $n_f=2+1$ (black squares) lattice-QCD calculations by ETMC~24 \cite{ExtendedTwistedMass:2024nyi}, BMW~24 \cite{Boccaletti:2024guq}, RBC/UKQCD~23 \cite{Blum:2023qou}, Mainz/CLS~22 \cite{Ce:2022kxy}, Aubin {\it et al.}~22 \cite{Aubin:2022hgm}, $\chi$QCD 22 \cite{Wang:2022lkq}, ETMC~22 \cite{Alexandrou:2022amy} and Lehner \& Meyer~20 \cite{Lehner:2020crt}. Our previous result, Fermilab/HPQCD/MILC 23, is shown in light red. BMW~21 \cite{Borsanyi:2020mff},  Aubin {\it et al.}~19 \cite{Aubin:2019usy} and RBC/UKQCD~18 \cite{RBC:2018dos}, shown in gray, have been superseded. The inner error bar shown for our result is from Monte Carlo statistics.  Also shown is a data-driven evaluation of $\amuLW$ using $e^+e^-$ cross section data (green triangle) by Benton {\it et al.}~23 \cite{Benton:2023dci,Benton:2024kwp}.}
    \label{fig:WIndividualCompare}
\end{figure}

\begin{figure}
    \centering
    \includegraphics[scale=0.72]{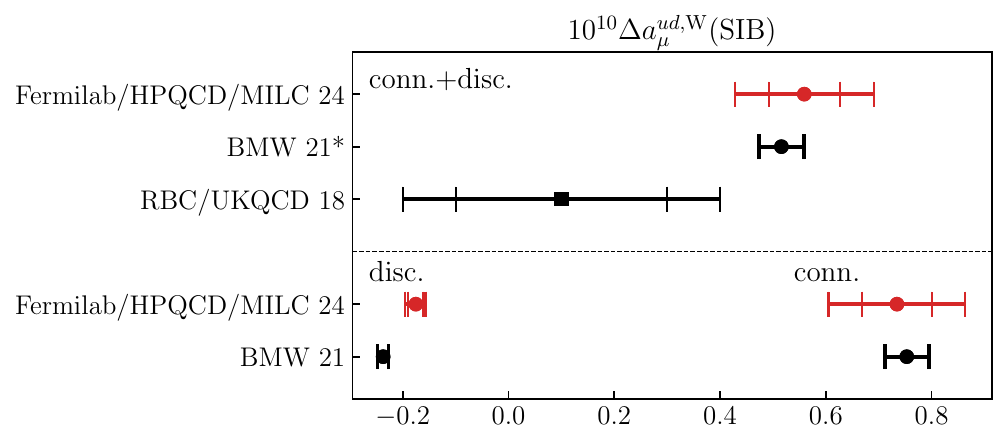}
    \vspace{-2mm}
    \caption{Comparison of our lattice determinations of $\Delta a_{\mu}^{ud,\mathrm{W}}(\mathrm{SIB,conn.})$ and $\Delta a_{\mu}^{ud,\mathrm{W}}(\mathrm{SIB,disc.})$ (red circles) labeled ``Fermilab/HPQCD/MILC~24'' with $n_f=2+1$ (black squares) and $n_f=2+1+1$ (black circles) lattice-QCD calculations by RBC/UKQCD~18 \cite{RBC:2018dos} and BMW~21 \cite{Borsanyi:2020mff}, respectively.  The ``BMW~21*" value is computed assuming the connected and disconnected contributions are uncorrelated.  The inner error bars are from Monte Carlo statistics.
}
    \label{fig:WSIBCompare}
\end{figure}

Our final determinations for the individual contributions to $\amuHVP$, in the intermediate-distance window, are 
\begin{align}
    \amuLWRes \label{eqn:amuLWRes}\, ,\\
    \amuSWRes \label{eqn:amuSWRes}\, ,\\
    \amuCWRes \label{eqn:amuCWRes}\, ,\\
    \amuDWRes \label{eqn:amuDWRes}\, ,\\
    \amuSIBCWRes \label{eqn:amuSIBCWRes}\, , \\
    \amuSIBDWRes \label{eqn:amuSIBDWRes}\, .
\end{align}
The respective error budgets for these contributions are given in \cref{table:WIndividualUncertainty}. We compare these results with previous determinations in \cref{fig:WIndividualCompare,fig:WSIBCompare}. We obtain $\amuLW$ with a relative error of $0.3\%$, a significant improvement over our previous result's 0.5\% precision \cite{FermilabLatticeHPQCD:2023jof}. Our previous leading uncertainty, from the continuum limit, is decreased from $0.34\% \to 0.18\%$. This significant improvement is a result of the introduction of a second vector-current discretization and a finer lattice spacing at $0.04$~fm. For similar reasons, our uncertainty from Monte-Carlo statistics is decreased by $0.19\% \to 0.07\%$. In addition, with the new scale setting determination from $M_\Omega$, our scale uncertainty is decreased from $0.21\% \to 0.15\%$. The systematic uncertainty from the finite-volume correction is largely unchanged as it is determined through the same approach as in Ref.~\cite{FermilabLatticeHPQCD:2023jof}. In \cref{fig:WIndividualCompare}, we observe complete consistency with our previous determination as well all the other lattice determinations \cite{Borsanyi:2020mff,Lehner:2020crt,Alexandrou:2022amy,Wang:2022lkq,Aubin:2022hgm,Ce:2022kxy,Blum:2023qou,Boccaletti:2024guq}. However, our calculation of $\amuLW$ differs from the data-driven evaluation of Ref.~\cite{Benton:2024kwp} by $6.2\sigma$. 

For the heavy-flavor contributions, $\amuSW$ and $\amuCW$, we find the uncertainties, as broken down in \cref{table:WIndividualUncertainty}, are dominated by scale-setting and the continuum limit. For $\amuSW$, we find consistency with all previous determinations, except for a slight tension with the Mainz/CLS~22 result~\cite{Ce:2022kxy}. For $\amuCW$, we see some tension with the most recent determinations of this quantity, while we observe good agreement with the BMW~21 result. Discretization errors are a significant source of uncertainty in our calculation of  $\amuCW$, and we plan to update our calculation by including data at $a\approx 0.04$~fm, which will allow us to test and refine our continuum extrapolations.


Our evaluation of the disconnected contribution to $\amuW$, given in \cref{eqn:amuDWRes}, is consistent with all previous determinations (right of \cref{fig:WIndividualCompare}) albeit with a larger error than the most recent determinations. We find the systematic uncertainty due to variations in continuum extrapolations completely dominates all other uncertainty sources. This calculation includes data at lattice spacings $a\gtrapprox 0.09$~fm, and we expect that adding data from $a\approx 0.06$~fm ensemble would yield a significant reduction in the uncertainty.

For the strong-isospin breaking corrections to $\amuW$, we obtain a result for the connected contribution, \cref{eqn:amuSIBCWRes}, that is consistent with the single previous result from BMW 21 (\cref{fig:WSIBCompare} bottom right), albeit with larger errors. For the disconnected contribution, \cref{eqn:amuSIBDWRes}, we see some tension with the BMW result (bottom left). We do find the combination of the two contributions is consistent with BMW 21 while being in mild tension with the RBC/UKQCD 18 estimate (top of \cref{fig:WSIBCompare}).

For the bottom-quark connected contribution to the intermediate window observable, $\amuBW$, which is almost negligible at our current level of precision, we use the perturbative estimate from \cref{eqn:amuBottomW}. We assign a $100\%$ error to account for the tensions found in \cref{subsec:pQCD_analysis} stemming from the quark-mass scheme in pQCD, yielding
\begin{align}
    \amuBW = 0.004(4). \label{eqn:amuBWres}
\end{align}

Our ongoing, direct lattice-QCD calculation of the QED corrections \cite{Ray:2022ycg} to $\amuW$ will be presented in a separate paper. Meanwhile, we consider inputs from phenomenology and comparisons of previous lattice-QCD calculations \cite{RBC:2018dos,Borsanyi:2020mff,Ce:2022kxy,Hoferichter:2023sli,FermilabLattice:2022izv,Benton:2024kwp} to provide an estimate of $\amuQEDW$. The phenomenological estimate in Ref.~\cite{Hoferichter:2023sli} uses a data-driven evaluation based on exclusive-channel $e^+e^-$ cross section measurements to quote a value of $+2.3(9) \times 10^{-10}$, which can be regarded as a conservative bound on the scale of the effect. The phenomenological analysis of QED and strong-isospin-breaking effects in Ref.~\cite{FermilabLattice:2022izv} finds a smaller correction, consistent with the lattice-QCD calculations of Refs.~\cite{RBC:2018dos,Borsanyi:2020mff,Ce:2022kxy}, and we base our estimate on this result. We note that Ref.~\cite{FermilabLattice:2022izv} presented some indications of cancellations between QED and strong-isospin-breaking effects. Coupled with the expectation that QED effects are expected to be smaller in windowed quantities, we allow for a range of $\pm 0.2 \times 10^{-10}$. Following Ref.~\cite{FermilabLattice:2022izv}, we do not apply a correction to the central value of $\amuQEDW$ and take the above range as the uncertainty, yielding
\begin{align}
   \amuQEDWRes. \label{eqn:amuQEDWres}
\end{align}

\section{Summary and outlook}\label{sec:conclusion}

In this paper, we present complete results for the short- and intermediate-distance window observables $\amuSD$ and $\amuW$ from our ongoing project to compute $\amuHVP$ in lattice QCD at few permille-level precision. The two observables together comprise about 45\% of the total HVP. We first compute the individual contributions to these observables, defined in \cref{eq:amuBreakdown,eq:sib-def,eq:amuQED}. This includes the contributions from quark-line connected contractions $\amuLSD$, $\amuLW$, $\amuSSD$, $\amuSW$, $\amuCSD$, $\amuCW$; quark-line-disconnected contractions $\amuDSD$, $\amuDW$; and isospin-breaking corrections $\amuSIBSD$, and $\amuSIBW$. Each contribution is obtained at the physical point and in the continuum and infinite volume limits, after a comprehensive systematic error analysis which employs BMA to obtain the total statistical and systematic uncertainties as well as estimates of the approximate breakdown of the error budgets. Our final results for the individual contributions to the short- and intermediate-distance observables are summarized in \cref{eqn:amuLSDRes,eqn:amuSSDRes,eqn:amuCSDRes,eqn:amuDSDRes,eqn:amuSIBCSDRes,eqn:amuSIBDSDRes} and \cref{eqn:amuLWRes,eqn:amuSWRes,eqn:amuCWRes,eqn:amuDWRes,eqn:amuSIBCWRes,eqn:amuSIBDWRes}, respectively. 
The bottom-quark connected and QED contributions are not computed directly in this work. Instead, we use a mix of previous  lattice-QCD results and phenomenological estimates to obtain the results listed in \cref{eqn:amuBSDres,eqn:amuQEDSDres} and \cref{eqn:amuBWres,eqn:amuQEDWres}, respectively.
Finally, we employ a generalized BMA procedure (see \cref{sec:bma_cov}) to compute the statistical and systematic correlations between the individual contributions listed in \cref{table:correlationsSD,table:correlationsW}. Summing up the individual contributions with these correlations, we obtain the following results for the complete determinations of $\amuSD$ and $\amuW$, respectively:  
\begin{align}
    \amuSDResABS, \label{eqn:TotalSD}\\
     \amuWResABS. \label{eqn:TotalW}
\end{align}
The approximate systematic error breakdown for each observable is shown in \cref{table:TotalUncertainty}. 
\begin{table}
\centering
\caption{Approximate relative error budgets (in \%) for the total $\amuSD$ and $\amuW$. 
\vspace{1mm}}
\label{table:TotalUncertainty}
\begin{tabularx}{\linewidth}{lcCcCcccccR}
\hline \hline
Contrib. &  stat.    &$a\to0,~\Delta_{\textrm{TB}}$   &  $\Delta_{\textrm{FV}}$,~$\Delta_{M_{\pi}}$   &  $a$  & $Z_V$  & $\amuB$ & $\amuQED$ & $m_u/m_d$ & Total\\ 
\hline
$\delta \amuSD$  & 0.02 & 0.24 & --- & 0.13 & 0.12 & 0.02 & 0.04 & 0.00 & 0.31\\
$\delta \amuW$ & 0.07 & 0.18 & 0.22 & 0.17 & 0.07 & 0.00 & 0.09 & 0.01 & 0.36\\
\hline \hline
\end{tabularx}
\end{table}
With a relative uncertainty of $0.31\%$, our calculation of $\amuSD$ is the most precise to date. Our $\amuW$ result includes an updated calculation of the light-quark connected contribution ($\amuLW$), which is significantly more precise than our previous result \cite{FermilabLatticeHPQCD:2023jof}, thanks to including correlation function data at a lattice spacing of $a\approx 0.04$~fm, the finest to date, as well as improved statistics, the addition of a second vector-current discretization and an improved scale setting determination from the $\Omega$ baryon mass. Compared with all other lattice-QCD calculations, it also currently has the smallest statistical uncertainty. 

\begin{figure}
\centering
\includegraphics[scale=1.05]{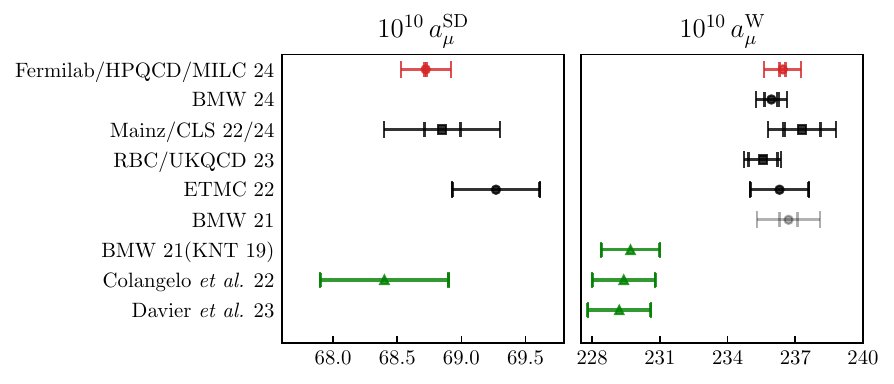}
\vspace{-5mm}
\caption{Comparison of our lattice determination of the total $\amuSD$ and $\amuW$ (red circles) labeled ``Fermilab/HPQCD/MILC~24'' with $n_f=2+1+1$ (black circles) and $n_f=2+1$ (black squares) lattice-QCD calculations by BMW~24 \cite{Boccaletti:2024guq}, Mainz/CLS~22/24 \cite{Ce:2022kxy,Kuberski:2024bcj}, RBC/UKQCD~23 \cite{Blum:2023qou} (W only) and ETMC~22 \cite{Alexandrou:2022amy}. The previous $\amuW$ result of BMW~21 \cite{Borsanyi:2020mff} is shown in gray. Also shown are data-driven evaluations of $\amuSD$ and $\amuW$ using $e^+e^-$ cross section data (green triangles) by Colangelo {\it et al.}~22 \cite{Colangelo:2022vok}, BMW~21 (using the KNT19 dataset) \cite{Borsanyi:2020mff}, and Davier {\it et al.}~23 \cite{Davier:2023cyp}.}
\label{fig:TotalCompare}
\end{figure}
As illustrated in \cref{fig:TotalCompare}, our $\amuSD$ and $\amuW$ results are in good agreement with previous lattice-QCD calculations \cite{Alexandrou:2022amy,Ce:2022kxy,Kuberski:2024bcj,Blum:2023qou,Borsanyi:2020mff,Boccaletti:2024guq}. This can also be seen in the detailed comparisons between our and previous results for the individual $\amuSD$ ($\amuW$) contributions shown in \cref{fig:SDIndividualCompare}
(\cref{fig:WIndividualCompare,fig:WSIBCompare}). 
We note that the sum of $\amuSD$ and $\amuW$ in \cref{eqn:TotalSD,eqn:TotalW} is consistent with the evaluation in Ref.~\cite{FermilabLattice:2022gku} of the one-sided window observable $a_\mu^{\textrm{win}_1(1,\,0.15)}$ (see \cref{eq:amuTintWin1}).  
  
Comparing our results with the (pre-2023) data-driven evaluations of Ref.~\cite{Colangelo:2022vok}, also shown in \cref{fig:TotalCompare}, we see that our $\amuSD$ determination is consistent with Ref.~\cite{Colangelo:2022vok} at $0.6\sigma$, while our $\amuW$ result differs from Ref.~\cite{Colangelo:2022vok} by $4.3\sigma$. These observations are in line with similar comparisons shown in Refs.~\cite{Alexandrou:2022amy,Blum:2023qou,Borsanyi:2020mff,Boccaletti:2024guq,Benton:2023dci,Benton:2023fcv,Davier:2023cyp,Benton:2024kwp}, and indicate that the difference between lattice-QCD and (pre-2023) data-driven evaluations stems from the low-energy region, which is dominated by the two-pion channel in the data-driven method or, correspondingly, by the light-quark-connected contribution in lattice-QCD calculations. This is also where the new cross section measurement by the CMD-3 collaboration \cite{CMD-3:2023alj} disagrees with the measurements included in the pre-2023 KNT19 \cite{Keshavarzi:2019abf} and DHMZ19 \cite{Davier:2019can} compilations. The CMD-3 measurement of the two-pion cross section implies a higher value for $\amuW$ that is compatible with lattice-QCD determinations \cite{Benton:2023dci,Boccaletti:2024guq,Benton:2024kwp}. To gain further insights from such comparisons, it is important first to understand or resolve the differences between the experimental measurements.  

Our calculation of $\amuSD$ benefits from the fact that the local vector-current used to generate the connected correlation functions is protected from log-enhanced contributions (see \cref{sec:logEffects}), enabling precise control over continuum-limit extrapolations. Still, as can be seen in the error budgets for $\amuSD$ in \cref{table:TotalUncertainty,table:SDIndividualUncertainty}, continuum limit extrapolations are the dominant source of uncertainty, where $\amuCSD$ contributes the largest error.  We note that the strange- and charm-connected contributions are computed on ensembles with lattice-spacings $a\gtrapprox 0.06$~fm. We are in the process of generating strange- and charm-connected correlation functions on the ensemble at the finest lattice spacing of $a \approx 0.04$~fm, which will refine our calculation of the corresponding contributions. 

Examining the systematic error budgets for $\amuW$ in \cref{table:TotalUncertainty,table:WIndividualUncertainty}, we see that the dominant sources of uncertainty are due to finite-volume corrections, the continuum limit, and scale setting. These are also dominant or important sources of error in the long-distance window observables, and will be addressed in our ongoing and planned computations. In particular, we note that the disconnected contributions are computed on ensembles with lattice-spacings $a\gtrapprox 0.09$~fm. We plan to extend this analysis to a finer lattice spacing of $a\approx 0.06$~fm. Our calculation of the light-quark connected contribution to $\amuLD$ and $\amuHVP$ is described in the companion paper \cite{Bazavov:2024eou}.

At $3.1\permil$ and $3.6\permil$, respectively, our results for the total $\amuSD$ and $\amuW$ are already close to our precision goal for $\amuHVP$. 
However, considerable exascale computing resources are needed to reach this precision for the full $\amuHVP$. In particular, we plan to continue to generate correlation functions on our finest ensemble ($a \approx 0.04$~fm) to complete the data set. This will improve statistical and systematic errors (due to discretization errors) for all observables. 
The finite-volume corrections will be quantified more precisely in a direct finite-volume study, for which we are analyzing a new gauge-field ensemble with a spatial extent of $L=11$~fm at a lattice spacing of $a\approx 0.09$~fm. 

We note that the $\amuSD$ and $\amuW$ results presented here rely on phenomenological estimates of the QED corrections. The results from our ongoing direct lattice-QCD calculation \cite{Ray:2022ycg} are deferred to a separate paper. While this calculation uses quenched QED, we are also working on computations of the disconnected and sea-quark QED corrections, which make use of our LMA set-up (see \cref{subsec:lattice}). In general, less is known about the QED corrections \cite{RBC:2018dos,Borsanyi:2020mff,Ce:2022kxy} compared with other contributions, especially in the long-distance region. While their contributions to $\amuHVP$ are small, they are relevant at the desired few permille precision level.

Finally, the well-known signal-to-noise problem is a limiting factor specific to HVP observables at long Euclidean time distances. 
Here, a recent pilot study \cite{Lahert:2024vvu} showed that spectral reconstructions of the vector current correlation function at large Euclidean times obtained from direct computation of the two-pion contributions can address this problem also for the case of staggered fermions. We plan to extend this study to finer lattice spacings, where the statistical gains will have a larger impact on the continuum extrapolated results.

\acknowledgments

We thank Claude Bernard, Urs Heller, Paul Mackenzie, Bob Sugar, and Doug Toussaint for their scientific leadership and collaboration.
In particular, we are grateful to Bob and Paul for their tireless efforts to obtain computational resources, to Claude for guidance on chiral perturbation theory, to Doug for his invaluable expertise in creating so many of our gauge-field ensembles, and to Urs for essential contributions to previous projects that formed the basis for this work. 
Computations for this work were carried out in part with computing and long-term storage resources provided by the USQCD Collaboration, the National Energy Research Scientific Computing Center (Cori), the Argonne Leadership Computing Facility (Mira) under the INCITE program, and the Oak Ridge Leadership Computing Facility (Summit) under the Innovative and Novel Computational Impact on Theory and Experiment (INCITE) and the ASCR Leadership Computing Challenge (ALCC) programs, which are funded by the Office of Science of the U.S.\ Department of Energy. 
This work used the Extreme Science and Engineering Discovery Environment (XSEDE) supercomputer Stampede 2 at the Texas Advanced Computing Center (TACC) through allocation TG-MCA93S002. The XSEDE program is supported by the National Science Foundation under grant number ACI-1548562.
Computations on the Big Red II+, Big Red 3, and Big Red 200 supercomputers were supported in part by Lilly Endowment, Inc., through its support for the Indiana University Pervasive Technology Institute.
The parallel file system employed by Big Red II+ was supported by the National Science Foundation under Grant No.~CNS-0521433.
This work utilized the RMACC Summit supercomputer, which is supported by the National Science Foundation (awards ACI-1532235 and ACI-1532236), the University of Colorado Boulder, and Colorado State University. The Summit supercomputer is a joint effort of the University of Colorado Boulder and Colorado State University.
Some of the computations were done using the Blue Waters sustained-petascale computer, which was supported by the National Science Foundation (awards OCI-0725070 and ACI-1238993) and the state of Illinois. Blue Waters was a joint effort of the University of Illinois at Urbana-Champaign and its National Center for Supercomputing Applications.
We also used the Cambridge Service for Data Driven Discovery (CSD3), part of which is operated by the University of Cambridge Research Computing Service on behalf of the Science and Technology Facilities Council (STFC) DiRAC HPC Facility. The DiRAC component of CSD3 was funded by BEIS capital funding via STFC capital grants ST/P002307/1 and ST/R002452/1 and STFC operations grant ST/R00689X/1.

This work was supported in part by the U.S.~Department of Energy, Office of Science, under Awards
No.~DE-SC0010005 (E.T.N. and J.W.S.),
No.~DE-SC0010120 (S.G.), 
No.~DE-SC0015655 (A.X.K., S.L., M.L., A.T.L.),
No.~DE-SC0009998 (J.L.),
the ``High Energy Physics Computing Traineeship for Lattice Gauge Theory'' No.~DE-SC0024053 (J.W.S.),
and the Funding Opportunity Announcement Scientific Discovery through Advanced Computing: High Energy Physics, LAB 22-2580 (D.A.C., C.T.P., L.H., M.L., S.L.); by the Exascale Computing Project (17-SC-20-SC), a collaborative effort of the U.S. Department of Energy Office of Science and the National Nuclear Security Administration (H.J.); by the National Science Foundation under Grants Nos.~PHY20-13064 and PHY23-10571 (C.D., D.A.C., S.L., A.V.), PHY23-09946 (A.B.),  and Grant No. 2139536 for Characteristic Science Applications for the Leadership Class Computing Facility (L.H., H.J.);
by the Simons Foundation under their Simons Fellows in Theoretical Physics program (A.X.K.); 
by the Universities Research Association Visiting Scholarship awards 20-S-12 and 21-S-05 (S.L.);  
by MICIU/AEI/10.13039/501100011033 and FEDER (EU) under Grant PID2022-140440NB-C21 (E.G.);
by Consejeria de Universidad, Investigaci\'on e Innovaci\'on and Gobierno de Espa\~na and EU--NextGenerationEU, under Grant AST22 8.4 (E.G.);
by AEI (Spain) under Grant No.\ RYC2020-030244-I / AEI / 10.13039/501100011033 (A.V.);
and by U.K. Science and Technology Facilities Council under Grant ST/T000945/1 (C.T.H.D).
A.X.K. and E.T.N. are grateful to the Pauli Center for Theoretical Studies and the ETH Z\"urich for support and hospitality. A.X.K, A.S.K, and E.T.N are grateful to the Kavli Institute for Theoretical Physics (KITP) for hospitality and support during the ``What is Particle Theory?'' program. The KITP is supported in part by the National Science Foundation under Grant PHY-2309135.
This document was prepared by the Fermilab Lattice, HPQCD, and MILC Collaborations using the resources of the Fermi National Accelerator Laboratory (Fermilab), a U.S. Department of Energy, Office of Science, HEP User Facility.
Fermilab is managed by Fermi Research Alliance, LLC (FRA), acting under Contract No.~DE-AC02-07CH11359.

\appendix

\section{Discretization effects at short-distances}
\label{sec:logEffects}

Discretization effects in the short-distance window---and therefore in any window starting at $t=0$---require a careful examination, because they explicitly include the first few time slices in $C(t)$.
For $a\ll t$ still, the Symanzik effective theory suggests~\cite{DellaMorte:2008xb,Ce:2021xgd,Alexandrou:2022amy,Sommer:2022wac}
\begin{equation}
    C_\text{lat}(t) = C(t)\left[1 + b_2 \frac{a^2}{t^2} + \cdots \right] .
\end{equation}
If $t$ is shorter than any hadronic or quark Compton wavelength, the correlator behaves as $C(t)\sim t^{-3}$ (by power counting), so the lattice artifact term has enhanced sensitivity to short times.

For present purposes, it is enough to consider a sharp cutoff on the window,
\begin{equation}
    {\amuSD}_\text{lat} = 4 \alpha^2 \int_a^T \tilde{K}(t) C_\text{lat}(t) \dd{t}.
\end{equation}
At short times, the QED kernel $\tilde{K}\sim t^4$, so in the end
\begin{equation}
    {\amuSD}_\text{lat} - \amuSD \sim b_2 a^2 \log(T/a) ,
\end{equation}
while the omitted terms (in this simple argument) are all of order~$a^2$.

This feature can be made more concrete by working out the lowest-order contribution to the vector-vector correlator in perturbation theory,
\begin{equation}
    C_V(t) = N_ca^3\third \sum_i \sum_{\bm{x}} V_i(x)S(x,0)V_i(0)S(0,y),
\end{equation}
where $N_c$ is the number of colors (3), $x=(\bm{x},t)$, $V_i$ is the vector current, and the position-space quark propagator $S(x,y)$ is obtained by Fourier transforming in the time variable.
Below, the combination
\begin{equation}
    \Sigma_4[C_V] = a\sum_{t/a=0}^\infty t^4 C_V(t)
    \label{eq:SigmaCV}
\end{equation}
is evaluated for unimproved~\cite{Kawamoto:1981hw,Sharatchandra:1981si} and improved staggered fermions~\cite{Naik:1986bn}, the latter being relevant to the asqtad and HISQ actions.

The details are tedious, so we do not show the full calculation but build up the result in steps.
The correlator is obtained by writing the quark propagator as a function of $(t,\bm{p})$, sewing two propagators together, and using the sum over $\bm{x}$
(integration over $\dd[3]x$ in the continuum) to set the three-momentum of the two quarks equal and opposite.
The first step picks up poles in the propagator, and the second is just a trace over Dirac matrices.
In the continuum, the result is
\begin{equation}
    C_V(t) = N_c \int \frac{\dd[3]{p}}{(2\pi)^3} \frac{\e^{-2|t|E}}{E^2} \left( m_0^2 + \third \bm{p}^2 + E^2 \right) ,
    \label{eq:Ccont}
\end{equation}
where $E^2=m_0^2+\bm{p}^2$.
It is possible to simplify this expression further, but it is written this way as a reminder of terms coming from the mass, $\gamma_i$, and $\gamma_0$ terms in the numerators of the quark propagators.
The time dependence is isolated, so it is convenient to integrate over time first,
\begin{equation}
    \int_0^\infty t^4 \e^{-2tE} \dd{t} = \frac{3}{4E^5}, 
    \label{eq:int}
\end{equation}
and then integrate over $\dd[3]{p}$ to obtain $\Sigma_4[C_V]$.

With staggered quarks, a similar structure emerges albeit with several technical changes.
The momentum integration is over the half-Brioullin zone, $\mathcal{B}_0=\{\bm{p}|-\pi/2\le p_ia<\pi/2\}$, as a consequence of how a Dirac-matrix structure emerges in momentum space~\cite{vandenDoel:1983mf}.
Because of doubling, if there is a pole at $p_0=\iunit E$, there is also one at $p_0=\iunit E+\pi$, leading to an oscillating contribution multiplying $(-1)^{|t|/a}$.
The relation between energy and momentum is more complicated, and in the integrand $\sinh Ea$ and $\sin p_ia$ appear.
The result for the \emph{local} current is
\begin{align}
    C_{V}(t) &= 4N_c \int_{\bm{\mathcal{B}}_0}\frac{\dd^3 p}{(2\pi)^3} \frac{a^2 \e^{-2|t|E}}{\cosh[2](Ea) \sinh[2](Ea)}
            \left[m_0^2 + \third \bm{S}^2 + \sinh[2](Ea)/a^2
        \right. \nonumber \\ & \hspace*{1em} + \left.
            (-1)^{|t|/a} \left(m_0^2 + \third \bm{S}^2 - \sinh[2](Ea)/a^2 \right) \right],
    \label{eq:Cstag}
\end{align}
where $aS_i(p)=\sin(p_ia)$, $\sinh^2(Ea)=a^2(\bm{S}^2+m_0^2)$ comes from finding the pole, and the factor of 4 corresponds to the four tastes in the loop.
Note how the structure is similar to \cref{eq:Ccont} while the sign of the energy term in the oscillating contribution changes.
The \emph{one-link} current is the same except that $m_0^2$ and $\sinh[2](Ea)$ are multiplied by $\third\sum_i\cos(p_ia)^2=1-\third a^2\bm{S}^2$, and $a^2\third\left[2\sum_iS_i^4-(\bm{S}^2)^2\right]$ is added to $\third\bm{S}^2$.

For improved staggered actions (asqtad and HISQ), the propagator has three poles in the upper half plane with real part close to zero, another three in the upper half plane with real part close to $\pi$, and then complex conjugate poles in the lower half plane.
They solve
\begin{equation}
    \sinh(Ea) \left[1-\sixth\sinh[2](Ea)\right] = a\Sh(\bm{p}) \equiv a\left[\tilde{\bm{S}}^2(\bm{p})+m_0^2\right]^{1/2} ,
    \label{eq:cubic}
\end{equation}
where $a\tilde{S}_i(\bm{p})=\sin(p_ia)[1+\sixth\sin[2](p_ia)]$.
As in any solution to a cubic equation, the expressions for the energies, which can be labeled $0$, $+$, and~$-$, involve (hyper)trigonometric functions and their inverses.
The most notable feature here is that the residue of one of the solutions, $-$, has opposite sign from the other two.
The result for the \emph{local} current is then
\begin{align}
    C_{V}(t) &= 4N_c \sum_{r,s} \int_{\bm{\mathcal{B}}_0}\frac{\dd^3 p}{(2\pi)^3} \frac{\e^{-|t|(E_r+E_s)}}{\Ch_r\Ch_s\Sh^2}
            \left[s_rs_sm_0^2 + \third s_rs_s\tilde{\bm{S}}^2 + \Sh^2
        \right. \nonumber \\ & \hspace*{1em} + \left.
            \e^{\iunit\pi t/a} \left(s_rs_sm_0^2 + \third s_rs_s\tilde{\bm{S}}^2 - \Sh^2 \right) \right],
    \label{eq:Cstag-Naik}
\end{align}
where $r,s$ range over $0,+,-$, $\Ch_r=\cosh(E_ra) \left[1-\half\sinh[2](E_ra)\right]$, and $s_-=-1$ while the others are $+1$.
The \emph{one-link} current is the same except that $m_0^2$ and $\Sh^2$ are multiplied by $\third\sum_i\cos(p_ia)^2=1-\third a^2\bm{S}^2$, and $a^2\third\left(2\sum_iS_i^2\tilde{S}_i^2-\bm{S}^2\tilde{\bm{S}}^2\right)$ is added to $\third\tilde{\bm{S}}^2$.

\begin{figure}
    \includegraphics[width=0.48\textwidth]{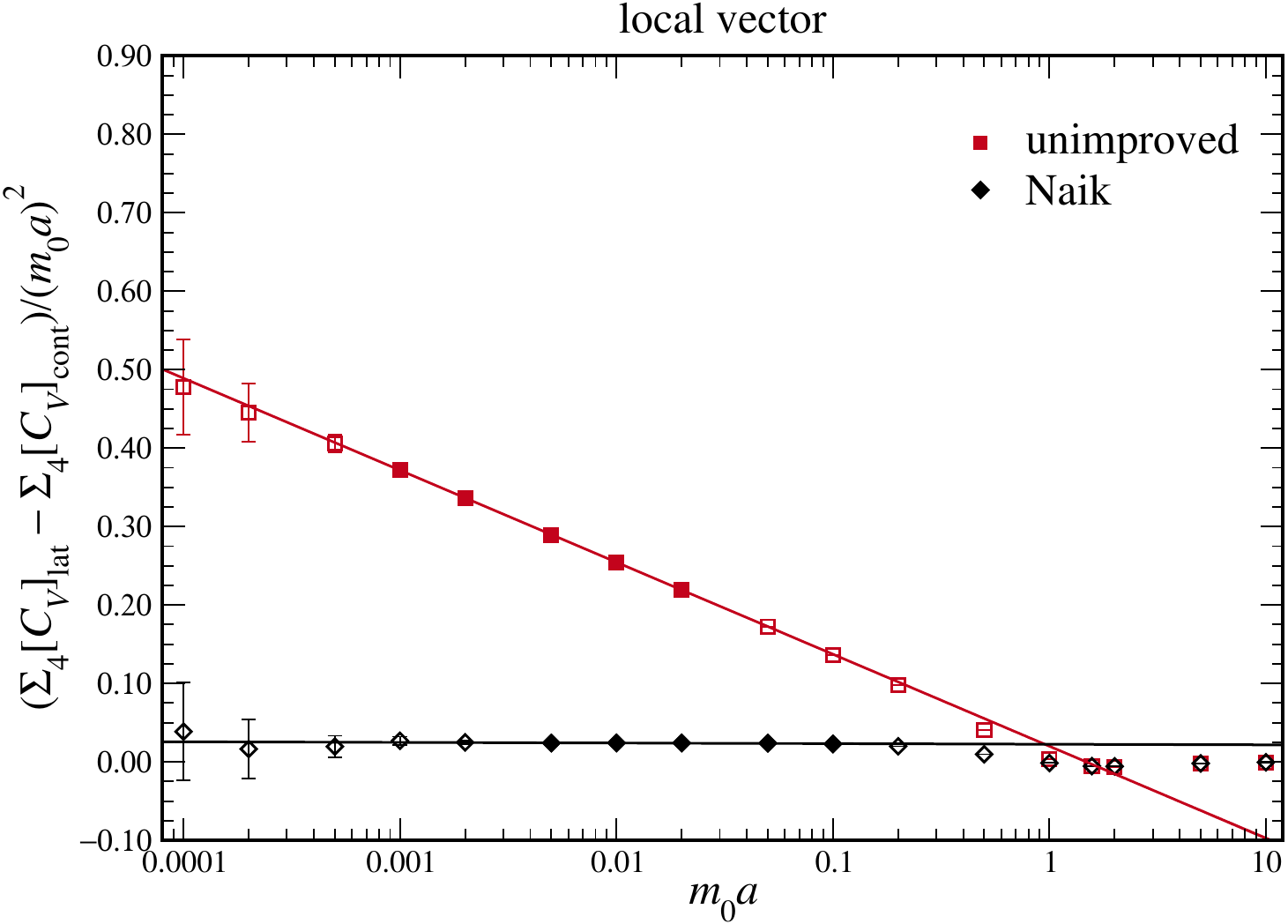} \hfill
    \includegraphics[width=0.48\textwidth]{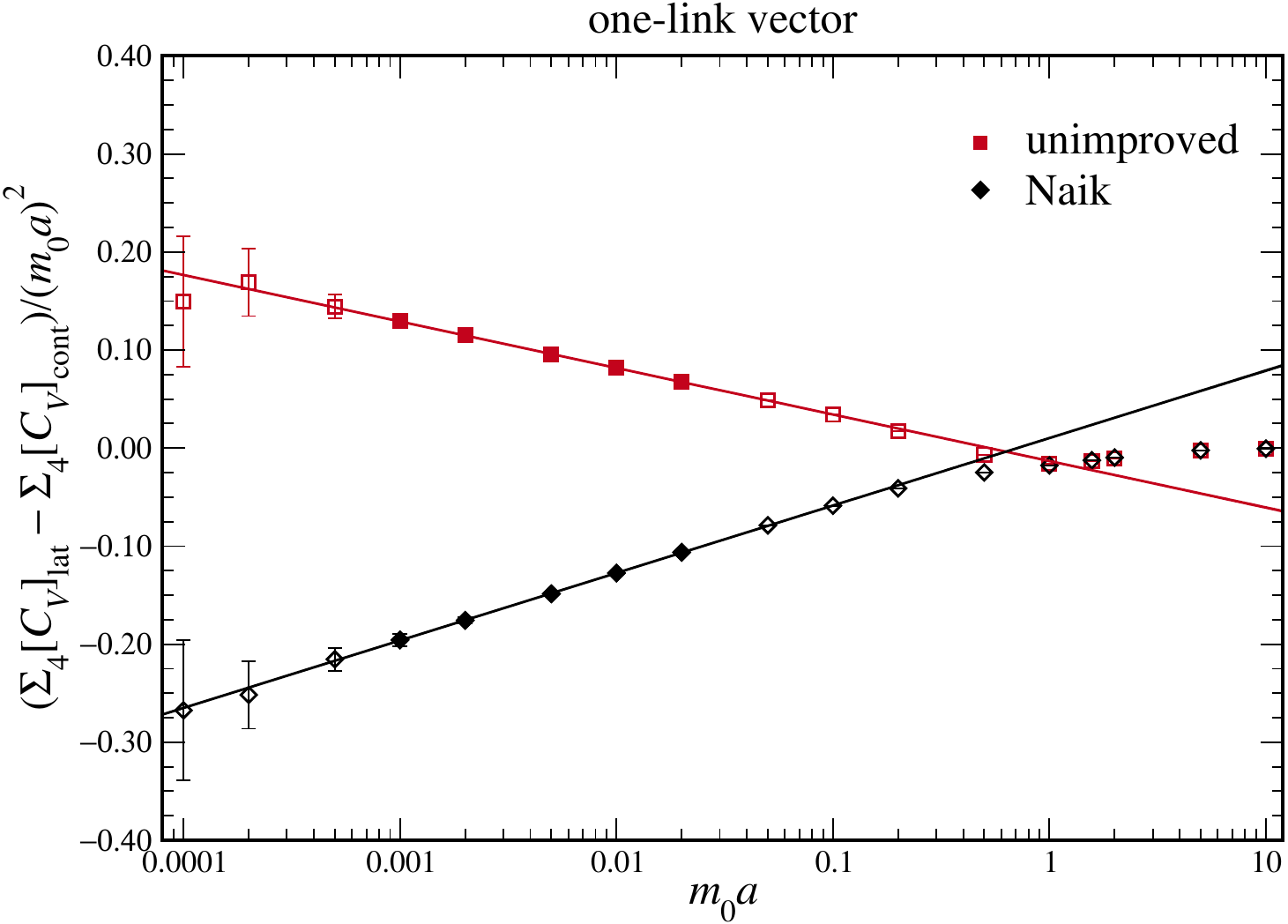}
    \caption[fig:ask:plots]{Lattice artifact versus\ $m_0a$ for the local (left) and one-link (right) vector correlator.
        Red points show the unimproved action and black, the Naik-improved case.
        The solid points are fit to the form $d+b\log m_0a$, yielding the lines.
        The approximate equality of unimproved and improved results for $m_0a>1$ is not understood but also not important.}
    \label{fig:ask:plots}
\end{figure}

In \cref{eq:Cstag,eq:Cstag-Naik}, the time dependence is again isolated.
Summing the exponentials over $t$ yields
\begin{align}
    a\sum_{t/a=0}^\infty t^4 \e^{-2tE} &=
        a^5\frac{\cosh(Ea) \{1 + \half \cosh[2](Ea)\}}{2\sinh[5](Ea)} ,
    \label{eq:sum} \\
    a\sum_{t/a=0}^\infty t^4 \e^{-2tE} (-1)^{t/a} &=
        a^5\frac{\sinh(Ea) \{1 - \half \sinh[2](Ea)\}}{2\cosh[5](Ea)} .
    \label{eq:alt}
\end{align}
The right-hand sides differ from the continuum by terms of order~$a^6$.

Combining \cref{eq:sum,eq:alt} with \cref{eq:Cstag,eq:Cstag-Naik} and dividing by 4 provides $\Sigma_4[C_V]_\text{lat}$, while combining \cref{eq:int} with \cref{eq:Ccont} provides $\Sigma_4[C_V]_\text{cont}$.
\Cref{fig:ask:plots} shows the quantities
\begin{equation}
    \frac{1}{(m_0a)^2} \left(\Sigma_4[C_J]_\text{lat} - \Sigma_4[C_J]_\text{cont} \right)
\end{equation}
for the local (left) and one-link (right) vector correlators.
This quantity is designed to behave as $d+b\log m_0a$ as $a\to0$, so a log-enhanced discretization effect appears as a significant slope on a log-linear plot.
Such effects are clearly present for the unimproved local current and the one-link current in any case.
The Naik term clearly removes the log-enhancement for the local current: the slope $b$ is much smaller than for the other cases and consistent with zero.

\section{Oscillating contributions in the short-distance window}
\label{sec:oscEffects}

Correlation functions constructed from staggered-quark operators, which are local in time, contain temporal oscillations from states of opposite parity. In the case at hand, the oscillating states are $1^{+-}$ states, which are heavier than the $1^{--}$ states, so their contribution to the correlator dies out after a few timeslices. This is observed in the $\amuHVP$ integrands for the different connected contributions we compute in this work; see \cref{fig:window06Data}. Depending on the relative sizes of the oscillating and non-oscillating state's energies and their couplings to the local-time staggered-quark operator, the oscillations can be quite pronounced relative to the size of the correlation function. In particular, we observe the charm integrand in \cref{fig:window06Data} to have much larger relative oscillation sizes than the light or strange. The oscillating contribution to the integrand and hence $\amu$ is, however, a discretization effect, with the expectation that the contribution in the continuum limit is zero.

In Ref.~\cite{FermilabLatticeHPQCD:2023jof}, we examined this expectation for the case of $\amuW$ and $\amuWTwo=a_\mu^{\mathrm{win}_2(1.5, 1.9, 0.15)}$. Our analysis procedure to remove the oscillating contribution involved fitting the light-quark connected correlation functions in the respective window regions. These fits then enabled a high-precision, high-fidelity reconstruction of the correlation function. Using the reconstruction, the observables were then computed with and without the oscillating contribution. In that work, we found a non-zero correlated difference for $\amuW$ between the oscillating and non-oscillating result on our coarsest two ensembles. On our two finest ensembles, the difference was statistically zero. We also found that the difference approached zero faster than the leading discretization effects in our continuum extrapolation fit function, namely $a^2 \alpha_s$. For $\amuWTwo$, as expected, we found that the difference was zero on all ensembles, as the oscillating state contributions had largely died off in the later time region. In our final Bayesian model averaged result of Ref.~\cite{FermilabLatticeHPQCD:2023jof}, we include model variations of $\amuHVP$ observables computed without the oscillating contribution, which we found to be consistent with the result that includes oscillations.

\begin{table}
\centering
\caption{Correlated differences between $\amu$ observables computed from the raw correlation functions and values computed from the fit reconstruction, including both oscillating and non-oscillating terms.}
\vspace{1mm}
\label{table:oscCorrDifferences}
\setlength{\tabcolsep}{6pt}
\newcommand{\mi}{\phantom{-}}
\begin{tabularx}{\linewidth}{LCCR}
\hline \hline
$\approx a$ (fm) & ~$\Delta \amuLSD$       & ~$\Delta \amuSSD$           & ~$\Delta \amuCSD$ \\ 
\hline
    0.15         & $\mi0.00055(52)$        &  $ - 7.7(8.9)\times10^{-8}$ & $\mi7(12)\times10^{-10}$   \\
    0.12         & $\mi0.000053(56)$       &  $ - 9(10)\times10^{-8}$    & $\mi1(13)\times10^{-10}$   \\
    0.09         & $ - 0.000012(56)$       &  $ - 0.0006(286)$           & $\mi1.8(3.4)\times10^{-8}$ \\
    0.06         & $\mi4(34)\times10^{-6}$ &  $\mi4.9(7.0)\times10^{-6}$ & $ - 3.7(4.2)\times10^{-8}$ \\
    0.04         & $ - 0.0018(60)$         &                             & $ - 0(15)\times10^{-5}$\\
    0.03         &                         &                             & $\mi0.00007(31)$           \\
\hline \hline
\end{tabularx}
\end{table}

For the short-distance window, it is not sensible to fit the earliest timeslices, so fit reconstructions of $\amuLSD$, $\amuSSD$, $\amuCSD$ cannot be included in the BMA. Instead, we construct a modified version of Eq.~(A3) from Ref.~\cite{FermilabLatticeHPQCD:2023jof},
\begin{align}
    C^\prime_{\textrm{no osc.}}(t) &= 
    \begin{cases}
        \sum^{N_{\textrm{states}}}_n Z^2_n e^{-E_n t} &\quad \textrm{  if $t/a > 1$,} \\
        C(t) &\quad \textrm{  if $t/a\le1$.} 
    \end{cases} \label{eqn:SDOscRecon}
\end{align}
The parameters $Z_n$ and $E_n$ come from a fit to the correlator using the Ans\"atze function in Eq.~(A2) of Ref.~\cite{FermilabLatticeHPQCD:2023jof}. We perform these fits for each flavor on all ensembles using $t_{\textrm{min}}/a =2$. We choose $t_{\textrm{max}}$ to be roughly 0.7~fm (the RHS of the SD window boundary plus $2\Delta$). We use $4+4$ states in our fits in all cases, except where increasing(decreasing) the number of non(-oscillating) states yields a smaller correlated difference between $\amu$ computed from the raw data and $\amu$ computed from the fit reconstruction with the oscillating contribution. For the light, strange, and charm fits, we select the lowest energy state prior from the PDG values of the $\rho$, $\phi$ and $J/\psi$ meson masses, respectively. The other prior selections follow from Ref.~\cite{FermilabLatticeHPQCD:2023jof}. The fit qualities, as measured by the $\chidof$, are all between $0.3$ and $1.3$. The correlated differences between the $\amu$ observables constructed from the raw correlator versus the correlator in (A2) in Ref.~\cite{FermilabLatticeHPQCD:2023jof} are given in \cref{table:oscCorrDifferences}. We find that they are all zero within uncertainties, indicating very good reproduction of the raw correlation function from the fit.

\begin{figure}
\centering
\includegraphics[scale=0.9]{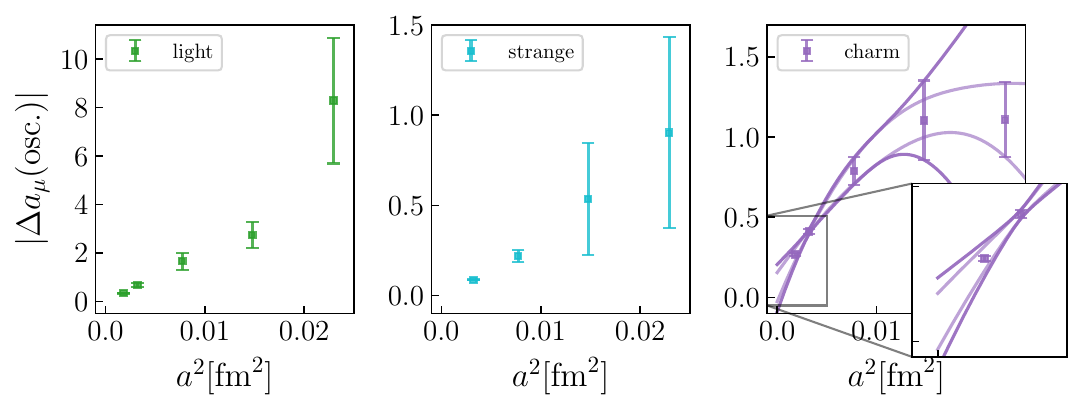}
\vspace{-5mm}
\caption{Plot of the correlated difference $\Delta a_\mu (\textrm{osc.})$ between $\amu$ computed from the raw correlation function and the modified non-oscillating correlation function of \cref{eqn:SDOscRecon} versus $a^2$.  This difference isolates the lattice-artifacts due to the oscillating contributions; 
we observe that they fall off rapidly compared to the leading discretization effects already included in our continuum fits. }
\label{fig:oscDiffLSC}
\end{figure}

The results from plugging
\begin{align}
    \Delta C_{\textrm{osc}}(t) = C(t) - C^\prime_{\textrm{no osc.}}(t) \label{eqn:DeltaNoOsc}
\end{align}
into \cref{eq:amuTintWin} for the short-distance window to obtain $| \Delta a_\mu (\textrm{osc.}) |$ are shown in \cref{fig:oscDiffLSC} for $\amuLSD$, $\amuSSD$, $\amuCSD$. We emphasize that due to not having a $t/a=1$ data point with the oscillating contribution, the curvature is reduced relative to the true $\| \Delta a_\mu (\textrm{osc.}) \|$ as it is a larger contribution at coarser spacings. For the light and strange contributions, we observe straightforwardly that the oscillating contribution falls off faster than the leading discretization terms in our continuum fits. For charm, this is not apparent for the physical ensembles used in this work. To resolve this, we have studied additional charm data generated on non-physical light quark ensembles~\cite{Hatton:2020qhk}, with lattice spacings of approximately $0.04$~fm and $0.03$~fm. With this extended dataset, we observe the oscillating contribution does indeed fall faster than the leading discretization terms in our continuum fit. 
In summary, we find that the discretization effects due to the oscillating contributions are small enough that they don't need to be explicitly removed before performing the continuum extrapolations. 

\section{Covariance and Bayesian model averaging} \label{sec:bma_cov}

Here we derive the formulas used to estimate covariance between different observables in the context of BMA.  We follow the notation of Ref.~\cite{Jay:2020jkz}.  This derivation is a simple generalization of the formula for the variance derived in that reference.  Our derivation of the BMA statistical covariance matches a similar result from Ref.~\cite{Davier:2023cyp}.

We consider two observables $a$ and $b$, which are specified over a joint space of $N_M$ models $\{M_i\}$.  The sample estimator of covariance between these two quantities is defined as
\begin{equation}
    \cov[a,b] = \ev{ab} - \ev{a}\ev{b},
\end{equation}
where $\ev{...}$ denotes the expectation over the posterior distribution. Each expectation value can be expanded out as a weighted average over the space of models \cite{Jay:2020jkz}:
\begin{equation}
    \cov[a,b] = \sum_{i=1}^{N_M} \ev{ab}_{i} \pr(M_i|D) - \sum_{i=1}^{N_M} \sum_{j=1}^{N_M} \ev{a}_i \ev{b}_j \pr(M_i | D) \pr(M_j | D),
\end{equation}
where $\pr(M_i | D)$ are the model weights as defined in \cref{sec:analysis}, and the expectation $\ev{a}_i$ denotes the expectation with respect to the specific model $M_i$.  Next, we manipulate this formula to expose the individual model covariances $\cov_i[a,b]$ with respect to each model $M_i$:
\begin{align}
    \cov[a,b] &= \sum_{i=1}^{N_M} ( \ev{ab}_i - \ev{a}_i \ev{b}_i) \pr(M_i | D) \nonumber \\
    &\quad + \sum_{i=1}^{N_M} \ev{a}_i \ev{b}_i \pr(M_i|D) -\sum_{i=1}^{N_M} \sum_{j=1}^{N_M} \ev{a}_i \ev{b}_j \pr(M_i | D) \pr(M_j | D) \\
    &= \sum_{i=1}^{N_M} \cov_i[a,b] \pr(M_i|D) + \sum_{i=1}^{N_M} \ev{a}_i \ev{b}_i \pr(M_i|D) - \ev{a} \ev{b}. \label{eq:bma_cov}
\end{align}
Similar to the analogous formula for the variance, the first term represents a weighted average of the ``statistical covariance" between $a$ and $b$ within each of the models, while the second two terms together encode a ``systematic covariance'' over the space of models. Note that the ``statistical covariance" also includes correlations from the use of shared parameters ({\it e.g.}, $M_\Omega$~(GeV), $aM_\Omega$, $Z_V$). 

The formula in \cref{eq:bma_cov} is fully general, but in the present context we are mainly interested in using it for the combination of different components of $\amu$.  Since these components are produced in separate analyses,
we can decompose the joint model space $\{M_i\}$ into the direct product of $\{M_i\} = \{M_{m_i}^A\} \times \{M_{n_i}^B\}$, where $\{M_m^A\},\{M_n^B\}$ represent the sets of models used in each analysis for estimating the individual quantities $a$ and $b$.  The double subscripts ``$m_i$'' and ``$n_i$'' denote the two models chosen from $\{M_m^A\}$ and $\{M_n^B\}$ that together comprise the joint model $M_i$.  Thus, \cref{eq:bma_cov} becomes
\begin{align}
    \cov[a,b]=& \sum_{m=1}^{N_{M^A}} \sum_{n=1}^{N_{M^B}} \cov_{mn}[a,b] \pr(M_m^{A},M_n^{B}|D) \nonumber\\
    &+ \sum_{m=1}^{N_{M^A}} \sum_{n=1}^{N_{M^B}} \ev{a}_m \ev{b}_n \pr(M_m^{A},M_n^{B}|D) - \ev{a} \ev{b}, \label{eq:bma_decompose}
\end{align}
where $N_{M^A}, N_{M^B}$ are the number of models in each respective model space, so that the total number of models is $N_M = N_{M^A} N_{M^B}$.

A useful way to rewrite this expression is to marginalize over one of the model spaces, giving a formula in terms of a model average over the other model space.  Without loss of generality, we average over the model space for $b$ by defining the quantities
\begin{align}
\cov_m[a,\bar{b}] &\equiv \sum_{n=1}^{N_{M^B}} \cov_{mn}[a,b] \frac{\pr(M_m^A, M_n^B | D)}{\pr(M_m^A | D)}, \label{eq:bma_cov_marg}\\
\ev{\bar{b}}_m &\equiv \sum_{n=1}^{N_{M^B}} \ev{b}_n \frac{\pr(M_m^A, M_n^B | D)}{\pr(M_m^A | D)}, \label{eq:bma_mean_marg}
\end{align}
where by definition
\begin{equation}
\pr(M_m^A | D) = \sum_{n=1}^{N_{M^B}} \pr(M_m^A, M_n^B | D),
\end{equation}
and thus rewrite \cref{eq:bma_decompose} as
\begin{equation}
    \cov[a,b]= \sum_{m=1}^{N_{M^A}} \cov_{m}[a,\bar{b}] \pr(M_m^{A}|D) 
    + \sum_{m=1}^{N_{M^A}} \ev{a}_m \ev{\bar{b}}_m \pr(M_m^{A}|D) - \ev{a} \ev{b}. \label{eq:bma_decompose_marg}    
\end{equation}
In thinking of this as a marginalization, the quantity $\ev{\bar{b}}_m$  may be counterintuitive; it is an average over the $\{M_n^B\}$ space, but with an index $m$ explicitly selecting a model from the other space $\{M_m^A\}$.  If one imagines doing a sequential model average in which $b$ is averaged over first, then the dependence of $\ev{\bar{b}}_m$ on the choice of model $M_m^A$ seems completely backwards.  The awkwardness of this notation is a consequence of the fact that, without additional simplifications, the idea of model averaging over $b$ first is not really well-defined; instead, one must consider the full model space in its entirety.  The choice of model $M_n^B$ influences the relative weight of different model choices in space $\{M_m^A\}$, altering the joint probability $\pr(M_m^A, M_n^B | D)$ and contributing to the covariance between $a$ and $b$ even in a sequential analysis where $b$ is estimated first.

An important simplifying case occurs when the dependence of the quantity $\ev{\bar{b}}_m$ on model space $\{M_m^A\}$ vanishes, {\it i.e.}, when $\ev{\bar{b}}_m = \ev{b}$.  In this case, the last two terms in \cref{eq:bma_decompose_marg} cancel exactly, leaving only the averaged statistical covariance.  A relevant case under which we will find $\ev{\bar{b}}_m = \ev{b}$ is when the model spaces are independent, so that $\pr(M_m^A, M_n^B | D) = \pr(M_m^A|D) \pr(M_n^B|D)$; this factorization immediately reduces \cref{eq:bma_mean_marg} to $\ev{b}$.  Other situations where $\ev{\bar{b}}_m = \ev{b}$ include when the model space $\{M_n^B\}$ is trivial, {\it i.e.}, when only one model with non-zero probability is present or when the expectation value $\ev{b}_n$ is independent of $n$.

In general, evaluation of the joint model probabilities $\pr(M_m^A, M_n^B | D)$ for all combinations across both model spaces is impractical.  In \cref{subsec:bma_cov_stat}, we explore how the further assumption of independence between the model spaces leads to relatively simple estimators for statistical and parametric covariance.  In \cref{subsec:bma_cov_syst}, we consider a conservative way to estimate systematic covariance terms in cases where they are present but difficult to estimate directly.  We adopt the latter estimate for the dominant source of systematic correlations in this work, namely that from shared EFT and EFT-inspired finite volume models.

\subsection{Independent model spaces and statistical covariance}\label{subsec:bma_cov_stat}

Here, we explore the consequences of the assumption of systematic independence between the two model spaces $\{M_m^A\}$ and $\{M_n^B\}$.  By systematic independence, we mean that the joint model weights factorize into the independent model weights calculated within the two separate analyses, {\it i.e.},
\begin{equation}\label{eq:indModels}
    \pr(M_i | D) = \pr(M_{m_i}^{A} | D) \pr(M_{n_i}^{B} | D).
\end{equation}
As noted above, this assumption reduces the marginalized $\ev{\bar{b}}_m$ defined in \cref{eq:bma_mean_marg} to the expectation value $\ev{b}$ over $\{M_n^B\}$, causing the systematic model-variation covariance terms to vanish.  The statistical and parametric correlations between $a$ and $b$ are retained in full.  In particular, \cref{eq:bma_cov_indep} immediately reduces to a simple average over statistical covariances:
\begin{equation} \label{eq:bma_cov_indep}
    \cov[a,b] = \cov^{\rm stat}[a,b] = \sum_{m=1}^{N_{M^A}} \sum_{n=1}^{N_{M^B}} \cov_{mn}[a,b] \pr(M_m^{A}|D) \pr(M_n^{B}|D).
\end{equation}
The individual model weights for $M_m^A$ and $M_n^B$ are estimated using the BAIC within each separate analysis, while the statistical covariances are estimated as described in the text.

For use in a series of model averages, it is useful to marginalize over one model space as in \cref{eq:bma_decompose_marg}. Under the assumption of independent model spaces, the marginalized covariance \cref{eq:bma_cov_marg} reduces to
\begin{align}\label{eq:bma_cov_marg_indep}
\cov_{m}[a,\overline{b}]=\cov_{m}[a,\ev{b}]\equiv\sum_{n=1}^{N_{M^B}}\cov_{mn}[a,b]\pr(M_n^{B}|D),
\end{align}
which allows us to rewrite the full statistical covariance as
\begin{equation} \label{eq:bma_cov_stat_indep}
\cov^{\rm stat}[a,b] = \sum_{m=1}^{N_{M^A}} \cov_{m}[a,\ev{b}] \pr(M_m^{A}|D),
\end{equation}
where $\cov_{m}[a,\ev{b}]$ are the statistical (and parametric) covariances between $\ev{a}_m$ (the individual model results from $\{M_m^A\}$) and $\ev{b}$ (the model average over $\{M_n^B\}$).

An important alternative case under which \cref{eq:bma_cov_stat_indep} holds is for parametric inputs in which case the model space $\{M_n^B\}$ can be viewed as consisting of only a single model.  In this case, the sum in \cref{eq:bma_cov_marg_indep} collapses, so that $\cov_m[a,\bar{b}] = \cov_m[a,b]$, the ordinary statistical covariance between $a$ and $b$ given model $M_m^A$.  
An alternate derivation of \cref{eq:bma_cov_stat_indep} first appeared in the Supplemental Material of Ref.~\cite{Bazavov:2024eou}.

\subsection{Systematic covariance}\label{subsec:bma_cov_syst}

Model independence, which leads to zero model-systematic covariance, is a reasonable assumption for most sources of systematic error between most observables; however, the shared use of EFT and EFT-inspired schemes among different observables may violate this assumption and lead to non-negligible systematic correlation between these observables.  Correlations associated with $M_{\pi}$ corrections are negligible since $\Delta_{M_{\pi}}$ is small in any case where these corrections are applied.  TB corrections do not lead to correlations as they are used only in the final results for one observable, namely $\amuDW$.  In contrast, FV corrections do lead to large systematic errors in multiple observables. Here, we derive a conservative upper bound on the systematic covariance associated with the shared use of FV correction schemes.

As argued above, the systematic covariance of interest entirely consists of that which comes from FV corrections, {\it i.e.},
\begin{align}
    \cov^{\rm syst}[a,b] = \cov^{\rm FV}[a,b].
\end{align}
In the current context, this covariance cannot be determined exactly as the joint model probabilities $\pr(M_i|D)$ are too cumbersome to compute.  Recall, however, that the covariance matrix is diagonally dominant, which means
\begin{align}\label{eq:corrFVBound}
    \abs{\cov^{\rm FV}[a,b]} \leq\sigma^{\rm FV}_a\sigma^{\rm FV}_b,
\end{align}
where the variances are evaluated in the usual way including appropriate factors of the model weights in each subset,
\begin{align}
    \sigma^{\rm FV}_a = \sqrt{\sum_{k=1}^{N_{S_{\mathrm{FV}}^A}}\ev{a}_{S_{\mathrm{FV},k}^A}^2\pr(S_{\mathrm{FV},k}^A|D)-\ev{a}_{\rm FV}^2},
\end{align}
with $\pr(S_{\mathrm{FV},k}^A|D)$ as defined in \cref{eqn:ssProb} and
\begin{align}
    \ev{a}_{\rm FV} &= \sum_{k=1}^{N_{S_{\mathrm{FV}}^A}}\ev{a}_{S_{\mathrm{FV},k}^A}\pr(S_{\mathrm{FV},k}^A|D),\\
    \ev{a}_{S_{\mathrm{FV},k}^A} &= \frac{1}{\pr(S_{\mathrm{FV},k}^A|D)}\sum_{M_m^A\in S_{\mathrm{FV},k}^A}\ev{a}_{m} \pr(M_m^A|D),
\end{align}
and similarly for $b$.  For emphasis, the subsets $\{S_{\mathrm{FV},k}^A\}$ partition the model space $\{M^A_m\}$ for a single observable into subsets (one for each FV correction scheme), which is distinct from the factorization of the joint model space for two observables as $\{M_i\}=\{M_{m_i}^A\}\times\{M_{n_i}^B\}$.  Note that in most cases $\ev{a}_{\rm FV}=\ev{a}$ as defined in \cref{eq:BMAMean}; the only exception is for $\amuDW$, where we average only over the variations without TB corrections, to disentangle these two effects.

In summary, what we have shown is
\begin{align}
\cov[a,b]&=\cov^{\rm stat}[a,b]+\cov^{\rm syst}[a,b]\\
&=\cov^{\rm stat}[a,b]+\cov^{\rm FV}[a,b]\\
&\leq \cov^{\rm stat}[a,b]+\abs{\cov^{\rm FV}[a,b]}\\
&\leq \cov^{\rm stat}[a,b]+\sigma^{\rm FV}_a\sigma^{\rm FV}_b.
\end{align}
$\cov^{\rm stat}[a,b]$ is as defined in \cref{eq:bma_cov_stat_indep}, and the additional term is used to account for the systematic correlations between $\amuLW$, $\amuDW$, and $\amuSIBCW$.  We conservatively assume that the covariance between our observables saturates this bound; this assumption is equivalent to assuming 100\% correlation between observables for the systematic error due to FV effects.

\subsection{Correlation matrices}\label{subsec:bma_corr}

\begin{table}
\centering
\caption{Correlation matrix for the individual contributions computed in this work in the short-distance window~SD.}
\vspace{1mm}
\label{table:correlationsSD}
\newcommand{\mi}{\phantom{-}}
\begin{tabularx}{\linewidth}{lcccccccccccc}
\hline\hline
{}         &  $\amuL$  &  $\amuS$  &  $\amuC$  &  $\amuD$ & $\amuSIBC$ & $\amuSIBD$ \\
\hline
$\amuL$    & $\mi1.00$ & $\mi0.19$ & $ - 0.01$ & $ \mi 0.00$ & $-0.01$ & $\mi0.00$  \\
$\amuS$    & $\mi0.19$ & $\mi1.00$ & $\mi0.17$ & $\mi0.00$ & $ \mi0.00$ & $\mi0.00$  \\
$\amuC$    & $ - 0.01$ & $\mi0.17$ & $\mi1.00$ & $\mi0.00$ & $ \mi 0.01$ & $\mi0.00$  \\
$\amuD$    & $ \mi 0.00$ & $\mi0.00$ & $\mi0.00$ & $\mi1.00$ & $  \mi0.00$ & $\mi0.00$  \\
$\amuSIBC$ & $ - 0.01$ & $ \mi0.00 $& $ \mi 0.01$ & $ \mi 0.00$ & $\mi1.00$ & $\mi0.00$  \\
$\amuSIBD$ & $\mi0.00$ & $\mi0.00$ & $\mi0.00$ & $\mi0.00$ & $\mi0.00$ & $\mi1.00$  \\
\hline
\hline
\end{tabularx}
\end{table}

\begin{table}
\centering
\caption{Correlation matrix for the individual contributions computed in this work in the intermediate-distance window~W.}
\vspace{1mm}
\label{table:correlationsW}
\begin{tabularx}{\linewidth}{lcccccccccccc}
\hline
\hline
{} &  $\amuL$ &  $\amuS$ &  $\amuC$ &  $\amuD$ &  $\amuSIBC$ &  $\amuSIBD$\\
\hline
$\amuL$   &      1.00    & 0.32  & 0.24 & 0.09  &   0.44   &  0.00 \\
$\amuS$   &    0.32   &  1.00  & 0.43 & 0.00   &  0.00   &  0.00 \\
$\amuC$ &     0.24    & 0.43  & 1.00 & 0.00  &   0.02    & 0.00\\
$\amuD$ &     0.09    & 0.00  & 0.00 & 1.00  &   0.11    & 0.06 \\
$\amuSIBC$ &   0.44   &  0.00  & 0.02 & 0.11  &   1.00    & 0.00 \\
$\amuSIBD$ &   0.00   &  0.00  & 0.00  &0.06   &  0.00    & 1.00 \\
\hline
\hline
\end{tabularx}
\end{table}

Using this formalism, our correlation matrices for $\amuSD$ and $\amuW$ are given in \cref{table:correlationsSD,table:correlationsW}, respectively.


\hfill

\bibliographystyle{apsrev4-2}
\bibliography{refs}

\end{document}